\documentstyle[preprint,prd,eqsecnum,aps,epsf]{revtex}

\tightenlines 
\begin{document}

    \newcommand{\DSC}{D\hspace{-0.25cm}\slash_{\bot}}
    \newcommand{\DSP}{D\hspace{-0.25cm}\slash_{\|}}   
    \newcommand{\DS}{D\hspace{-0.25cm}\slash}   
    \newcommand{\DC}{D_{\bot}}
    \newcommand{\DSCX}{D\hspace{-0.20cm}\slash_{\bot}}
    \newcommand{\DSPX}{D\hspace{-0.20cm}\slash_{\|}}  
    \newcommand{\DP}{D_{\|}}
    \newcommand{\QV}{Q_v^{+}}
    \newcommand{\QVB}{\bar{Q}_v^{+}}
    \newcommand{\QVP}{Q^{\prime +}_{v^{\prime}} }
    \newcommand{\QVBP}{\bar{Q}^{\prime +}_{v^{\prime}} }
    \newcommand{\QVHZ}{\hat{Q}^{+}_v}
    \newcommand{\QVHZB}{\bar{\hat{Q}}_v{\vspace{-0.3cm}\hspace{-0.2cm}{^{+}} } }    
    \newcommand{\QVPHZB}{\bar{\hat{Q}}_{v^{\prime}}{\vspace{-0.3cm}\hspace{-0.2cm}{^{\prime +}}} }    
    \newcommand{\QVPHFB}{\bar{\hat{Q}}_{v^{\prime}}{\vspace{-0.3cm}\hspace{-0.2cm}{^{\prime -}} } }    
    \newcommand{\QVPHB}{\bar{\hat{Q}}_{v^{\prime}}{\vspace{-0.3cm}\hspace{-0.2cm}{^{\prime}} }   }
    \newcommand{\QVHF}{\hat{Q}^{-}_v}
    \newcommand{\QVHFB}{\bar{\hat{Q}}_v{\vspace{-0.3cm}\hspace{-0.2cm}{^{-}} }}    
    \newcommand{\QVH}{\hat{Q}_v}
    \newcommand{\QVHB}{\bar{\hat{Q}}_v}
    \newcommand{\VS}{v\hspace{-0.2cm}\slash} 
    \newcommand{\MQ}{m_{Q}}
    \newcommand{\MQP}{m_{Q^{\prime}}}
    \newcommand{\QVHPMB}{\bar{\hat{Q}}_v{\vspace{-0.3cm}\hspace{-0.2cm}{^{\pm}} }}    
    \newcommand{\QVHMPB}{\bar{\hat{Q}}_v{\vspace{-0.3cm}\hspace{-0.2cm}{^{\mp}} }  }  
    \newcommand{\QVHPM}{\hat{Q}^{\pm}_v}
    \newcommand{\QVHMP}{\hat{Q}^{\mp}_v}

\draft
\preprint{AS-ITP-99-06}  
\title{ Weak Matrix Elements and $|V_{cb}|$  \\ in New Formulation of 
Heavy Quark Effective Field Theory}
\author{ W.Y. Wang, Y.L. Wu and Y.A. Yan  }
\address{Institute of Theoretical Physics, Chinese Academy of Sciences, Beijing 100080, China } 
\date{ylwu@itp.ac.cn}
\maketitle

\begin{abstract} 
  The heavy quark effective field theory (HQEFT) containing both effective `quark fields' and 
  `antiquark fields' is investigated in detail. By integrating out (but not neglecting) the 
  effective antiquark fields, 
  we present a new formulation of effective theory which differs from the usual heavy quark 
  effective theory (HQET) and exhibits 
  valuable features because of the inclusion of the contributions from the antiquark fields. 
  Matrix elements of vector and axial vector heavy quark 
  currents between pseudoscalar and vector mesons containing a heavy quark (b or c) are then evaluated 
  systematically up to the order of $1/m^2_Q$ and parameterized by a set of universal form factors. 
  With a consistent normalization condition between the effective heavy hadron states, 
  the form factors at zero recoil are related to the ground state meson masses,
  which enables us to estimate the values of form factors at zero recoil. In particular, the Luke's 
  theorem comes out automatically in the new formulation of HQEFT without the need of imposing the 
  equation of motion $iv\cdot D Q^{+}_{v} =0$. Consequently, the differential decay rates of both 
  $B\rightarrow D^{\ast}l\nu$ and $B\rightarrow D l\nu$ do not receive $1/m_{Q}$ order corrections at 
  zero recoil, 
  which is not the case in the usual HQET. 
  Thus we quote that the Cabibbo-Kobayashi-Maskawa matrix element $\vert V_{cb}\vert$ can nicely 
  be extracted from either of these two exclusive semileptonic decays at the order of $1/m^2_Q$. 
  Our estimates for $|V_{cb}|$ are presented.
\end{abstract}

\section{Introduction}\label{int}

     It is well known that, because of the heavy quark spin-flavor 
  symmetry \cite{shur,nuss,isgu} in the infinite quark mass limit revealed by QCD, 
 the dynamics of hadrons consisting of one heavy quark and any number of light quarks 
 simplifies greatly. Consequently, the effective theories for heavy quarks\cite{volo}-\cite{wise} 
  have developed rapidly since the observation of spin-flavor symmetry, 
and achieved great success. The effective theories manifestly exhibit the spin-flavor symmetry 
of hadrons containing a single heavy quark at the leading order.
This enables one to separate the long distance dynamics from the short one in a reliable way. 
In particular, the Luke's theorem \cite{luke} has shown that the order $1/{\MQ}$ corrections 
  to the transition matrix elements of heavy 
  meson decays vanish at zero recoil. These features make weak transitions 
  such as $B\rightarrow D^{\ast}l\bar{\nu}$ rather suitable for the 
  determination of the Cabibbo-Kobayashi-Maskawa matrix element 
  $\vert V_{cb}\vert$ \cite{neub}. 

    In the usual framework of heavy quark effective theory (HQET), one generally 
 decouples the `quark fields' and `antiquark fields' and treats only one of them 
 independently. Strictly speaking, in quantum field theory particle 
 and antiparticle decouple completely only in 
 the infinite heavy quark mass limit ${\MQ\rightarrow \infty}$. 
 To consider the finite quark mass correction, it is necessary to include the 
contributions from the components of the antiquark fields. For that, one can simply 
extend the usual HQET to a heavy quark effective field theory with keeping 
both effective quark and antiquark fields (which may be called for simplicity as HQEFT 
so as to distinguish with the usual HQET). This was actually first pointed out 
by one of us \cite{main}, where a new formulation of heavy 
  quark effective Lagrangian was derived from full QCD. Its form permits 
  an expansion in powers of the heavy quark momentum characterizing its 
  off-shellness divided by its mass.  In this paper we will provide a more detailed 
study on the new formulation of the HQEFT\cite{main} and apply it to evaluate the weak transition
matrix elements between the heavy hadrons containing a single heavy quark.  

Our paper is organized as follows: In section \ref{rev}, 
  based on the construction of a new formation of HQEFT in ref.\cite{main}, 
  and using the equations of motion, we derive an effective Lagrangian with integrating out 
  antiparticles.
  And the $1/{\MQ}$ expansion for an arbitrary heavy quark current 
  is also obtained. 
  Alternatively, these can be achieved through the approach of 
  the functional integral method, which is briefly presented in apprendix \ref{app:functional}.
  The functional integral method has also been adopted in 
  Ref.\cite{robe}. But the HQET in \cite{robe} only considered the 
  components of heavy `particle fields' without including `antiparticle fields'. 
  Though the new effective Lagrangian and current derived in section \ref{rev} of 
  this paper are in terms of only effective quark fields, they include additional 
  contributions from antiquark components, so that they differ from the usual HQET 
  counterparts. 
  In section \ref{mat}, within the new framework, a consistent normalization 
  condition between two heavy hadrons is introduced to reflect spin-flavor symmetry. 
  Transition matrix elements of heavy hadrons are generally investigated up to the 
  order of $1/{m^2_Q}$. 
  In section \ref{fac}, a set of universal form 
  factors are introduced and the transition matrix elements of the 
  vector and axial vector currents between pseudoscalar as well as vector 
  ground state mesons are evaluated in detail up to the order of $1/m^2_Q$. 
  In particular, it is explicitly shown that the Luke's 
  theorem automatically holds within the framework of the new formulation of HQEFT. 
  With the new normalization condition, it is found that the form factors at zero 
  recoil are related to the meson masses. The decomposition of some Lorentz tensors 
  used in this section are given in apprendix \ref{app:decomp}. 
  And apprendix \ref{app:factor} contains the general results for meson form factors.
In section \ref{vcb}, as the differential decay rates of both $B\rightarrow D^{\ast}l\nu$ 
  and $B\rightarrow D l\nu$ have no $1/m_{Q}$ order corrections at zero recoil in the new 
formulation of HQEFT, and the most important relevant form factors at zero recoil 
can be fitted from the ground state meson masses, the 
  Cabibbo-Kobayashi-Maskawa matrix element $\vert V_{cb}\vert$ are extracted from 
  these two exclusive semileptonic decays at the order of $1/m^2_Q$. 
  A brief summary is presented in section \ref{sum}. 
  In this paper we restrict our discussions to the tree level.

  \section{HQEFT with Keeping Both Particles and Antiparticles}\label{rev}

  We now briefly describe the new formulation of heavy quark effective field Lagrangian 
that contains both effective quark and antiquark fields\cite{main}.
 Firstly,  denote the heavy quark field as
     \begin{equation}
     \label{eq:7}
       Q=Q^{+}+Q^{-} ,
     \end{equation}
  with $Q^{+}$ and $Q^{-}$ the components of `quark field' and `antiquark 
  field' respectively. More strictly speaking, they are corresponding to 
  the two solutions of the Dirac equation.
  
  Defining
    \begin{equation}
    \label{eq:9}
      \hat{Q}^{\pm}_v\equiv \frac{1{\pm}v\hspace{-0.2cm}\slash}{2}Q^{\pm},\hspace{1.5cm}
      R^{\pm}_v\equiv \frac{1{\mp}v\hspace{-0.2cm}\slash}{2}Q^{\pm}
    \end{equation} 
  with $v^{\mu}$ an arbitrary four-vector satisfying $v^2=1$, 
  the original fields $Q$ and $\bar{Q}$ can be expressed by the new field 
  variables $\hat{Q}_v$ and $\bar{\hat{Q}}_v$,
     \begin{equation}
     \label{eq:11}
       \left\{
        \begin{array}{l}
         Q=[1+(1-\frac{i\DSPX+\MQ}{2\MQ})^{-1}
           \frac{i\DSCX}{2\MQ}] \hat{Q}_{v}\equiv\hat{\omega} \hat{Q}_v ,\\
         \bar{Q} = \bar{\hat{Q}}_{v}[1+\frac{-i\stackrel{\hspace{-0.1cm}\leftarrow}
         {\DSCX}}{2\MQ}(1-\frac{-i\stackrel{\hspace{-0.1cm}\leftarrow}
         \DSPX+\MQ}{2\MQ})^{-1}]\equiv \bar{\hat{Q}}_v\stackrel{\leftarrow}
           {\hat{\omega}}.
         \end{array}
        \right.  
     \end{equation}
  The QCD Lagrangian becomes
    \begin{equation}
    \label{eq:13}
       L_{QCD} = L_{light}+L_{Q,v} ,
    \end{equation}
  where $L_{light}$ represents the part of Lagrangian containing no heavy 
  quarks, and 
     \begin{equation}
     \label{eq:15}
        L_{Q,v} = L^{(++)}_{Q,v}+L^{(--)}_{Q,v}+L^{(+-)}_{Q,v}+L^{(-+)}_{Q,v}
     \end{equation}
  with 
    \begin{eqnarray}
    \label{eq:16}
      L^{(\pm \pm)}_{Q,v} &=& \QVHPMB [i\DSP-\MQ 
           + \frac{1}{2\MQ}i\DSC (1-\frac{i\DSP+\MQ}{2\MQ})^{-1}
           i\DSC ] \QVHPM 
        \equiv  \QVHPMB \hat{A} \QVHPM \nonumber \\ 
      L^{(\pm \mp)}_{Q,v} &=& \QVHPMB [ \frac{1}{2\MQ}i\DSC (1-\frac{i\DSP
         +\MQ}{2\MQ})^{-1}(i\DSP-\MQ)
         +\frac{1}{4\MQ^2}(-i\stackrel{\hspace{-0.2cm}\leftarrow}{\DSC}) \nonumber\\
         &&\times (1-\frac{-i\stackrel{\hspace{-0.1cm}\leftarrow}{\DSP}+\MQ}{2\MQ})
         ^{-1}i\DSC  (1-\frac{i\DSP+\MQ}{2\MQ})^{-1}
         i\DSC ] \QVHMP \nonumber \\
         &\equiv & \QVHPMB \hat{B}  \QVHMP .
    \end{eqnarray}
  $\stackrel{\hspace{-0.1cm}\leftarrow}{D^\mu}$, $\DSP$ and $\DSC$ are 
  defined as
      \begin{equation}
      \label{eq:12}
       \left\{
        \begin{array}{l}
           \int\kappa\stackrel{\hspace{-0.1cm}\leftarrow}{D^\mu}\varphi 
   \equiv -\int\kappa D^{\mu}\varphi, \\
           \DSP \equiv v\hspace{-0.2cm}\slash(v\cdot D), \\
           \DSC \equiv \DS-\VS (v\cdot D).
       \end{array}
         \right.
      \end{equation}
    
  When quark fields and antiquark fields decouple completely, it is reasonable 
  to deal with only section $L^{(++)}_{Q,v}$ or $L^{(--)}_{Q,v}$ independently. 
  This is just the case considered in the framework of the usual HQET. In this paper, 
  we will consider the complete effective lagrangian and investigate its possible 
  new effects on the physical observables.
 
    Note that $L_{Q,v}$ in eq.(\ref{eq:15}) accounts 
  for only one flavor of heavy quarks moving with velocity $v^{\mu}$. 
  If there are $N_f$ flavors of heavy quarks and they move in different velocities, 
  the heavy quark Lagrangian should generalize to $\sum_{Q} L_{Q,v_Q}$. But in this paper 
  we use $L_{Q,v}$ for simplicity. It should also be mentioned that 
  the effective Lagrangian eq.(\ref{eq:15}) as shown in ref. \cite{main} is automatically 
  velocity reparametrization invariant and Lorentz invariant.

  The Lagrangian given by eqs.(\ref{eq:15}) and (\ref{eq:16}) contains 
  both quark fields and antiquark fields manifestly. 
  But in many processes it is reasonable and more convenient to deal with either 
  quarks or antiquarks since the initial and final states contain no antiquarks
  or quarks. For this reason, we will derive an effective 
  Lagrangian which on one hand contains only heavy quark fields, and on the other hand, 
  takes the contributions of antiquarks into account. Two methods can achieve 
  this. One is to replace the antiquark fields by quark fields with the help 
  of the equations of motion, and the other is to apply the functional integral 
  method to `integrate out' the antiquark fields. 

  It is much simpler to do that through the equations 
  of motion. The following equations of motion can be easily yielded from 
  eqs.(\ref{eq:15}) and (\ref{eq:16}),
  \newcommand{\LEFT}[1]{\stackrel{\leftarrow}{\hat{#1}}}
  \begin{eqnarray}
  \label{eq:new1}
    &\hat{A}\QVHF + \hat{B} \QVHZ =0 & \;\;\;\QVHFB \LEFT{A} + \QVHZB \LEFT{B} = 0 , \nonumber\\
    \mbox{or:}\;\;\;&  \QVHF=-(\hat{A})^{-1} \hat{B} \QVHZ & \;\;\; \QVHFB=-\QVHZB \LEFT{B} (\LEFT{A})^{-1},
  \end{eqnarray}
  where $\LEFT{A}$ and $\LEFT{B}$ are just $\hat{A}$ and $\hat{B}$ but with 
  replacing $D^{\mu}$ by $-\stackrel{\hspace{-0.1cm}\leftarrow}{D^{\mu}}$. 
  By inserting $\QVHF$ and $\QVHFB$ in eq.(\ref{eq:new1}) into eq.(\ref{eq:16}), 
  we then arrive at 
   \begin{eqnarray}
   \label{eq:20}
      L^{(++)}_{eff}&=&\QVHZB(\hat{A}-\hat{B}\hat{A}^{-1}\hat{B})\QVHZ \nonumber\\
      &=&\QVHZB \{i\DSP-\MQ+\frac{1}{2\MQ}i\DSC (1-\frac{i\DSP+\MQ}{2\MQ})^{-1}i\DSC 
        \nonumber \\
      && -[i\stackrel{\hspace{-0.1cm}\leftarrow}{\DSC}+\frac{1}{4\MQ^2}
(-i\stackrel{\hspace{-0.1cm}\leftarrow}
      {\DSC})(1-\frac{-i\stackrel{\hspace{-0.1cm}\leftarrow}{\DSP}+\MQ}{2\MQ})^{-1}(-i\stackrel
      {\hspace{-0.1cm}\leftarrow}{\DSC}) (1-\frac{-i\stackrel{\hspace{-0.1cm}
\leftarrow}{\DSP}+\MQ}{2\MQ})^{-1}\nonumber\\
      &&\times(-i\stackrel{\hspace{-0.1cm}\leftarrow}{\DSC})]{[i\DSP-\MQ+\frac{1}{2\MQ}i\DSC 
      (1-\frac{i\DSP+\MQ}{2\MQ})^{-1}i\DSC]}^{-1} \nonumber \\
      &&\times [-i\DSC+\frac{1}{4\MQ^2}i\DSC (1-\frac{i\DSP+\MQ}{2\MQ})^{-1}i\DSC
      (1-\frac{i\DSP+\MQ}{2\MQ})^{-1}i\DSC]\}\QVHZ .
    \end{eqnarray}

  One can make a further transformation to eliminate the mass terms in 
  eq.(\ref{eq:20}) through introducing new field variables $Q_v$ and $\bar{Q}_v$:
    \begin{equation}
    \label{eq:21}
       Q_v=e^{iv\hspace{-0.15cm}\slash \hat{m}_Q v\cdot x}\QVH ,\hspace{1.5cm}
       \bar{Q}_v=\QVHB e^{-i v\hspace{-0.15cm}\slash \hat{m}_Q v \cdot x} .
    \end{equation}  
  Since eq.(\ref{eq:21}) is nothing more than a field variable redefinition, $\hat{m}_Q$ 
  may be any parameter with mass dimension. Here we write for convenience 
     \begin{equation}
     \label{eq:22}
       \hat{m}_Q \equiv \MQ+\Lambda
     \end{equation}
 It will be found to be useful for taking 
 $\Lambda$ as a binding energy $\bar{\Lambda}$ 
  which is independent of the heavy flavors and reflects 
  the contributions of all light degrees of freedom in the hadron. 
    
  Noticing the fact that $\VS$ commutes with $\DSP$ but anticommutes with $\DSC$  ,
  it is not difficult to construct a new form of Lagrangian in terms of $\QV$ and 
  $\QVB$.
  When the mass of a heavy quark is much larger than the QCD scale $\Lambda_{QCD}$, 
this effective Lagrangian can be expanded in inverse powers of 
the quark mass and be straightforwardly written as follows
    \begin{eqnarray}
    \label{eq:24}
      L^{(++)}_{eff}&=&\QVB \{i\DSP+\Lambda+\sum\limits^{\infty}_{n=1}\frac{1}{(2\MQ)^n}
        i\DSC (i\DSP-\Lambda)^{n-1}i\DSC \} \frac{1}{i\DSP+\Lambda}
        \{i\DSP+\Lambda \nonumber \\
       &&+\sum\limits^{\infty}_{l=1}\frac{1}{(2\MQ)^l}
        i\DSC (i\DSP-\Lambda)^{l-1}i\DSC \}\QV 
       \equiv L^{(0)}_{eff}+L^{(1/\MQ)}_{eff}  
    \end{eqnarray}
  with
    \begin{eqnarray}
    \label{eq:25}
      L^{(0)}_{eff} & \equiv & \QVB(i\DSP+\Lambda)\QV ,  \\
    \label{eq:26}
      L^{(1/\MQ)}_{eff} & \equiv & \frac{1}{\MQ}\QVB(i\DSC)^2\QV + 
   \frac{1}{2\MQ^2}\QVB i\DSC (i\DSP-\Lambda)i\DSC \QV \nonumber \\
           &&+ \frac{1}{4\MQ^2}\QVB (i\DSC)^2 \frac{1}{i\DSP+\Lambda}(i\DSC)^2 \QV 
           +O(\frac{1}{\MQ^3}) , 
    \end{eqnarray}  
  where $L^{(0)}_{eff}$ possesses spin-flavor symmetry, whereas $L^{(1/\MQ)}_{eff}$ 
  contains symmetry breaking terms and suppressed by the $1/m_{Q}$. 
Note that $L^{(++)}_{eff}$ also automatically 
preserves the velocity reparametrization invariance as well as
 Lorentz invariance without the need of summing over the velocity.
 
 As the resulting
  improved effective Lagrangian eq.(\ref{eq:24}) describe interactions concerning `quark fields', 
  it is convenient to be applied to physical processes in which 
the initial and final states contain only quarks. 
  As quark and antiquark appear simultaneously in any full quantum field theory, 
  a similar effective Lagrangian $L^{(--)}_{eff}$ for antiquark fields can be 
  simply obtained by replacing the effective quark field variables $\QV$ and 
  $\QVB$ in eq.(\ref{eq:24}) with the effective antiquark field variables $Q^{-}_v$ and 
  $\bar{Q}^{-}_v$. 
 
   The effective heavy quark currents can also be derived easily via the $1/{\MQ}$ 
  expansion approach. The general form of a heavy quark current may be written as 
    \begin{equation}
    \label{eq:28}
       J(x)=\bar{Q}^{\prime}(x)\Gamma  Q(x)
    \end{equation}
  with $\Gamma$ an arbitrary Dirac matrix. Using eq.(\ref{eq:7}) this current 
  can be decomposed into four parts,
    \begin{eqnarray}
    \label{eq:29}
       J=\bar{Q}^{\prime +}\Gamma Q^{+}+\bar{Q}^{\prime +}\Gamma Q^{-}+
       \bar{Q}^{\prime -}\Gamma Q^{+}+\bar{Q}^{\prime -}\Gamma Q^{-} ,
    \end{eqnarray}
where $Q^{\pm}$ are related to $\QVHPM$ by eq.(\ref{eq:11}), 
whereas $\QVHZ$ and $\QVPHZB$ are related to new variables $\QV$ and $\QVBP$ by 
eq.(\ref{eq:21}).

  The equations of motion in eq.(\ref{eq:new1}) can be used to eliminate the field 
  variables $\QVHF$ and \vspace{0.35cm} $\QVHFB$. The four parts in eq.(\ref{eq:29}) 
  then get the following form in the framework of the new formulation of HQEFT 
    \begin{eqnarray}
    \label{eq:31}
       <\bar{Q}^{\prime +}\Gamma Q^{+}>&=&\QVPHZB 
           \stackrel{\leftarrow}{\hat{\omega}} \Gamma\hat{\omega} \QVHZ , \nonumber\\
       <\bar{Q}^{\prime +}\Gamma Q^{-}>&=&\QVPHZB 
           \stackrel{\leftarrow}{\hat{\omega}} \Gamma\hat{\omega} 
           \QVHF =\QVPHZB \stackrel{\leftarrow}{\hat{\omega}}
           \Gamma\hat{\omega} (-\hat{A}^{-1} \hat{B}) \QVHZ , \nonumber\\
        <\bar{Q}^{\prime -}\Gamma Q^{+}>&=&\QVPHFB
          \stackrel{\leftarrow}{\hat{\omega}} \Gamma \hat{\omega} \QVHZ   
           =\QVPHZB (-\LEFT{B} (\LEFT{A})^{-1})\stackrel
           {\leftarrow}{\hat{\omega}} \Gamma\hat{\omega} \QVHZ , \nonumber\\
       <\bar{Q}^{\prime -}\Gamma Q^{-}>&=&\QVPHFB
           \stackrel{\leftarrow}{\hat{\omega}} \Gamma
           \hat{\omega} \QVHF
           =\QVPHZB (-\LEFT{B} (\LEFT{A})^{-1})\stackrel
           {\leftarrow}{\hat{\omega}} \Gamma
           \hat{\omega} (-\hat{A}^{-1} \hat{B}) \QVHZ  .
    \end{eqnarray}       
  The operators $\hat{\omega}$, $\stackrel{\leftarrow}{\hat{\omega}}$ and 
  $\hat{A}$, $\hat{B}$ have been defined in eqs.(\ref{eq:11}) and (\ref{eq:16}). 
  Here we have used the symbol $<\hat{j}>$ to represent the effective 
  operator of $\hat{j}$ within the framework of the new formulation of HQEFT. 

  Expanding the formulae in eq.(\ref{eq:31}) in powers of $1/{\MQ}$ and summing them up, one then obtains 
  the effective heavy quark current
    \begin{eqnarray}
    \label{eq:34}
      J^{(++)}_{eff} &\equiv &<\bar{Q}^{\prime} \Gamma Q > = e^{i(\hat{m}_{Q^{\prime}} v^{\prime}
        -\hat{m}_Q v)\cdot x}
       \{ \QVBP \Gamma \QV+\frac{1}{2\MQ}\QVBP\Gamma\frac{1}{i\DSP+\Lambda}(i\DSC)^2 \QV  \nonumber\\
        &&+\frac{1}{2\MQP}\QVBP (-i\stackrel{\hspace{-0.1cm}\leftarrow}{\DSC})^2\frac{1}
        {-i\stackrel{\hspace{-0.1cm}\leftarrow}{\DSP}+\Lambda}\Gamma\QV+\frac{1}{4\MQ^2}\QVBP 
        \Gamma\frac{1}{i\DSP+\Lambda}
        i\DSC (i\DSP-\Lambda)  \nonumber\\
        &&\times i\DSC \QV 
        +\frac{1}{4\MQP^2}\QVBP (-i\stackrel{\hspace{-0.1cm}\leftarrow}{\DSC})
        (-i\stackrel{\hspace{-0.1cm}\leftarrow}{\DSP}-\Lambda)(-i\stackrel{\hspace{-0.1cm}\leftarrow}{\DSC})
        \frac{1}{-i\stackrel{\hspace{-0.1cm}\leftarrow}{\DSP}+\Lambda}\Gamma \QV \nonumber\\
        &&+\frac{1}{4\MQP \MQ}\QVBP (-i\stackrel{\hspace{-0.1cm}\leftarrow}{\DSC})^2 
        \times \frac{1}{-i\stackrel{\hspace{-0.1cm}\leftarrow}{\DSP}+\Lambda}\Gamma     
        \frac{1}{i\DSP+\Lambda}(i\DSC)^2 \QV
        +O(\frac{1}{m_{Q^{(\prime)}}^3}) \} \nonumber\\
        & \equiv & J^{(0)}_{eff}+J^{(1/m_Q)}_{eff}
    \end{eqnarray}
  with $J^{(0)}_{eff}$ the leading term $J^{(0)}_{eff}=e^{i(\hat{m}_{Q^{\prime}} v^{\prime}
  -\hat{m}_Q v)\cdot x}\QVBP \Gamma \QV $ and $J^{(1/m_Q)}_{eff}$ the remaining 
  terms in $J^{(++)}_{eff}$. 
  In appendix \ref{app:functional} we also present the derivation of $L^{(++)}_{eff}$ and $J^{(++)}_{eff}$ 
  from the functional integral method. We see that the approaches of 
  the functional integral and the equations of motion are indeed equivalent. 
    
  One can see from eqs.(\ref{eq:15}), (\ref{eq:16}) and (\ref{eq:21}) 
  that, if we neglect the antiquark contributions and choose $\Lambda = 0$, the 
  Lagrangian becomes 
    \begin{eqnarray}
    \label{eq:5}
      L_{eff}&=&\bar{Q}(i\DS-\MQ)Q=\bar{Q}_v iv\cdot D Q_v + \bar{Q}_vi\DSC 
             \frac{1}{2\MQ +iv\cdot D} i\DSC Q_v   \nonumber\\
    &=&\bar{Q}_v iv\cdot D Q_v + \frac{1}{2\MQ}\bar{Q}_v (i\DSC)^2 Q_v  
          -\frac{1}{4\MQ^{2}} \bar{Q}_v i\DSC iv\cdot D i\DSC Q_v +O(\frac{1}{\MQ^3}).
    \end{eqnarray}
  which is just the Lagrangian used in the usual HQET, 
  whose starting point is separating the quark field $Q$ into `large' 
  and `small' component fields $Q_v$ and $\chi_v$ as follows, 
    \begin{equation}
    \label{eq:1}
      Q(x)=e^{-i\MQ v \cdot x}[Q_v(x)+\chi_v(x)]  ,
    \end{equation}
  where
    \begin{equation}
    \label{eq:2}
      Q_v(x) \equiv  e^{i\MQ v\cdot x}\frac{1+\VS}{2}Q(x) ,\hspace{1.5cm}
      \chi_v(x) \equiv  e^{i\MQ v\cdot x}\frac{1-\VS}{2}Q(x).
    \end{equation} 
  In this case, the arbitrary current $\bar{Q}^{\prime}\Gamma Q$ turns into the form in the usual HQET, 
  i.e,
    \begin{eqnarray}
    \label{eq:6}
      J&=&\bar{Q}^{\prime}\Gamma Q=e^{i(\MQP v^{\prime}-\MQ v)\cdot x}
      \{\bar{Q}^{\prime}_{v^{\prime}}\Gamma Q_v
      +\frac{1}{2\MQ}\bar{Q}^{\prime}_{v^{\prime}}\Gamma i\DSC Q_v
      +\frac{1}{2\MQP}\bar{Q}^{\prime}_{v^{\prime}}
      (-i\stackrel{\hspace{-0.1cm}\leftarrow}{\DSC})\Gamma Q_v \nonumber\\
 && -\frac{1}{4\MQ^2}\bar{Q}^{\prime}_{v^{\prime}}\Gamma (iv\cdot D)i\DSC Q_v 
      -\frac{1}{4\MQP^2}\bar{Q}^{\prime}_{v^{\prime}}
      (-i\stackrel{\hspace{-0.1cm}\leftarrow}{\DSC})
      (-iv^{\prime}\cdot \stackrel{\hspace{-0.1cm}\leftarrow}{D})\Gamma Q_v\nonumber\\
&&   +\frac{1}{4\MQP \MQ}\bar{Q}^{\prime}_{v^{\prime}}
      (-i\stackrel{\hspace{-0.1cm}\leftarrow}{\DSC})\Gamma i\DSC Q_v +O(\frac{1}{m_{Q^{(\prime)}}^3})\} .
    \end{eqnarray}
  Clearly, in this usual framework of HQET only quark fields appear. The antiquarks have 
  been omitted from the starting point in eqs.(\ref{eq:1}) and (\ref{eq:2}).
    
  The above forms of the effective heavy quark Lagrangian (eq.(\ref{eq:24})) 
  and effective current (eq.(\ref{eq:34})) are quite different from that in the usual HQET. 
 Such differences arise from the additional contributions of antiquark 
 fields in the new formulation of HQEFT. As is 
  expected, antiquarks give contributions 
  to the current from the order of $1/{\MQ}$. Especially, the terms 
  $\frac{1}{2\MQ}\QVBP \Gamma i\DSC \QV$ and $\frac{1}{2\MQP}\QVBP 
  (-i\stackrel{\hspace{-0.15cm}\leftarrow}{\DSC})\Gamma \QV$ appearing in the 
  usual HQET current eq.(\ref{eq:6}) are absent. This is because they are exactly 
  cancelled by the additional contributions arising from the intermediate antiquark 
  fields. As a consequence, the operator forms in the Lagrangian $L^{(++)}_{eff}$ 
  (eq.(\ref{eq:24})) and those in the effective current (eq.(\ref{eq:34})) become similar. 
  Such an interesting feature resulting from the new formulation of HQEFT
 becomes remarkable in evaluating the hadronic matrix elements.  Firstly, 
 fewer form factors are needed in this framework than in the usual 
  framework of HQET, as will be shown explicitly in section \ref{fac}. Secondly, the Luke's theorem 
  automatically holds without using the equation of motion $iv\cdot D Q^{+}_v =0$. Thirdly, 
  the differential decay rate of the exclusive semileptonic $B\rightarrow D l \nu$ decay 
  receives no contribution from the $1/m_{Q}$ order at zero recoil, which is not the case in the 
  usual HQET. 

  \section{Weak Transition Matrix Elements}\label{mat}

  In the framework of the effective theory, the hadronic matrix element in full QCD theory 
  can be expanded into a series of matrix elements in terms of $1/{\MQ}$. 
  We will study in this section the hadronic matrix element up to the order 
  of $1/{m_Q^2}$.
  
  Using the following conventional relativistic normalization
      \begin{equation}
      \label{eq:35}
         <H(p^{\prime})\vert H(p)>=2p^{0}(2\pi)^3 \delta^{3}({\vec{p}}^{\hspace{0.1cm}\prime}
         -\vec{p}),
      \end{equation}  
  the conservation of the vector current $\bar{Q}\gamma^{\mu}Q$ leads to
      \begin{equation}
      \label{eq:36}
          <H(p)\vert \bar{Q}\gamma^{\mu}Q \vert H(p)>=2p^{\mu}=2m_H v^{\mu},
      \end{equation}  
  where $\vert H>$ denotes a hadron state in QCD and $p^{\mu}=m_H v^{\mu}$ 
  is the momentum of the heavy hadron $H$.

  In the effective theory, it is better to introduce an effective heavy hadron state 
  $\vert H_v>$ which exhibits a manifest spin-flavor symmetry. Thus, the 
  hadron state $\vert H_v>$ should be related to the state $\vert H>$ by the following 
  relation
    \begin{equation}
    \label{eq:37}    
     \frac{1}{\sqrt{m_{H^{\prime}}m_{H}}} <H^{\prime}\vert Q^{\prime} \Gamma Q \vert H>= 
\frac{1}{\sqrt{\bar{\Lambda}_{H^{\prime}} \bar{\Lambda}_H}} <H^{\prime}_{v^{\prime}}\vert
          J^{(++)}_{eff} e^{i\int d^4x L^{(1/{\MQ})}_{eff}}\vert H_v>.
    \end{equation}  
  Here $\bar{\Lambda}_{H}$ and $\bar{\Lambda}_{H'}$ are binding energies defined as 
    \begin{equation}
    \label{eq:38}
       \bar{\Lambda}_{H} \equiv m_{H}-m_{Q} ,\hspace{1.5cm}
       \bar{\Lambda}_{H^{\prime}} \equiv m_{H^{\prime}}-m_{Q^{\prime}}.
    \end{equation}
  The flavor-dependent factor $\frac{1}{\sqrt{\bar{\Lambda}_{H^{\prime}}
  \bar{\Lambda}_H}}$ appears due to the different normalizations of the 
  two hadron states $\vert H>$ and $\vert H_v>$. Here the normalization of $\vert H_v>$ 
   is taken so as to exhibit a manifest spin-flavor symmetry, i.e., 
\begin{equation}
     \label{eq:39}
         <H_{v} \vert \QVB \gamma^{\mu} \QV \vert H_v> = 2\bar{\Lambda} v^{\mu} 
     \end{equation}
with 
\[ \bar{\Lambda} = \lim_{m_{Q}\to \infty} \bar{\Lambda}_H \]
being a heavy flavor-independent binding energy that reflects the 
effects of the light degrees of freedom in the heavy hadron.
  
   In the heavy quark expansion, $1/{m_Q}$ corrections to the hadronic matrix 
   elements can be classified into three parts\cite{lqhq,fane}: ({\rm I}). 
   corrections from purely effective current $J^{(1/{\MQ})}_{eff}$; 
   ({\rm II}). corrections 
   from purely effective Lagrangian $L^{(1/{\MQ})}_{eff}$; and ({\rm III}). 
   mixed corrections from both $J^{(1/{\MQ})}_{eff}$ and $L^{(1/{\MQ})}_{eff}$. 
   
   The leading matrix element is simply given by
     \begin{equation} 
     \label{eq:40}
       {\cal A}^{({\rm 0})} \equiv <H^{\prime}_{v^{\prime}} \vert J^{(0)}_{eff}\vert H_v>
       =<H^{\prime}_{v^{\prime}} \vert \QVBP \Gamma \QV \vert H_v>.
     \end{equation}
   
   The three types of corrections are found to be 
   \begin{eqnarray}
   \label{eq:41}
     {\cal A}^{({\rm I})} &\equiv& <H^{\prime}_{v^{\prime}} \vert J^{(1/{\MQ})}_{eff} \vert H_v> =
     \frac{1}{2\MQ}<H^{\prime}_{v^{\prime}}
     \vert \QVBP O_{1}(\Gamma) \QV \vert H_v> \nonumber \\
     &&+\frac{1}{2\MQP}<H^{\prime}_{v^{\prime}}\vert \QVBP O_{1}^{\prime}(\Gamma)
     \QV \vert H_v> +\frac{1}{4\MQ^2}<H^{\prime}_{v^{\prime}}\vert \QVBP O_{2}(\Gamma)\QV \vert 
     H_v> \nonumber\\
     &&+\frac{1}{4\MQP^2}<H^{\prime}_{v^{\prime}}\vert \QVBP O_{2}^{\prime}(\Gamma)\QV 
     \vert H_v> 
     +\frac{1}{4\MQP \MQ}<H^{\prime}_{v^{\prime}}\vert \QVBP O_{4}(\Gamma)\QV \vert H_v> \nonumber\\
     &&+O(\frac{1}{m_{Q^{(\prime)}}^3}) , \\
   \label{eq:42}
   {\cal A}^{({\rm II})} &\equiv& <H^{\prime}_{v^{\prime}}\vert J^{(0)}_{eff} 
    e^{i\int d^4x L^{(1/{\MQ})}_{eff}}\vert H_v>
     =-\frac{1}{\MQ}<H^{\prime}_{v^{\prime}}\vert \QVBP O_{1}(\Gamma) \QV \vert H_v>\nonumber\\
     &&-\frac{1}{\MQP}<H^{\prime}_{v^{\prime}}\vert \QVBP O_{1}^{\prime}(\Gamma)
     \QV \vert H_v> 
      -\frac{1}{2\MQ^2}<H^{\prime}_{v^{\prime}}\vert \QVBP O_{2}(\Gamma)\QV \vert H_v> \nonumber\\
     &&-\frac{1}{2\MQP^2}<H^{\prime}_{v^{\prime}}\vert \QVBP O_{2}^{\prime}(\Gamma)\QV \vert H_v> 
     +\frac{1}{4\MQ^2}<H^{\prime}_{v^{\prime}}\vert \QVBP O_{3}(\Gamma)\QV \vert H_v> \nonumber\\
     &&+\frac{1}{4\MQP^2}<H^{\prime}_{v^{\prime}}\vert \QVBP O_{3}^{\prime}(\Gamma)\QV \vert H_v>  
     +\frac{1}{\MQP \MQ}<H^{\prime}_{v^{\prime}}\vert \QVBP O_{4}(\Gamma)\QV \vert H_v>\nonumber\\
     &&+O(\frac{1}{m_{Q^{(\prime)}}^3})  ,  \\
    \label{eq:43}
    {\cal A}^{({\rm III})} &\equiv& <H^{\prime}_{v^{\prime}}\vert J^{(1/{\MQ})}_{eff} 
    e^{i\int d^4x L^{(1/{\MQ})}_{eff}}\vert H_v>
    = -\frac{1}{2\MQ^2}<H^{\prime}_{v^{\prime}}\vert \QVBP O_{3}(\Gamma)\QV \vert H_v> \nonumber\\
     &&-\frac{1}{2\MQP^2}<H^{\prime}_{v^{\prime}}\vert \QVBP O_{3}^{\prime}(\Gamma)\QV \vert H_v>
     -\frac{1}{\MQP \MQ}<H^{\prime}_{v^{\prime}}\vert \QVBP O_{4}(\Gamma)\QV \vert H_v>\nonumber\\
     &&+O(\frac{1}{m_{Q^{(\prime)}}^3}) ,
    \end{eqnarray}
  where the operators $O_{i}(\Gamma)$ and  $O'_{i}(\Gamma)$ are defined as follows
   \begin{eqnarray}
   \label{eq:44}
     O_1(\Gamma)& =&\Gamma\frac{1}{i\DSP+\Lambda}(i\DSC)^2 ,\nonumber\\
     O_1^{\prime}(\Gamma)&=&(-i\stackrel{\hspace{-0.1cm}\leftarrow}{\DSC})^2 
\frac{1}{-i\stackrel{\hspace{-0.1cm}\leftarrow}
        {\DSP}+\Lambda} \Gamma ,   \nonumber \\
     O_2(\Gamma)& =&\Gamma\frac{1}{i\DSP+\Lambda}(i\DSC)(i\DSP-\Lambda)i\DSC , \nonumber\\
     O_2^{\prime}(\Gamma)& =&(-i\stackrel{\hspace{-0.1cm}\leftarrow}{\DSC})
(-i\stackrel{\hspace{-0.1cm}\leftarrow}{\DSP}-\Lambda)
        (-i\stackrel{\hspace{-0.1cm}\leftarrow}{\DSC})
\frac{1}{-i\stackrel{\hspace{-0.1cm}\leftarrow}{\DSP}+\Lambda}
        \Gamma ,\nonumber\\
     O_3(\Gamma) &=&\Gamma\frac{1}{i\DSP+\Lambda}(i\DSC)^2 \frac{1}{i\DSP+\Lambda}
        (i\DSC)^2 , \nonumber\\
     O_3^{\prime}(\Gamma)& =&(-i\stackrel{\hspace{-0.1cm}\leftarrow}{\DSC})^2 
\frac{1}{-i\stackrel{\hspace{-0.1cm}\leftarrow}
        {\DSP}+\Lambda}(-i\stackrel{\hspace{-0.1cm}\leftarrow}{\DSC})^2 
\frac{1}{-i\stackrel{\hspace{-0.1cm}\leftarrow}
        {\DSP}+\Lambda} \Gamma ,\nonumber\\
     O_4(\Gamma)& =&(-i\stackrel{\hspace{-0.1cm}\leftarrow}{\DSC})^2 
\frac{1}{-i\stackrel{\hspace{-0.1cm}\leftarrow}
        {\DSP}+\Lambda}\Gamma\frac{1}{i\DSP+\Lambda}(i\DSC)^2.
   \end{eqnarray}
  The first type of corrections ${\cal A}^{(\rm I)}$ can be read from eq.(\ref{eq:34}). In obtaining 
  the second and the third types of corrections ${\cal A}^{(\rm II)}$ and ${\cal A}^{(\rm III)}$ , 
  we have contracted the heavy quark pair and used the heavy quark propagator 
    $ i/(i\DSP+\Lambda)  $.

   The hadronic matrix element is then given by 
   \begin{eqnarray}
   \label{eq:45}
   {\cal A} &\equiv &<H^{\prime}_{v^{\prime}}\vert J^{(++)}_{eff} e^{i\int d^4x L^{(1/{\MQ})}_{eff}}
        \vert H_v> = {\cal A}^{(\rm 0)}+{\cal A}^{(\rm I)}+{\cal A}^{(\rm II)}+{\cal A}^{(\rm III)} 
         \nonumber\\
        &=& <H^{\prime}_{v^{\prime}}\vert \QVBP \Gamma \QV \vert H_v> 
        -\frac{1}{2\MQ}<H^{\prime}_{v^{\prime}}\vert \QVBP O_{1}(\Gamma) \QV \vert H_v>\nonumber\\
        &&-\frac{1}{2\MQP}<H^{\prime}_{v^{\prime}}\vert \QVBP O_{1}^{\prime}(\Gamma)
        \QV \vert H_v>
        -\frac{1}{4\MQ^2}<H^{\prime}_{v^{\prime}}\vert \QVBP O_{2}(\Gamma)\QV 
        \vert H_v>  \nonumber\\
        &&-\frac{1}{4\MQP^2}<H^{\prime}_{v^{\prime}}\vert \QVBP O_{2}^{\prime}(\Gamma)\QV
        \vert H_v> 
        -\frac{1}{4\MQ^2}<H^{\prime}_{v^{\prime}}\vert \QVBP O_{3}(\Gamma)\QV \vert H_v> 
       \nonumber  \\
        &&-\frac{1}{4\MQP^2}<H^{\prime}_{v^{\prime}}\vert \QVBP O_{3}^{\prime}(\Gamma)\QV
        \vert H_v>  
        +\frac{1}{4\MQP \MQ}<H^{\prime}_{v^{\prime}}\vert \QVBP O_{4}(\Gamma)\QV \vert H_v> 
      \nonumber \\
         &&+O(\frac{1}{m_{Q^{(\prime)}}^3}) .
   \end{eqnarray}
   
  From eqs.(\ref{eq:37}) and (\ref{eq:45}) as well as the hadron state normalization conditions in 
  eqs.(\ref{eq:36}) and (\ref{eq:39}), one can extract the hadron mass by setting $v^{\prime}=v$. 
  Explicitly, one has 
    \begin{eqnarray}
    \label{eq:46}
       2m_H v^{\mu}&=&<H\vert Q\gamma^{\mu} Q \vert H>=\frac{m_H}{\bar{\Lambda}_H}
         \{2\bar{\Lambda} v^{\mu}-\frac{1}{\MQ}<H_v\vert \QVB O_{1}(\gamma^{\mu})
         \QV \vert H_v> \nonumber\\
       &&-\frac{1}{2\MQ^2}<H_v\vert \QVB (O_{2}(\gamma^{\mu})+O_{3}(\gamma^{\mu}))
         \QV \vert H_v> \nonumber\\
       &&+\frac{1}{4\MQ^2}<H_v\vert \QVB O_{4}(\gamma^{\mu})\QV \vert H_v>
         +O(\frac{1}{m_{Q}^3})  \} ,
    \end{eqnarray}
  which leads to the following relation
    \begin{eqnarray}
    \label{eq:47}
       \bar{\Lambda}_{H}&=&\bar{\Lambda}-\frac{1}{2\MQ}<H_v\vert \QVB O_{1}(\VS)
        \QV \vert H_v>  
       -\frac{1}{4\MQ^2}<H_v\vert \QVB (O_{2}(\VS)+O_{3}
         (\VS)\QV \vert H_v> \nonumber\\
      &&+\frac{1}{8\MQ^2}<H_v\vert \QVB O_{4}(\VS)\QV \vert H_v> 
        +O(\frac{1}{\MQ^3}) .
    \end{eqnarray}
  The heavy hadron mass is then given by 
    \begin{equation}  
    \label{eq:48}
      m_H=m_Q +\bar{\Lambda}_H=m_Q +\bar{\Lambda}+ O(1/{m_Q}),
    \end{equation}
  which coincides with the fact that the mass of a hadron equals the sum of
  the heavy quark mass $m_Q$, the binding energy $\bar{\Lambda}$ due to light 
  degrees of freedom and terms suppressed by $1/{\MQ}$.
  
  Let us consider the case that both the initial and final states are pseudoscalar 
  mesons (or both vector mesons as well). An interesting example is the 
  $B\rightarrow D$ transition matrix element of vector current. 
  From eqs.(\ref{eq:37}) and (\ref{eq:45}), we obtain 
  \newcommand{\CVP}{\bar{c}^{+}_{v^{\prime}}}
  \newcommand{\BV}{b_v^{+}}
   \begin{eqnarray}
   \label{eq:49}
      <D\vert \bar{c} \gamma^{\mu} b \vert B &&> =
      \sqrt{\frac{m_D m_B}{\bar{\Lambda}_D \bar{\Lambda}_B}}
        \{  <D_{v^{\prime}}\vert \CVP \gamma^{\mu} \BV \vert B_v>
        -\frac{1}{2m_b}<D_{v^{\prime}}\vert \CVP O_{1}(\gamma^{\mu}) \BV \vert B_v>\nonumber\\
       && -\frac{1}{2m_c}<D_{v^{\prime}}\vert \CVP O_{1}^{\prime}(\gamma^{\mu})
        \BV \vert B_v> -\frac{1}{4m_b^2}<D_{v^{\prime}}\vert \CVP O_{2}(\gamma^{\mu})\BV 
        \vert B_v> \nonumber\\
       &&-\frac{1}{4m_c^2}<D_{v^{\prime}}\vert \CVP O_{2}^{\prime}(\gamma^{\mu})\BV
        \vert B_v> 
        -\frac{1}{4m_b^2}<D_{v^{\prime}}\vert \CVP O_{3}(\gamma^{\mu})\BV \vert B_v> \nonumber \\
       && -\frac{1}{4m_c^2}<D_{v^{\prime}}\vert \CVP O_{3}^{\prime}(\gamma^{\mu})\BV
        \vert B_v>  
       +\frac{1}{4m_c m_b}<D_{v^{\prime}}\vert \CVP O_{4}(\gamma^{\mu})\BV \vert B_v> \nonumber\\
       && +O(\frac{1}{m_{b(c)}^3})  \} .
   \end{eqnarray}
   
  From spin-flavor symmetry, one can use the following relations for operator 
  $O_i$
    \begin{equation}
    \label{eq:50}
      <B_v \vert \bar{b}^{+}_{v}O_i b^{+}_v\vert B_v>
        =<D_v \vert \bar{c}^{+}_{v}O_i c^{+}_v 
      \vert D_v> \equiv <P_v \vert \QVB O_i \QV \vert P_v>. 
    \end{equation} 
  With these relations and eq.(\ref{eq:47}), eq.(\ref{eq:49}) can be simplified 
  to the following form at zero recoil $v^{\prime}=v$,
    \begin{eqnarray}  
    \label{eq:51}
      <D\vert \bar{c}\gamma^{\mu} b\vert B>|_{q^{2}=q^{2}_{max}}
         &=& 2\sqrt{m_B m_D}v^{\mu}\{1+\frac{1}{32\bar{\Lambda}^{2}}
      (\frac{1}{m_b}-\frac{1}{m_c})^2 {<P_v\vert \QVB O_{1}(\VS)\QV \vert P_v>}^2 \nonumber\\
      && -\frac{1}{16\bar{\Lambda}}(\frac{1}{m_b}-\frac{1}{m_c})^2 
       <P_v\vert \QVB O_{4}(\VS)\QV \vert P_v> +O(\frac{1}{m_{b(c)}^3}) \} ,
    \end{eqnarray}
  which explicitly shows that when working with the normalizations in 
  eqs.(\ref{eq:36}), (\ref{eq:37}) and (\ref{eq:39}), 
  the transition matrix elements of heavy quark vector current between two pseudoscalar 
  mesons receive no corrections of order $1/{\MQ}$ at zero recoil. The same 
  result holds for the transition matrix elements of heavy quark vector current 
  between two vector mesons. 
  For axial vector current matrix elements or vector current matrix elements 
  between heavy meson states with different spins, it is not so manifest and 
  one needs to analyze the concrete Lorentz structures 
  of currents and meson states. This is the object of the next 
  section.
  
  \section{Meson Form Factors and Luke's Theorem}\label{fac}

  The HQET has been applied fruitfully to study heavy meson decays. In particular, 
  the Luke's theorem \cite{luke} plays an important role in the determination 
  of the Cabibbo-Kobayashi-Maskawa matrix element $\vert V_{cb} \vert$ from 
  exclusive $B\rightarrow D^{\ast} l \bar{\nu}$ decay. This is because the order 
  $1/{\MQ}$ corrections to the $b\rightarrow c$ transition matrix element 
  of weak current are absent at zero recoil. As the Luke's 
  theorem was deduced from the heavy quark effective Lagrangian and 
  currents in the usual HQET, it is interesting to check whether the Luke's 
  theorem remains valid within the framework of the new formulation of HQEFT. 
  At the same time, the order $1/m^2_Q$ corrections, though expected to be small, are worthy 
  to be considered since they can provide us information about 
  the convergence of the $1/m_{Q}$ expansion. 
  And as we will illustrate, some useful features can be observed because the new formulation 
  of the heavy quark effective Lagrangian (i.e. eq.(\ref{eq:24})) and currents 
  (eq.(\ref{eq:34})) are different from the usual ones due to new contributions 
  from integrating out antiparticles. 

 Generally, matrix elements of vector and axial vector currents 
  between pseudoscalar and vector ground 
  meson states are described by 18 meson form factors defined as follows 
  \begin{eqnarray}
  \label{eq:delhi}
   &&\hspace{-0.7cm}<D(v^{\prime})\vert \bar{c}\gamma^{\mu} b \vert B(v)> 
     =\sqrt{m_D m_B}[h_{+}(\omega)(v+v^{\prime})^{\mu}+h_{-}(\omega)(v-v^{\prime})^{\mu}] ,\nonumber\\
   &&\hspace{-0.7cm}<D^{\ast}(v^{\prime},\epsilon^{\prime})\vert \bar{c}\gamma^{\mu} b \vert B(v)>
         = i \sqrt{m_{D^{\ast}} m_B} 
         h_{V}(\omega) \epsilon^{\mu \nu \alpha \beta} \epsilon^{\prime \ast}_{\nu}
         v^{\prime}_{\alpha} v_{\beta} ,\nonumber\\
   &&\hspace{-0.7cm}<D^{\ast}(v^{\prime},\epsilon^{\prime})\vert \bar{c}\gamma^{\mu} \gamma^{5}b \vert B(v)>
         = \sqrt{m_{D^{\ast}} m_B}
         [h_{A_1}(\omega)(1+\omega) \epsilon^{\prime \ast \mu} 
       -h_{A_2}(\omega)(\epsilon^{\prime \ast} \cdot v)v^{\mu} \nonumber \\
  &&\hspace{1cm} -h_{A_3}(\omega)(\epsilon^{\prime \ast} \cdot v)v^{\prime\mu}], \nonumber\\
  &&\hspace{-0.7cm}<D^{\ast}(v^{\prime},\epsilon^{\prime})\vert \bar{c}\gamma^{\mu}b\vert
         B^{\ast}(v,\epsilon)>
       =\sqrt{m_{D^{\ast}} m_{B^{\ast}} } 
    \{-(\epsilon\cdot \epsilon^{\prime\ast})[h_{1}(\omega)(v+v^{\prime})^{\mu}
       +h_{2}(\omega)(v-v^{\prime})^{\mu}] \nonumber\\
  &&\hspace{1cm} +h_{3}(\omega)(\epsilon^{\prime\ast}\cdot v)\epsilon^{\mu}   
 +h_{4}(\omega)(\epsilon\cdot v^{\prime})\epsilon^{\prime\ast\mu}
       -(\epsilon\cdot v^{\prime})(\epsilon^{\prime\ast}\cdot v)  
    [h_{5}(\omega)v^{\mu}+h_{6}(\omega)v^{\prime\mu}] \} ,\nonumber \\
&&\hspace{-0.7cm}<D^{\ast}(v^{\prime},\epsilon^{\prime})\vert \bar{c}\gamma^{\mu} \gamma^{5}b\vert
        B^{\ast}(v,\epsilon)>
        =i\sqrt{m_{D^{\ast}} m_{B^{\ast}}}  
        \{\epsilon^{\mu\nu\alpha\beta}\{\epsilon_{\alpha}\epsilon^{\prime\ast}_{\beta}
        [h_{7}(\omega)(v+v^{\prime})_{\nu}\nonumber\\
   &&\hspace{1cm}+h_{8}(\omega)(v-v^{\prime})_{\nu}]
   +v^{\prime}_{\alpha}v_{\beta}[h_{9}(\omega)(\epsilon^{\prime\ast}\cdot v)  
       \epsilon_{\nu}+h_{10}(\omega)(\epsilon\cdot v^{\prime})
        \epsilon^{\prime\ast}_{\nu}] \} \nonumber\\
  &&\hspace{1cm}+\epsilon^{\alpha\beta\gamma\delta}\epsilon_{\alpha}
        \epsilon^{\prime\ast}_{\beta}v_{\gamma}v^{\prime}_{\delta} 
       [h_{11}(\omega)v^{\mu}+h_{12}(\omega)v^{\prime\mu} ] \} .
 \end{eqnarray}

   \newcommand{\DCL}{\stackrel{\leftarrow}{D}_{\bot}} 
   \newcommand{\DL}{\stackrel{\leftarrow}{D}} 
   \newcommand{\DCLS}{\stackrel{\hspace{-0.02cm}\leftarrow}{D_{\bot}^2}}

  In order to relate explicitly the form factors $h_{i}(\omega)$ to matrix 
  elements of operators, eq.(\ref{eq:45}) can be rewritten as 
  \begin{eqnarray}
  \label{eq:52}
     {\cal A} &\equiv&<H^{\prime}_{v^{\prime}}\vert J^{(++)}_{eff} e^{i\int d^4x L^{(1/{\MQ})}_{eff}}
        \vert H_v> = {\cal A}^{(\rm 0)}+{\cal A}^{(\rm I)}+{\cal A}^{(\rm II)}+{\cal A}^{(\rm III)} 
         \nonumber \\
      &=& <H^{\prime}_{v^{\prime}}\vert \QVBP \Gamma \QV \vert H_v> 
        -\frac{1}{2\MQ}<H^{\prime}_{v^{\prime}}\vert \QVBP \Gamma 
        \frac{-1}{\Lambda+iv\cdot D}P_{+} [ D^2_{\bot} 
        +\frac{i}{2}\sigma_{\alpha\beta} 
        F^{\alpha\beta} ] \QV \vert H_v>   \nonumber\\
       &&
        -\frac{1}{2m^{\prime}_Q}<H^{\prime}_{v^{\prime}}\vert \QVBP 
        \DCLS   +\frac{i}{2}\sigma_{\alpha\beta} 
        F^{\alpha\beta} ] P^{\prime}_{+} 
        \frac{-1}{\Lambda-iv^{\prime}\cdot {\stackrel{\leftarrow}{D}}}\Gamma
         \QV \vert H_v>\nonumber\\
       &&
        -\frac{1}{4m^2_Q}<H^{\prime}_{v^{\prime}}\vert \QVBP \Gamma 
        \frac{1}{\Lambda+iv\cdot D}P_{+}[\Lambda D^2_{\bot}+iD^2_{\bot}(v\cdot D)
        -iD^{\alpha} v^{\beta}F_{\alpha\beta}+D^2_{\bot}\frac{1}{\Lambda+iv\cdot 
        D}P_{+} D^2_{\bot} \nonumber\\
       &&
        + \Lambda\frac{i}{2}\sigma_{\alpha
        \beta}F^{\alpha\beta}-\frac{1}{2}\sigma_{\alpha\beta}F^{\alpha\beta}
        (v\cdot D) -\sigma^{\sigma\alpha}v^{\beta}D_{\sigma}F_{\alpha\beta} 
        +D^2_{\bot}\frac{1}{\Lambda+iv\cdot D}P_{+}\frac{i}{2}
        \sigma_{\gamma\sigma}F^{\gamma\sigma}   \nonumber\\
       && 
        +\frac{i}{2}\sigma_{\alpha\beta}
        F^{\alpha\beta}\frac{1}{\Lambda+iv\cdot D}P_{+}D^2_{\bot}         
        -\frac{1}{4}\sigma_{\alpha\beta}F^{\alpha\beta}
        \frac{1}{\Lambda+iv\cdot D} P_{+} \sigma_{\gamma\sigma}F^{\gamma\sigma}
        ]  \QV \vert H_v> \nonumber\\
      &&
        -\frac{1}{4\MQP^2}<H^{\prime}_{v^{\prime}}\vert \QVBP 
        [\Lambda \DCLS+i(v^{\prime}\cdot \DL)\DCLS 
        -iF_{\alpha\beta}\DL^{\alpha}v^{\prime\beta}+\DCLS P^{\prime}_{+}
        \frac{1}{\Lambda-iv^{\prime} \cdot \DL}\DCLS  \nonumber\\
      &&
        +\Lambda\frac{i}{2}\sigma_{\alpha\beta}F^{\alpha\beta}
        -\frac{1}{2}(v^{\prime}\cdot \DL)\sigma_{\alpha\beta}F^{\alpha\beta}
        -F_{\alpha\beta}\DL_{\sigma}v^{\prime\beta}\sigma^{\sigma\alpha}         
        +\frac{i}{2}\sigma_{\gamma\sigma}F^{\gamma\sigma}P^{\prime}_{+}
        \frac{1}{\Lambda-iv^{\prime} \cdot \DL}\DCLS \nonumber\\
      &&
        +\DCLS P^{\prime}_{+}\frac{1}{\Lambda-iv^{\prime}\cdot \DL}\frac{i}{2}
        \sigma_{\alpha\beta}F^{\alpha\beta}  
        -\frac{1}{4}\sigma_{\gamma\sigma}F^{\gamma\sigma}P^{\prime}_{+}         
        \frac{1}{\Lambda-iv^{\prime}\cdot \DL}\sigma_{\alpha\beta}F^{\alpha\beta}
        ]P^{\prime}_{+}\frac{1}{\Lambda-iv^{\prime}\cdot 
        \DL}\Gamma  \QV \vert H_v>  \nonumber\\
      && +\frac{1}{4\MQP\MQ}<H^{\prime}_{v^{\prime}}\vert \QVBP [\DCLS 
       P^{\prime}_{+}\frac{1}{\Lambda-iv^{\prime}\cdot \DL}\Gamma\frac{1}{\Lambda+iv\cdot D}
       P_{+}\DC^2   \nonumber\\
       &&+\DCLS P^{\prime}_{+}\frac{1}{\Lambda -iv^{\prime}\cdot \DL}\Gamma\frac{1}{\Lambda+iv\cdot D}
       P_{+}\frac{i}{2}\sigma_{\gamma\sigma}F^{\gamma\sigma} 
       +\frac{i}{2}\sigma_{\alpha\beta}F^{\alpha\beta}P^{\prime}_{+}
       {\frac{1}{\Lambda-iv^{\prime} \cdot \DL}}
       \Gamma {\frac{1}{\Lambda+iv \cdot D}} P_{+}\DC^2  \nonumber\\
       &&
       -\frac{1}{4}\sigma_{\alpha\beta}F^{\alpha\beta}P^{\prime}_{+}
       {\frac{1}{\Lambda-iv^{\prime} \cdot \DL} } 
       \Gamma\frac{1}{\Lambda+v\cdot D}P_{+}\sigma_{\gamma\sigma}
       F^{\gamma\sigma} ] \QV \vert H_v> 
   \end{eqnarray}
  with $F^{\alpha \beta}=[D^{\beta},D^{\alpha}]$ the field strength of gluons 
  tensor, $\sigma^{\alpha \beta}=\frac{i}{2}[\gamma^{\alpha},\gamma^{\beta}]$, 
  $P_{+}=\frac{1+v\hspace{-0.15cm}\slash}{2}$ and 
  $P^{\prime}_{+}=\frac{1+{v\hspace{-0.15cm}\slash}^{\prime}}{2}$.
  
  A simple way to evaluate the hadronic matrix elements is to associate the 
  spin wave functions adopted in refs.\cite{lqhq,fane,neub}. In the framework of new 
formulation of HQEFT with the normalization condition eq. (\ref{eq:39}), 
the spin wave functions have the following form
     \begin{eqnarray}
     \label{eq:53}
       {\cal M}(v)=\sqrt{\bar{\Lambda}}P_{+}
         \left\{
           \begin{array}{cl}
              -\gamma^{5}, & \mbox{pseudoscalar meson} \; \; P\\
              \epsilon\hspace{-0.15cm}\slash, & \mbox{vector meson} \; \; V
           \end{array}
         \right.
     \end{eqnarray}
  with meson states $\vert M_v>$ defined in the HQEFT. Here $\epsilon^{\mu}$ is the 
  polarization vector of the vector meson.

  Lorentz invariance enables us to parameterized the relevant matrix elements by 
  the following trace formulae
    \begin{eqnarray}
    \label{eq:54}
  <M^{\prime}_{v^{\prime}}\vert \QVBP\Gamma &&\QV \vert M_v> =-\xi(\omega)
     Tr[\bar{\cal M}^{\prime}\Gamma {\cal M}],
   \nonumber   \\ 
  <M^{\prime}_{v^{\prime}}\vert \QVBP\Gamma &&\frac{-1}{\Lambda+iv\cdot D} P_{+} \DC^2 
      \QV \vert M_v> =-\kappa_1(\omega) \frac{1}{\bar{\Lambda}} Tr[\bar{\cal M}^{\prime}\Gamma {\cal M}],
\nonumber      \\
  <M^{\prime}_{v^{\prime}}\vert \QVBP\Gamma &&\frac{-1}{\Lambda+iv\cdot D} P_{+} \frac{i}{2}
      \sigma_{\alpha\beta}F^{\alpha\beta}\QV \vert M_v> =\frac{1}{\bar{\Lambda}}Tr[\kappa_{\alpha\beta}(v,v^{\prime})
      \bar{\cal M}^{\prime}\Gamma P_{+}\frac{i}{2}\sigma^{\alpha\beta}{\cal M}] , \nonumber\\
  <M^{\prime}_{v^{\prime}}\vert \QVBP\Gamma &&\frac{1}{\Lambda+iv\cdot D} P_{+} 
     [i\DC^2(v\cdot D)-iD^{\alpha}v^{\beta} F_{\alpha\beta}]\QV \vert M_v> 
      =-\varrho_1(\omega)\frac{1}{\bar{\Lambda}}Tr[\bar{\cal M}^{\prime}\Gamma {\cal M}] ,\nonumber \\
  <M^{\prime}_{v^{\prime}}\vert \QVBP\Gamma &&\frac{1}{\Lambda+iv\cdot D} P_{+} 
      [-\frac{1}{2}\sigma_{\alpha\beta}F^{\alpha\beta}(v\cdot D)-\sigma^{\sigma\alpha}
      v^{\beta}D_{\sigma} F_{\alpha\beta}]\QV \vert M_v> \nonumber\\
     && =\frac{1}{\bar{\Lambda}}Tr[\varrho_{\alpha\beta}(v,v^{\prime})\bar{\cal M}^{\prime}\Gamma P_{+}
      \frac{i}{2}\sigma^{\alpha\beta}{\cal M}] ,\nonumber \\
  <M^{\prime}_{v^{\prime}}\vert \QVBP\Gamma&& \frac{1}{\Lambda+iv\cdot D} P_{+} \DC^2 
     \frac{1}{\Lambda+iv\cdot D}P_{+}\DC^2  \QV \vert M_v> 
      =-\chi_1(\omega)\frac{1}{\bar{\Lambda}^2} Tr[\bar{\cal M}^{\prime}\Gamma {\cal M}] , \nonumber\\
  <M^{\prime}_{v^{\prime}}\vert \QVBP\Gamma &&\frac{1}{\Lambda+iv\cdot D} P_{+} 
      [\DC^2 \frac{1}{\Lambda+iv\cdot D}P_{+} \frac{i}{2} \sigma_{\alpha\beta} 
      F^{\alpha\beta}  
      + \frac{i}{2}\sigma_{\alpha\beta}F^{\alpha\beta} \frac{1}{\Lambda+iv\cdot D} 
      P_{+} \DC^2 ] \QV \vert M_v>  \nonumber\\
        && =\frac{1}{\bar{\Lambda}^2} Tr[\chi_{\alpha\beta}(v,v^{\prime})\bar{\cal M}^{\prime}\Gamma P_{+}
      \frac{i}{2}\sigma^{\alpha\beta}{\cal M}],   \nonumber\\
  <M^{\prime}_{v^{\prime}}\vert \QVBP\Gamma&& \frac{1}{\Lambda+iv\cdot D} P_{+} 
    \frac{i}{2}\sigma_{\alpha\beta} F^{\alpha\beta} \frac{1}{\Lambda+iv\cdot D} P_{+}
    \frac{i}{2}\sigma_{\gamma\sigma}F^{\gamma\sigma}\QV \vert M_v> \nonumber\\
  &&  =- \frac{1}{\bar{\Lambda}^2} Tr[\chi_{\alpha\beta\gamma\sigma}(v,v^{\prime})\bar{\cal M}^{\prime}\Gamma P_{+}
    \frac{i}{2}\sigma^{\alpha\beta} P_{+} \frac{i}{2}\sigma^{\gamma\sigma} {\cal M}] ,  \nonumber \\
  <M^{\prime}_{v^{\prime}}\vert \QVBP &&\DCLS P^{\prime}_{+} \frac{1}{\Lambda-iv^{\prime} \cdot \DL} 
     \Gamma \frac{1}{\Lambda+iv\cdot D} P_{+} \DC^2 \QV \vert M_v> 
     =-\eta_{1}(\omega) \frac{1}{\bar{\Lambda}^2} Tr[\bar{\cal M}^{\prime}\Gamma {\cal M}] ,\nonumber \\
  <M^{\prime}_{v^{\prime}}\vert \QVBP && \DCLS P^{\prime}_{+} \frac{1}{\Lambda
      -iv^{\prime}\cdot \DL} 
     \Gamma \frac{1}{\Lambda+iv\cdot D} P_{+} \frac{i}{2} \sigma_{\alpha\beta}
     F^{\alpha\beta} \QV \vert M_v> \nonumber\\
    && =\frac{1}{\bar{\Lambda}^2} Tr[\eta_{\alpha\beta}(v,v^{\prime})
      \bar{\cal M}^{\prime}\Gamma P_{+}\frac{i}{2}\sigma^{\alpha\beta}{\cal M}], \nonumber \\
  <M^{\prime}_{v^{\prime}}\vert \QVBP \frac{i}{2} &&\sigma_{\alpha\beta} F^{\alpha\beta}
     P^{\prime}_{+} \frac{1}{\Lambda-iv^{\prime} \cdot \DL} 
     \Gamma \frac{1}{\Lambda+iv\cdot D} P_{+} \frac{i}{2} \sigma_{\gamma\sigma}
     F^{\gamma\sigma} \QV \vert M_v>  \nonumber \\
    && =-\frac{1}{\bar{\Lambda}^2} Tr[\eta_{\alpha\beta\gamma\sigma}(v,v^{\prime})
      \bar{\cal M}^{\prime}\frac{i}{2}\sigma^{\alpha\beta} P^{\prime}_{+} \Gamma P_{+} 
      \frac{i}{2}\sigma^{\gamma\sigma}{\cal M}]  .
  \end{eqnarray}
  Note that all other matrix elements in eq.(\ref{eq:52}) can also be represented by 
the form factors introduced in eq.(\ref{eq:54}). For example, one can conjugate the second equality 
  in eq.(\ref{eq:54}) to get 
  \begin{equation}
  \label{eq:55}
  <M^{\prime}_{v^{\prime}}\vert \QVBP\frac{i}{2} \sigma_{\alpha\beta} 
      F^{\alpha\beta} P^{\prime}_{+} \frac{-1}{\Lambda-iv^{\prime} \cdot \DL} 
      \Gamma \QV \vert M_v> =\frac{1}{\bar{\Lambda}}Tr[\bar{\kappa}_{\alpha\beta}(v^{\prime},v)
      \bar{\cal M}^{\prime}(-\frac{i}{2}\sigma^{\alpha\beta}) P^{\prime}_{+} \Gamma 
      {\cal M}]   ,
  \end{equation}
  where $\omega=v\cdot v^{\prime}$, $\bar{\cal M}=\gamma^{0}{\cal M}^{\dagger}\gamma^{0}$ 
  and $\bar{\kappa}_{\alpha \beta}$, $\bar{\varrho}_{\alpha \beta}$, 
  $\bar{\chi}_{\alpha \beta}$, $\bar{\eta}_{\alpha \beta}$ are defined similarly. 
  The decomposition of the Lorentz tensors $\kappa_{\alpha \beta}(v,v^{\prime})$, $\varrho_{\alpha \beta}(v,v^{\prime})$, 
  $\chi_{\alpha \beta}(v,v^{\prime})$, $\eta_{\alpha \beta}(v,v^{\prime})$, 
  $\chi_{\alpha\beta\gamma\delta}(v,v^{\prime})$, 
  and $\eta_{\alpha\beta\gamma\delta}(v,v^{\prime})$ are presented in the appendix \ref{app:decomp}. 
  $\xi(\omega)$ is the Isgur-Wise function that 
  normalizes to unity at the point of zero recoil $\omega =1$, i.e., $\xi(1) = 1$ 
  \cite{luke,neub,wise}. 
  
   When taking the $\Lambda$ as the heavy flavor-independent binding energy $\bar{\Lambda}$, 
 the momentum $ \tilde{k} = P_{Q} - \hat{m}_Q v$ carried by the effective field $\QV$ 
within the heavy hadron is expected to be much smaller than the binding energy 
$\bar{\Lambda}$. So we can perform following expansion 
  \begin{equation}
  \label{eq:56}
  \frac{1}{\Lambda+iv\cdot D}\rightarrow \frac{1}{\bar{\Lambda}+iv\cdot D}=
\frac{1}{\bar{\Lambda}}\left(1+O(\frac{iv\cdot D}{\bar{\Lambda}})\right) 
  \sim \frac{1}{\bar{\Lambda}} . 
  \end{equation}
  Here $\bar{\Lambda}$ characterizes the effects of the light degrees of freedom in the heavy 
hadron. This may be understood as follows: In general a heavy quark 
within a hadron cannot truly be on-shell due to 
strong interactions among heavy quark and light quark as well as soft gluons. 
The off-shellness of the heavy quark in the heavy hadron 
is characterized by a residual momentum $k=\bar{\Lambda}v + \tilde{k}$. The total momentum $P_Q$ 
of the heavy quark in a hadron may be written 
as:  $P_Q=m_Q v + k = \hat{m}_Q v + \tilde{k} $. Thus the residual 
momentum $k = \bar{\Lambda}v + \tilde{k} $ of the heavy quark 
within a hadron is assumed to comprise the main contributions of the light degrees 
of freedom. Where $\tilde{k}$ is the part which depends on heavy flavor and is suppressed 
by $1/m_Q$. With this picture the heavy quark may be regarded as a 
`dressed heavy quark', and the heavy hadron containing a single heavy quark 
is more reliable to be considered as a dualized particle of a `dressed heavy quark'.
This differs from the picture in the usual HQET, where one mainly deals with 
the heavy quark and treats the light quark as a spectator. For this 
reason, the form factors defined in the ways of eq.(\ref{eq:54}) should 
have a very weak dependence on the light constituents of heavy hadrons.  Thus it is useful 
in the new formulation of HQEFT to define the `dressed heavy quark' mass 
as 
\begin{equation}
\label{addmass0}
\hat{m}_Q \equiv \lim_{m_{Q}\to \infty} m_{H} = m_Q +\bar{\Lambda}.
\end{equation}
  
  One can complete the trace calculation and write the matrix elements of vector and 
  axial vector currents between pseudoscalar and vector mesons in terms of Lorentz 
  scalar factors $\kappa_i$ ,$\varrho_{i}$, 
  $\chi_{i}$ and $\eta_{i}$. At the zero recoil point, we obtain 
    \begin{eqnarray}
    \label{eq:58}
     h_{+}(1)&=& \sqrt{\frac{1}{\bar{\Lambda}_D \bar{\Lambda}_B}} \bar{\Lambda} 
         \{\xi-(\frac{1}{2m_b \bar{\Lambda}}+\frac{1}{2m_c \bar{\Lambda}}
         -\frac{1}{4m^2_b}-\frac{1}{4m^2_c})(\kappa_{1}+3\kappa_{2}) \nonumber \\
  && -\frac{1}{4\bar{\Lambda}^2}(\frac{1}{m^2_b}+\frac{1}{m^2_c})
     (\varrho_1\bar{\Lambda}+3\varrho_{2}\bar{\Lambda}+\chi_1+3\chi_2
         -3\chi_{4}-9\chi_{5}-6\chi_{6}) \nonumber\\
&&         + \frac{1}{4m_b m_c \bar{\Lambda}^2}(\eta_{1}+6\eta_{3}-3\eta_{4}
         -9\eta_{5}-6\eta_{6})  \} ,\nonumber\\ 
  h_{A_{1}}(1)&=&\sqrt{\frac{1}{\bar{\Lambda}_{D^{\ast}}\bar{\Lambda}_B}} \bar{\Lambda}
         \{\xi-(\frac{1}{2m_b \bar{\Lambda}} -\frac{1}{4m_b^2})(\kappa_1+3\kappa_2)   
         -(\frac{1}{2m_c \bar{\Lambda}}-\frac{1}{4m_c^2}) 
         (\kappa_1-\kappa_2)  \nonumber \\
         &&-\frac{1}{4m^2_b \bar{\Lambda}^2}
         (\varrho_{1} \bar{\Lambda}+3\varrho_{2} \bar{\Lambda}
         +\chi_1+3\chi_2-3\chi_{4}  -9\chi_{5}-6\chi_{6}) \nonumber\\
  &&-\frac{1}{4m^2_c \bar{\Lambda}^2}
         (\varrho_{1} \bar{\Lambda}-\varrho_{2} \bar{\Lambda}+\chi_{1}-\chi_{2} 
         -3\chi_{4}-\chi_{5}+2\chi_{6}) \nonumber \\
&&       +\frac{1}{4m_b m_c \bar{\Lambda}^2}(\eta_{1}+2\eta_{2}
       +\eta_{4}+3\eta_{5}+2\eta_{6}) \} ,\nonumber \\
   h_{1}(1)&=&\sqrt{\frac{1}{\bar{\Lambda}_{D^{\ast}}
         \bar{\Lambda}_{B^{\ast}}}} \bar{\Lambda} \{\xi
         -(\frac{1}{2m_b \bar{\Lambda}}+\frac{1}{2m_c \bar{\Lambda}}
         -\frac{1}{4m^2_b} -\frac{1}{4m^2_c}) (\kappa_{1}-\kappa_{2})  \nonumber \\
&&       -\frac{1}{4\bar{\Lambda}^2}(\frac{1}{m^2_b}+\frac{1}{m^2_c})(\varrho_{1} \bar{\Lambda}
       -\varrho_{2} \bar{\Lambda}+\chi_1-\chi_2-3\chi_{4} 
      -\chi_{5}+2\chi_{6}) \nonumber\\
         &&  +\frac{1}{4m_b m_c \bar{\Lambda}^2}(\eta_{1}-2\eta_{2}-3\eta_{4}
       -\eta_{5}+2\eta_{6}) \}, \nonumber\\ 
 h_{7}(1)&=& -\sqrt{\frac{1}{\bar{\Lambda}_{D^{\ast}}\bar{\Lambda}_{B^{\ast}}}} 
       \bar{\Lambda} \{\xi-(\frac{1}{2m_b \bar{\Lambda}}+\frac{1}{2m_c \bar{\Lambda}} 
    -\frac{1}{4m^2_b} -\frac{1}{4m^2_c})(\kappa_{1}-\kappa_{2})  \nonumber\\
&&       -\frac{1}{4\bar{\Lambda}^2}(\frac{1}{m^2_b}+\frac{1}{m^2_c})
       (\varrho_{1} \bar{\Lambda}-\varrho_{2} \bar{\Lambda}+\chi_{1}-\chi_{2}
 -3\chi_{4}-\chi_{5}+2\chi_{6})   \nonumber\\
&&       +\frac{1}{4m_b m_c \bar{\Lambda}^2}[\eta_{1}-2\eta_{2}+\eta_{4}
       -\eta_{5}-2\eta_{6}]  \},
   \end{eqnarray}
  where $\xi(1)$ has been written as $\xi$ for simplicity, and similarly for 
  other form factors. 
        
  From the normalization condition eq.(\ref{eq:36}), the first two equalities of 
eq.(\ref{eq:58}) and the definition of eq. (\ref{eq:delhi}), one arrives at 
the following relations 
     \begin{eqnarray}
     \label{eq:59}
       \bar{\Lambda}_{D(B)}&=&
          \bar{\Lambda}-(\frac{1}{m_{c(b)}}-\frac{\bar{\Lambda}}{2m^2_{c(b)}})
          (\kappa_1+3\kappa_2)
          -\frac{1}{2m^2_{c(b)} \bar{\Lambda}}(\varrho_{1} \bar{\Lambda}
          +3\varrho_{2} \bar{\Lambda}+\chi_1+3\chi_2 -3\chi_{4}\nonumber\\
        &&  -9\chi_{5}-6\chi_{6})
          +\frac{1}{4m^2_{c(b)} \bar{\Lambda}}(\eta_{1}+6\eta_{2}-3\eta_{4}
          -9\eta_{5}-6\eta_{6})
          +O(\frac{1}{m_{c(b)}^3}) , \nonumber\\
       \bar{\Lambda}_{D^{\ast}(B^{\ast})}&=&
          \bar{\Lambda}-(\frac{1}{m_{c(b)}}-\frac{\bar{\Lambda}}{2m^2_{c(b)}})
          (\kappa_1-\kappa_2)
          -\frac{1}{2m^2_{c(b)} \bar{\Lambda}}(\varrho_{1} \bar{\Lambda}
          -\varrho_{2} \bar{\Lambda}+\chi_1-\chi_2 \\
         && -3\chi_{4}-\chi_{5}+2\chi_{6})
          +\frac{1}{4m^2_{c(b)} \bar{\Lambda}}(\eta_{1}-2\eta_{2}-3\eta_{4}
          -\eta_{5}+2\eta_{6}) +O(\frac{1}{m_{c(b)}^3})  ,
     \end{eqnarray}
  where the normalization of $\xi(\omega)$ at zero recoil: $\xi(\omega=1)=1$ has been used.

  Putting the above results for $\bar{\Lambda}_{D(B)}$ 
  and $\bar{\Lambda}_{D^{\ast}(B^{\ast})}$ back into eq.(\ref{eq:58}), 
  we get directly:
    \begin{eqnarray}
    \label{eq:60}
    h_{+}(1)&=&1+\frac{1}{8\bar{\Lambda}^2}(\frac{1}{m_b}-\frac{1}{m_c})^2
         (\kappa_{1}+3\kappa_{2})^2 
         -\frac{1}{8\bar{\Lambda}^2}(\frac{1}{m_b}-\frac{1}{m_c})^2 
         (\eta_{1}+3\eta_{2}-3\eta_{4}-9\eta_{5}-6\eta_{6}) ,\nonumber\\
   h_{A_{1}}(1)&=&1+\frac{1}{8\bar{\Lambda}^2}[\frac{1}{m_b}(\kappa_1+3\kappa_2)
         -\frac{1}{m_c}(\kappa_1  -\kappa_2)]^2
       -\frac{1}{8m^2_b \bar{\Lambda}^2}
         (\eta_{1}+3\eta_{2}-3\eta_{4}-9\eta_{5}-6\eta_{6}) \nonumber\\
  &&-\frac{1}{8m^2_c \bar{\Lambda}^2}
         (\eta_{1}-\eta_{2} -3\eta_{4} -\eta_{5}+2\eta_{6}) 
       +\frac{1}{4m_b m_c \bar{\Lambda}^2}
         (\eta_{1}+\eta_{2}+\eta_{4}+3\eta_{5}+2\eta_{6}) , \nonumber\\
     h_{1}(1)&=&1+\frac{1}{8\bar{\Lambda}^2}(\frac{1}{m_b}-\frac{1}{m_c})^2 
      (\kappa_{1}-\kappa_{2})^2 
       -\frac{1}{8\bar{\Lambda}^2}(\frac{1}{m_b}-\frac{1}{m_c})^2
       (\eta_1-\eta_2-3\eta_{4}-\eta_{5}+2\eta_{6}) , \nonumber \\ 
    h_{7}(1)&=&- \{1+\frac{1}{8\bar{\Lambda}^2}(\frac{1}{m_b}-\frac{1}{m_c})^2 
      (\kappa_{1}-\kappa_{2})^2
       -\frac{1}{8\bar{\Lambda}^2}(\frac{1}{m^2_b}+\frac{1}{m^2_c})
       (\eta_1-2\eta_2-3\eta_{4}-\eta_{5}+2\eta_{6})  \nonumber\\
&&     +\frac{1}{4m_b m_c \bar{\Lambda}^2}(\eta_1-2\eta_2+\eta_{4}-\eta_{5}
     -2\eta_{6}) \} .
   \end{eqnarray}
   
  Here we only give the form factors for the case at zero recoil point. The 
most general results at non-zero recoil are quite lengthy and we present them 
in the appendix \ref{app:factor}. Unlike the framework of the usual HQET, we observe that 
  $h_{-}(\omega)=h_{2}(\omega)=0$ in the new formulation of HQEFT. Such an interesting 
feature results from the fact that in the new framework of HQEFT, the operators in the effective 
  Lagrangian and effective current contain only terms with even powers of ${\DSC}^2$. 
  
  Generally, as shown in the appendix \ref{app:factor}, mesonic matrix elements up to second 
  order power corrections can be described by a set of 29 form factors, 
  which are universal functions of the kinematic variable $\omega=v\cdot v^{\prime}$. 
  Such a number is less than the one introduced in the usual HQET, 
  where 34 form factors are needed \cite{fane}. This is also due to the
  new structures of the effective Lagrangian eq.(\ref{eq:24}) 
  and the effective current eq.(\ref{eq:34}). 
  
  At zero recoil point, some of the form factors are kinematically suppressed, only
 15 universal form factors are needed to 
  describe the mesonic matrix elements up to order $1/m^2_Q$. Where $\kappa_1$ and 
  $\kappa_2$ characterize the contributions of the order $1/m_Q$ operators 
  at zero recoil. As shown in eq.(\ref{eq:59}), the first order corrections to 
  the meson mass arise from these two form factors. 
  They play the same roles as the parameters $\lambda_1$ and $\lambda_2$ defined in the 
  framework of the usual HQET \cite{fane}\cite{wise}-\cite{jeon} though the 
  corresponding operator forms are different. 
   
  It is seen that the order $1/{\MQ}$ corrections in the meson transition matrix elements 
  of weak currents are absent at zero recoil, though the forms of the effective 
  Lagrangian and currents in the new formulation of HQEFT are obviously different 
  from the ones in the usual HQET. This means that Luke's theorem 
  remains valid within the framework of the new formulation of HQEFT. 
  
  \section{Extraction of Form Factors and $|V_{cb}|$}\label{vcb}

  Since the discovery of spin-flavor symmetry in the heavy quark limit, great efforts 
   have been made to extract the CKM matrix element $\vert V_{cb}\vert$ from 
   the exclusive semileptonic decay modes $B\rightarrow D^{\ast}l\nu$ 
   and $B\rightarrow Dl\nu$ in the usual framework of HQET 
   \cite{fane}\cite{neub} \cite{man428}-\cite{neu238} 
   \cite{neu455}-\cite{shif}. 

   The differential decay rates of $B\rightarrow D^{\ast}l\nu$ and 
   $B\rightarrow Dl\nu$ decays can be simply formulated as follows
   \cite{neu84} \cite{neu455}
   \begin{eqnarray}
   \label{eq:61}
  \frac{d\Gamma(B\rightarrow D^{\ast}l\nu)}{d\omega} &=&\frac{G^2_F}{48\pi^3}(m_B-m_{D^{\ast}})^2 
    m^3_{D^{\ast}}\sqrt{\omega^2-1}(\omega+1)^2 \nonumber\\
    &&  \times  [1+\frac{4\omega}{\omega+1} 
    \frac{m^2_B-2\omega m_B m_{D^{\ast}}+m^2_{D^{\ast}}}
    {(m_B-m_{D^{\ast}})^2}]\vert V_{cb} \vert^2 {\cal F}^2(\omega) ,
   \end{eqnarray}
  and 
   \begin{equation}
   \label{eq:611}
    \frac{d\Gamma(B\rightarrow Dl\nu)}{d\omega} =\frac{G^2_F}{48\pi^3}(m_B+m_{D})^2 
    m^3_{D}(\omega^2-1)^{3/2} \vert V_{cb} \vert^2 {\cal G}^2(\omega)  
   \end{equation}
   with 
   \begin{eqnarray}
   \label{eq:62}
     {\cal F}(\omega) &=& \eta_{A} h_{A_1}(\omega),\\
   \label{eq:621}
     {\cal G}(\omega) &=& \eta_{V} [h_{+}(\omega)-\frac{m_B-m_D}{m_B+m_D}
      h_{-}(\omega)],
   \end{eqnarray}
   where the coefficients $\eta_{A}$ and $\eta_{V}$ characterize short distance QCD 
   corrections, whereas the functions $h_{A_1}(\omega)$, $h_{+}(\omega)$ and 
   $h_{-}(\omega)$ contain long distance dynamics. 

   Based on experimental measurements of the differential decay rate, one 
   can extract $\vert V_{cb} \vert {\cal F}(\omega)$ and 
   $\vert V_{cb} \vert {\cal G}(\omega)$ and then extrapolate 
   them to $\omega=1$ to obtain the following quantities
     \begin{eqnarray}
     \label{eq:64}
     \vert V_{cb} \vert {\cal F}(1)&=&\vert V_{cb} \vert \eta_A h_{A_1}(1) 
     =\vert V_{cb}\vert \eta_A (1+\delta^{\ast}),\\
     \label{eq:642}
     \mbox{and}\hspace{1cm} 
     \vert V_{cb} \vert {\cal G}(1)&=&\vert V_{cb} \vert \eta_V 
     [h_{+}(1)-\frac{m_B-m_D}{m_B+m_D}h_{-}(1)]
      =\vert V_{cb}\vert \eta_V (1+\delta). 
     \end{eqnarray}   

   In the framework of the usual HQET, it was shown that, from the 
   theoretical point of view, the $B\rightarrow D^{\ast}l\nu$ 
   decay is more favorable for the extraction of $\vert V_{cb}\vert$ 
   because its decay rate at zero recoil is strictly protected by Luke's 
   theorem against first-order power corrections in $1/m_Q$. 
   For decay channel $B\rightarrow Dl\nu$, though the transition 
   matrix element is protected by Luke's theorem, the decay rate do receive 
   order $1/m_Q$ corrections from the form factor $h_{-}(\omega)$, as can 
   be seen form eq.(\ref{eq:642}). 
   
   In the new framework of HQEFT, however, we have seen in last section that 
   $h_{-}(\omega)=0$. This means that both the differential decay rates of channels 
   $B\rightarrow Dl\nu$ and $B\rightarrow D^{\ast}l\nu$ receive 
   no order $1/m_Q$ corrections. 
   
   Since all quantities in eqs.(\ref{eq:61}) and (\ref{eq:611}) are the same as in 
   the usual HQET except for the form factors $h_{A_1}(\omega)$,
   $h_{+}(\omega)$ and $h_{-}(\omega)$, 
   we follow the same strategy and use 
     \begin{eqnarray}
     \label{eq:65}
     \vert V_{cb}\vert {\cal F}(1)&=&0.0352\pm 0.0026,\\
     \label{eq:651}
     \vert V_{cb}\vert {\cal G}(1)&=&0.0386\pm 0.0041,
     \end{eqnarray}
   which are average results of recent experimental data used 
   in Ref. \cite{neu269}. $\eta_A$ and $\eta_V$ have been calculated by many authors 
   \cite{neu238} \cite{pasc}-\cite{ac}.
   Here we use \cite{ac}
    \begin{eqnarray}
    \label{eq:66}
     \eta_A=0.960\pm 0.007,\nonumber \\
     \eta_V=1.022\pm 0.004.
     \end{eqnarray}
   
   The form factors $h_{A_1}(1)$ and $h_{+}(1)$ in eqs.(\ref{eq:64}) and 
   (\ref{eq:642}) are given in terms of 14 unknown scalar 
   form factors: $\kappa_1$, $\kappa_2$, $\varrho_{1}$, $\varrho_{2}$, 
   $\chi_1$, $\chi_{i}$ and $\eta_{i}$ $(i=1,2,4,5,6)$. 
   These form factors contain information of long-distance interaction in the 
   heavy hadron. While little is known about the long-distance dynamics, 
   evaluating these forms factors and extracting their values at zero recoil 
   are great challenges. Until now their values 
   can only be estimated by using either certain models or QCD sum rules. 
   Within the framework of new formulation of HQEFT, the ground state meson masses 
which have been measured to a relatively high 
   accuracy are correlated with the form factors via eq.(\ref{eq:59}). This allows us 
to figure out the most important form factors and to extract the important 
CKM matrix element $\vert V_{cb}\vert$. 
   
   As commented in last section, the number of form factors contributing 
   at zero recoil increases quickly as the power of $1/m_Q$ expansion 
   becomes higher. At the $1/m^2_Q$ order,  there are already 15 form factors that contribute to 
   hadronic matrix elements at zero recoil so that some simplifications 
   must be made before extracting $\vert V_{cb}\vert$.  
   
   Firstly, we note that operators $O_3(\Gamma)$, $O^{\prime}_3(\Gamma)$ 
   and $O_4(\Gamma)$ appear at order $1/m^2_Q$ 
   and have similar forms. Their contributions to hadronic matrix elements 
   should not be significant. At the zero recoil point $v=v^{\prime}$, supposing that residual momenta 
   of the heavy quarks are approximately equal, i.e, $k^{\prime 2}_{\bot} \approx
   k^{2}_{\bot} \sim \bar{\Lambda}$, then the left and right actions of the derivative 
operators on the heavy effective fields almost leads to the same results, i.e.,
   \begin{equation}
    \label{eq:68}
    -\stackrel{\leftarrow}{D}_{\mu} \sim D_{\mu}, \qquad 
     \frac{1}{i\DSP+\Lambda} \sim
     \frac{1}{-i\stackrel{\hspace{-0.1cm}\leftarrow}{\DSP}+\Lambda} \sim 
     \frac{1}{\bar{\Lambda}}.
    \end{equation}
   Under these considerations, we have $O_3(\gamma^{\mu})
   =O^{\prime}_3(\gamma^{\mu})=O_4(\gamma^{\mu})$, 
   which implies the following relations among the form factors,
   \begin{equation}
   \label{eq:69}
    \chi_1=\eta_{1}; \; \;  \chi_2=2\eta_{2}; \; \; \chi_{i}=\eta_{i}\; (i=4,5,6).
   \end{equation}
  
   As argued in Ref. \cite{fane}, the $1/m_Q$ corrections can be well 
   described by neglecting the form factors arising from the chromomagnetic 
   moment operator. They  
   corresponds to the fictitious limit of vanishing field strength, 
   $F^{\alpha \beta}\rightarrow 0$. Now we will use a similar treatment, 
   but only neglect the contributions arising from operators with two gluon field 
   strength tensors and still keep the contributions of operators containing 
   one gluon field strength tensor. This implies that the form factors 
   $\chi_{j}$ and $\eta_{j}(j=4,5,6)$ can be dropped out.
   
   From above considerations, there are only six form factors left. 
   eq.(\ref{eq:59}) is simplified to be 
     \begin{eqnarray}
      \label{eq:70}
       \bar{\Lambda}_{D(B)}&=&
          \bar{\Lambda}-(\frac{1}{m_{c(b)}}
          -\frac{\bar{\Lambda}}{2m^2_{c(b)}})(\kappa_1+3\kappa_2)
          -\frac{1}{4m^2_{c(b)} \bar{\Lambda}}(F_{1}+3F_{2})
          +O(\frac{1}{m_{c(b)}^3}) ,  \nonumber\\
       \bar{\Lambda}_{D^{\ast}(B^{\ast})}&=&
          \bar{\Lambda}-(\frac{1}{m_{c(b)}}-\frac{\bar{\Lambda}}
          {2m^2_{c(b)}})(\kappa_1-\kappa_2)
          -\frac{1}{4m^2_{c(b)} \bar{\Lambda}}(F_{1}-F_{2})
          +O(\frac{1}{m_{c(b)}^3}),
     \end{eqnarray}
   where $F_{1}$ and $F_{2}$ are defined as 
   \begin{equation}
   \label{eq:F12}
   F_{1}=\chi_{1}+2\bar{\Lambda}\varrho_{1} ,  \hspace{1.5cm}
   F_{2}=\chi_{2}+2\bar{\Lambda}\varrho_{2} .
   \end{equation}
   Thus the $ h_{A_{1}}(1)$ and $h_{+}(1)$ in eq.(\ref{eq:60}) turn into
    \begin{eqnarray}
    \label{eq:71}
    h_{A_{1}}(1)&=&1+\frac{1}{8\bar{\Lambda}^2}[\frac{1}{m_b}(\kappa_1+3\kappa_2)
     -\frac{1}{m_c}(\kappa_1-\kappa_2)]^2 
     -\frac{1}{8m^2_b\bar{\Lambda}^2}
     (F_{1}+3F_{2}  \nonumber\\
&&     -2\bar{\Lambda}\varrho_{1}-6\bar{\Lambda}\varrho_{2})
     -\frac{1}{8m_c^2\bar{\Lambda}^2}(F_{1}-F_{2}-2\bar{\Lambda}\varrho_{1}
     +2\bar{\Lambda}\varrho_{2}) \nonumber\\
&&     +\frac{1}{4m_b m_c \bar{\Lambda}^2}(F_{1}+F_{2}-2\bar{\Lambda}\varrho_{1}
     -2\bar{\Lambda}\varrho_{2}) ,
    \end{eqnarray}   
    \begin{equation}
    \label{eq:711}
    h_{+}(1)= 1+\frac{1}{8\bar{\Lambda}^2}(\frac{1}{m_b}-\frac{1}{m_c})^2 
     [(\kappa_1+3\kappa_2)^2-(F_1+3F_2) +2\bar{\Lambda}(\varrho_1+3\varrho_2)].
    \end{equation}   
   
   We will take the heavy quark masses $m_b$ and $m_c$ as well as the `dressed heavy quark' mass 
  $\hat{m}_{b}= m_{b} + \bar{\Lambda}$ ( or the heavy flavor-independent binding energy 
  $\bar{\Lambda}$ ) as three basic parameters of the theory. 
   Usually,  one fixes the value of the mass difference $m_b-m_c$,
   to extract $\vert V_{cb}\vert$ as functions of mass $m_{b}$ (or $m_{c}$) from either 
   exclusive or inclusive B decays\cite{fane} \cite{man428}-\cite{shif}. 
   In our present considerations, we permit the 
   value of $m_b-m_c$ change between 3.32GeV and 3.41GeV. 
      
   Based on eq.(\ref{eq:70}) and using ground state meson masses \cite{epjc} as input, 
   we can extract the four form factors $\kappa_1$, $\kappa_2$, $F_{1}$ and 
   $F_{2}$ as functions of $m_b$, $m_b-m_c$ and $m_b+\bar{\Lambda}$. Thus it allows us to calculate 
   the parameters $\delta^{\ast}$ and $\delta$ up to the $1/m^2_Q$ order corrections in the 
   $B\rightarrow D^{\ast}l\nu$ and $B\rightarrow D l\nu$ transition matrix elements at zero recoil.

   The form factors $\kappa_1$, $\kappa_2$, $F_1$ and $F_2$ as functions 
   of $m_b$, $m_b-m_c$ and $m_b+\bar{\Lambda}$ are plotted in Fig.1.  
   One sees from Figs.1a-1b that $\kappa_1$ is insensitive to 
   $m_b$ and $m_b-m_c$, but it is quite sensitive to $m_b+\bar{\Lambda}$ as explicitly 
   shown in Fig.1c.  From Figs.1d-1e, one sees that $\kappa_{2}$ only slightly 
   changes against $m_b$ and $m_b-m_c$ and is independent of  $m_b+\bar{\Lambda}$ as shown 
   from the straight lines in Fig.1f. The combined form factors $F_1$ and $F_2$ have only a 
   slight dependence on $m_b$ as shown in Figs.1g and 1j,  but both of them heavily 
  depends on $m_b-m_c$ and decrease as $m_b-m_c$ increases, which have been shown 
  in Figs.1h and 1k. They are sensitive to $m_b+\bar{\Lambda}$ as seen from Figs.1j and 1i, 
but their variations against $m_b+\bar{\Lambda}$ go to opposite direction, where $F_{1}$ decreases 
and $F_2$ increases as $m_b+\bar{\Lambda}$ goes to be large.  
As can be seen from these figures and eq.\ref{eq:71}, the values of $F_2$ are very small and its 
   contributions to $h_{A_{1}}(1)$ and $h_{+}(1)$ are negligible. Thus the main contributions 
to the form factors $h_{A_{1}}(1)$ and $h_{+}(1)$ arise from the 
form factors $\kappa_1$ and $\kappa_2$ at the order $1/m_Q$  as well as the form 
factor $F_1$ at the order $1/m^2_Q$. 
   
   To show typical values for the form factors $\kappa_1$, $\kappa_2$, 
   $F_1$ and $F_2$ at zero recoil, we take the central
   values of $m_b$ and $m_b-m_c$, namely, $m_b=4.7\mbox{GeV}$ and $m_b-m_c=3.36\mbox{GeV}$, 
   and $m_b+\bar{\Lambda}=5.21\mbox{GeV}$. The reliability of taking  $m_b+\bar{\Lambda}=5.21$ GeV
 will be seen explicitly below.  With these data, one can straightforwardly read from Figs.1a-1l
the following reasonable values for the form factors
   \begin{equation}
   \label{eq:76}
   \kappa_1 \sim -0.615 \mbox{GeV$^2$}, \;\;\; \kappa_2 \sim 0.056 \mbox{GeV$^2$} ,\;\;\; 
   F_{1} \sim 0.917\mbox{{GeV}$^4$}  ,\; \;\; F_{2} \sim 0.004\mbox{{GeV}$^4$}.
   \end{equation}
  With these typical values, $\delta^{\ast}$ and $\delta$ can be simply 
represented as functions of $\varrho_1$ and $\varrho_2$ 
   \begin{eqnarray}
   \label{eq:delpro}
   \delta^{\ast} &\approx& -0.045+0.14 \varrho_1 -0.362 \varrho_2 ,\nonumber\\
   \delta &\approx& -0.1+0.14(\varrho_1+3\varrho_2).
   \end{eqnarray}
 So $\delta^{\ast}$ and $\delta$ strongly depend on $\varrho_2$ and change in opposite 
direction. $\delta^{\ast}$ decreases whereas $\delta$ increases as 
 $\varrho_2$ goes up. Their dependence on $\varrho_1$ is relatively weak 
and has a similar behaviour.
   
   Up to now, the two form factors $\varrho_1$ and $\varrho_2$ have not yet 
   been determined. 
   But their contributions should be such that the values of 
   $\vert V_{cb}\vert$ estimated from the two channels 
   $B\rightarrow D^{\ast}l\nu$ and $B\rightarrow Dl\nu$ approximately equal 
   each other. From this consideration, we find that it is reliable to take
   \begin{equation}
   \label{eq:varrhoval}
  { \varrho_1 \approx 0.3 \mbox{GeV}^3, \hspace{1.5cm} \varrho_2 \approx 0.11 
   \mbox{GeV}^3 }.
   \end{equation}
 This can be explicitly seen from eqs.(\ref{eq:64}) and (\ref{eq:642}) and 
Fig.2. 
  $\vert V_{cb}\vert$ can be 
   extracted from either of the two exclusive B decay channels. 
 As shown in 
 Fig.2a, the $\vert V_{cb}\vert$ extracted from 
   $B\rightarrow D^{\ast}lv$ channel becomes large as $\varrho_2$ increases, 
   whereas the $\vert V_{cb}\vert$ extracted from 
   $B\rightarrow Dlv$ channel decreases as $\varrho_2$ increases as shown in 
 Fig.2b. 
Fig.2c shows $\vert V_{cb}\vert$ extracted from both of the channels as function of 
   $\varrho_2$. When taking $\varrho_2=0.08\mbox{GeV}^3$, the values of 
   $\vert V_{cb}\vert$ from the two channels differ obviously. While for 
   $\varrho_2=0.11\mbox{GeV}^3$, the two curves of $\vert V_{cb}\vert$ 
   almost coincide with each other. Their little difference may also be seen from 
Fig.2d.
   
   With the values of $\varrho_1$ and $\varrho_2$ given in eq.(\ref{eq:varrhoval}), 
   we obtain a reliable $\vert V_{cb} \vert $. The numerical results are listed in 
   Table. 1-3. 

   We also show in 
 Figs.3a-3b the parameter 
   $\delta^{\ast}$ as function of $m_b+\bar{\Lambda}$ for $m_b-m_c=$ 3.41GeV, 
   3.36GeV, 3.32GeV, respectively. The three lines in each figure correspond to  
   $m_b=4.6$GeV, $m_b=4.7$GeV and $m_b=4.8$GeV. Clearly, on each curve 
   there is a minimum around which the curve becomes relatively flat. 
   This means that the correction $\delta^{\ast}$ is not very sensitive to the 
   value of $m_b+\bar{\Lambda}$ around that minimal point. 
   As a consequence, the resulting values for $\vert V_{cb} \vert$ also becomes
   more stable around that minimal point of $m_b+\bar{\Lambda}$. 
   One can also see from Table. 1-3 or those figures in 
 Fig.3 that the minimal value 
   is near the point $m_b+\bar{\Lambda}=5.2$GeV, which has been taken as a reliable 
  value in getting the estimates in eq.(\ref{eq:76}). 
      
   It can be read from Fig.3 that within $4.6GeV 
   \leq m_b \leq 4.8GeV$ and $3.32GeV \leq m_b-m_c \leq 3.41GeV$, the values 
   of $\delta^{\ast}$ at the minimal point on a curve range 
   from $-$0.07 to $-$0.01. This gives an optimistic estimate: 
   \begin{equation}
   \label{eq:74}
   \delta^{\ast}=-0.04 \pm 0.03.
   \end{equation}
      
   To be more conservative, one may take a range of 
   $m_b+\bar{\Lambda}$ around the minimal point. For the 
   region $5.175\mbox{GeV}\leq m_b+\bar{\Lambda} \leq 5.275\mbox{GeV}$, 
the values of $\delta^{\ast}$ are found to be 
   \begin{equation}
   \label{eq:75}
   -0.07 <  \delta^{\ast} < 0.01.
   \end{equation}      
   
   The two figures in Fig.4 exhibit the $m_b$ and $m_b-m_c$ dependences of 
   $\delta^{\ast}$. It can be seen from Fig.4a that the variation of 
   $m_b$ (or $m_c$) only very slightly influences $\delta^{*}$ when $m_b-m_c$ is fixed. 
   On the contrary, Fig.4b shows that $\delta^{\ast}$ is 
   rather sensitive to the mass difference $m_b-m_c$ between the b and c quarks.
  Fig.5 shows the resulting $\vert V_{cb}\vert$ as function of $m_b$, 
   $m_b-m_c$ and $m_b+\bar{\Lambda}$. 
   
   By using the values of $\delta^{\ast}$ given in eqs.(\ref{eq:74}) 
   and (\ref{eq:75}), we then obtain from eq.(\ref{eq:64}) the important CKM matrix element
 $\vert V_{cb}\vert$ within the framework of the new formulation of HQEFT. Here we would like 
to present the numerical result for $\vert V_{cb}\vert$ from the eq.(\ref{eq:75})
   \begin{equation}
   \label{eq:78}
   \vert V_{cb} \vert = 0.0378 \pm 0.0028_{\mbox{exp}} \pm 0.0018_{\mbox{th}} .
   \end{equation}

  Now turn to the case of $B\rightarrow Dl\nu$ decay. 
  Following analogous analyses, we obtain
   \begin{eqnarray}
   &&\delta=-0.03 \pm 0.05 \;\;(\mbox{optimistic estimate}) ,\\
   &&-0.08 <  \delta < 0.06   \;\;(\mbox{conservative estimate}) ,
   \end{eqnarray} 
 and 
   \begin{equation}
   \vert V_{cb} \vert = 0.0382 \pm 0.0041_{\mbox{exp}} \pm 0.0028_{\mbox{th}} .
   \end{equation}    

\section{Summary}\label{sum}

  We have derived an effective Lagrangian in terms of effective heavy quark (or antiquark) 
fields by integrating out heavy antiquark (or quark) fields based on the
  framework of new formulation of HQEFT\cite{main}. The resulting heavy quark effective 
Lagrangian differs from the one in the usual HQET because of the 
additional contributions from the antiquark fields. An arbitrary heavy quark current is 
expanded into a power series of $1/{\MQ}$. Week matrix elements between heavy 
hadron states have been investigated in detail and explicitly evaluated up to the order 
of $1/m^2_Q$. The resulting special structures of the heavy quark effective Lagrangian 
and currents within the framework of new formulation of HQEFT results in some interesting 
features:  Firstly, it has been explicitly shown that Luke's theorem comes out automatically 
without the need of imposing the equation of motion $iv\cdot D Q_{v}^{+} =0$. 
	Secondly, the consistent normalization condition between two heavy hadrons naturally reflects 
spin-flavor symmetry, and the form factors at zero recoil are found to be related to the meson 
masses, so that the most important relevant form factors at zero recoil have been fitted 
from the ground state meson masses. Thirdly, we find $h_{-}(\omega)=0$, so the differential 
decay rates of both $B\rightarrow D^{\ast}l\nu$ and $B\rightarrow D l\nu$ have no $1/m_{Q}$ order corrections 
at zero recoil. This enables one to extract the CKM matrix element $\vert V_{cb}\vert$ 
from either of the two exclusive semileptonic decays at the order of $1/m^2_Q$. 
Finally, a set of 29 universal form factors up to the order 
  of $1/m^2_Q$, which is less than the one in the usual HQET, has been introduced to 
 describe transition matrix elements of weak currents 
  between ground state pseudoscalar and vector mesons. Note that at zero recoil 
  point, some of the form factors are kinematically suppressed and only 15 form 
  factors contribute to the weak matrix elements. By reasonable considerations, 
the zero recoil heavy meson transition matrix 
  elements can be approximately described by only 6 form factors. These form factors are 
  hadronic parameters concerning long distance dynamics and are hardly 
  evaluated, four of them have been estimated from ground state meson masses.
  Following the conventional strategy, we have extracted $\vert V_{cb} \vert$ from the exclusive 
  semileptonic decay channel $B\rightarrow D^{\ast} l\nu$ with the value  

  \[ \vert V_{cb}\vert=0.0378 \pm 0.0028_{\mbox{exp}} \pm 0.0018_{\mbox{th}}\  ,  \]
and from the exclusive semileptonic decay channel $B\rightarrow D l\nu$ with the value 
 
  \[ \vert V_{cb} \vert = 0.0382 \pm 0.0041_{\mbox{exp}} \pm 0.0028_{\mbox{th}}\  . \]
    
  In this paper, we have shown some interesting features in applying the new formulation of HQEFT to
 the exclusive semileptonic decays of heavy hadrons. More interesting features 
can be found in applying the new formulation of HQEFT to the inclusive decays 
of heavy hadrons\cite{WWY}, such as: 
the new formulation of HQEFT allows us to simply clarify the well known ambiguity of 
using the quark mass or hadron mass in the inclusive heavy hadron decays, as a consequence, 
the CKM matrix element $|V_{cb}|$ can also be well determined from the inclusive semileptonic 
decay rate, and the result is nicely consistent with the above one;  
the resulting lifetime differences between bottom mesons and baryons also agree well 
 with the experimental data.

\acknowledgments

We would like to thank professor Y.B. Dai for useful discussions. 
This work was supported in part by the NSF of China under the grant No. 19625514.

\appendix

\section{Functional integral method}\label{app:functional}
  Corresponding to eq.(\ref{eq:15}), the generating functional relevant to heavy quarks is as follows,
   \begin{eqnarray}
   \label{eq:17}
     \tilde{W}[\bar{\eta}^{+}_v,\eta^{+}_v&&,\bar{\eta}^{-}_v,\eta^{-}_v]
     =N\int D\QVHZB D \QVHZ D\QVHFB D\QVHF \exp\{i\int d^4 x[L_{Q,v}+\bar{\eta}^{+}_v 
      \QVHZ \nonumber \\
     && +\QVHZB \eta^{+}_v+\bar{\eta}^{-}_v \QVHF+ \QVHFB \eta^{-}_v]\} \nonumber\\
      =&& N \int D \QVHZB D \QVHZ D\QVHFB D \QVHF \exp\{i\int d^4 x
      [\QVHZB \hat{A} \QVHZ + \QVHFB \hat{A} \QVHF \nonumber\\
    && +\QVHZB \hat{B} \QVHF +\QVHFB \hat{B} \QVHZ +
      \bar{\eta}^{+}_v \QVHZ +\QVHZB \eta^{+}_v +\bar{\eta}^{-}_v \QVHF+ \QVHFB \eta^{-}_v ]  \}  ,
   \end{eqnarray}
  where $\bar{\eta}^{+}_v$, $\eta^{+}_v$ and $\bar{\eta}^{-}_v$, $\eta^{-}_v$ are external 
  sources of quark and antiquark fields, respectively. $N$ is a normalization 
  constant. The operators $\hat{A}$ and $\hat{B}$ have been defined in 
  eq.(\ref{eq:16}).

  Integrating over $\QVHFB$ and $\QVHF$, eq.(\ref{eq:17}) becomes
     \begin{eqnarray}  
     \label{eq:18}
      \tilde{W}[\bar{\eta}^{+}_v,\eta^{+}_v,\bar{\eta}^{-}_v,\eta^{-}_v]&=&N 
        \det(-i\hat{A}) \int D \QVHZB D\QVHZ \exp\{i\int d^4x
        [\QVHZB (\hat{A}-\hat{B} \hat{A}^{-1} \hat{B})\QVHZ  \nonumber\\
   && + (\bar{\eta}^{+}_v-\bar{\eta}^{-}_v\hat{A}^{-1}\hat{B})\QVHZ
        + \QVHZB(\eta^{+}_v-\hat{B}\hat{A}^{-1}\eta^{-}_v)
        -\bar{\eta}^{-}_v\hat{A}^{-1}\eta^{-}_v]  \}.
     \end{eqnarray} 
  And then setting the sources $\bar{\eta}^{-}_v$ and $\eta^{-}_v$ to be zero, we get
   \begin{eqnarray}
   \label{eq:19}
       { W[\bar{\eta}^{+}_v,\eta ^{+}_v]=N \det(-i\hat{A}) \int D \QVHZB
      D \QVHZ e^{i\int d^4 x[\QVHZB (\hat{A}-\hat{B}\hat{A}^{-1} \hat{B})\QVHZ 
      + \QVHZB \eta^{+}_v +\bar{\eta}^{+}_v \QVHZ]} }.
   \end{eqnarray} 
  The factor $N\det(-i\hat{A})$ contains no quark field and can be seen as 
  a new normalization factor. Then the effective Lagrangian eq.(\ref{eq:20}) 
  can be read from eq.(\ref{eq:19}).

To get the effective current, one can express $\QVHF$ and $\QVHFB$ in eq.(\ref{eq:29}) 
by the variationals of the generating functional eq.(\ref{eq:17}) over the corresponding 
external sources,
    \begin{eqnarray}
    \label{eq:30}
       \QVHF&=&(\frac{\delta}{i\delta \bar{\eta}^{-}_v}W[\bar{\eta}^{+}_v,
          {\eta}^{+}_v,\bar{\eta}^{-}_v,{\eta}^{-}_v])_{\bar{\eta}^{\pm}_v={\eta}^{\pm}_v
          =0,}  \nonumber\\
       \QVHFB&=&(\frac{\delta}{i\delta {\eta}^{-}_v}W[\bar{\eta}^{+}_v,
          {\eta}^{+}_v,\bar{\eta}^{-}_v,{\eta}^{-}_v])_{\bar{\eta}^{\pm}_v
          ={\eta}^{\pm}_v=0 .}  
    \end{eqnarray} 
  Combining eqs.(\ref{eq:11}), (\ref{eq:18}) and (\ref{eq:30}), eq.(\ref{eq:31}) is obtained 
  once more.

\section{Decomposition}\label{app:decomp}

  Using the identity $ v_\alpha P_{+} \sigma^{\alpha \beta}{\cal M}=0 $, the Lorentz tensors 
  $\kappa_{\alpha \beta}(v,v^{\prime})$, $\varrho_{\alpha \beta}(v,v^{\prime})$, 
  $\chi_{\alpha \beta}(v,v^{\prime})$, $\eta_{\alpha \beta}(v,v^{\prime})$, 
  $\chi_{\alpha\beta\gamma\delta}(v,v^{\prime})$, 
  and $\eta_{\alpha\beta\gamma\delta}(v,v^{\prime})$ 
  can be decomposed into the following general forms 
  in terms of Lorentz scalar factors 
     \begin{eqnarray}
     \label{eq:57}
        \kappa_{\alpha \beta}(v,&&v^{\prime})=i\kappa_2(\omega) \sigma_{\alpha \beta}
         +\kappa_3(\omega) (v^{\prime}_{\alpha}\gamma_{\beta}-v^{\prime}_{\beta} 
           \gamma_{\alpha}),\nonumber\\
  \varrho_{\alpha \beta}(v,&&v^{\prime})=i\varrho_{2}(\omega) \sigma_{\alpha \beta}
         + \varrho_3(\omega) (v^{\prime}_{\alpha}\gamma_{\beta}-v^{\prime}
           _{\beta} \gamma_{\alpha}),\nonumber\\
 \chi_{\alpha \beta}(v,&&v^{\prime})=i\chi_2(\omega) \sigma_{\alpha \beta}
         + \chi_3(\omega) (v^{\prime}_{\alpha}\gamma_{\beta}-v^{\prime}
           _{\beta} \gamma_{\alpha}), \nonumber\\
\eta_{\alpha \beta}(v,&&v^{\prime}) = i\eta_{2}(\omega) \sigma_{\alpha \beta}
          +\eta_3(\omega) (v^{\prime}_{\alpha}\gamma_{\beta}-v^{\prime}
           _{\beta} \gamma_{\alpha}), \nonumber\\
\chi_{\alpha\beta\gamma\delta}(v,&&v^{\prime}) = \chi_{4}(\omega)(g_{\alpha\gamma}
     g_{\beta\delta}-g_{\alpha\delta}g_{\beta\gamma})+\chi_{5}(\omega)
     \sigma_{\gamma\delta}\sigma_{\alpha\beta}+i\chi_{6}(\omega)(g_{\alpha\gamma}
     \sigma_{\beta\delta} -g_{\beta\gamma}\sigma_{\alpha\delta}\nonumber\\
&&      -g_{\alpha\delta} \sigma_{\beta\gamma}    
     + g_{\beta\delta}\sigma_{\alpha\gamma}) 
     +\chi_{7}(\omega)(v^{\prime}_{\gamma} \gamma_{\delta}
     -v^{\prime}_{\delta} \gamma_{\gamma})(v^{\prime}_{\alpha}\gamma_{\beta}
     -v^{\prime}_{\beta}\gamma_{\alpha}) 
     +\chi_{8}(\omega)(g_{\alpha\gamma} v^{\prime}_{\beta}v^{\prime}_{\delta} 
\nonumber      \\   
&&     -g_{\beta\gamma} v^{\prime}_{\alpha}v^{\prime}_{\delta}
     -g_{\alpha\delta} v^{\prime}_{\beta}v^{\prime}_{\gamma}
     +g_{\beta\delta} v^{\prime}_{\alpha}v^{\prime}_{\gamma})
     +\chi_{9}(\omega)(g_{\alpha\gamma} v^{\prime}_{\beta}\gamma_{\delta}
     -g_{\beta\gamma} v^{\prime}_{\alpha}\gamma_{\delta}
     -g_{\alpha\delta} v^{\prime}_{\beta}\gamma_{\gamma} \nonumber\\
&&     +g_{\beta\delta} v^{\prime}_{\alpha}\gamma_{\gamma})  
     +\chi_{10}(\omega)(g_{\alpha\gamma} \gamma_{\beta}v^{\prime}_{\delta}
     -g_{\beta\gamma} \gamma_{\alpha}v^{\prime}_{\delta}
     -g_{\alpha\delta} \gamma_{\beta}v^{\prime}_{\gamma}
     +g_{\beta\delta} \gamma_{\alpha}v^{\prime}_{\gamma}) 
    +i\chi_{11}(\omega) \nonumber \\
&&     \times (\sigma_{\alpha\gamma} v^{\prime}_{\beta}\gamma_{\delta}
     -\sigma_{\beta\gamma} v^{\prime}_{\alpha}\gamma_{\delta} 
     -\sigma_{\alpha\delta} v^{\prime}_{\beta}\gamma_{\gamma}   
     +\sigma_{\beta\delta} v^{\prime}_{\alpha}\gamma_{\gamma}) 
     +i\chi_{12}(\omega)(\sigma_{\alpha\gamma} \gamma_{\beta}v^{\prime}_{\delta}
     -\sigma_{\beta\gamma} \gamma_{\alpha}v^{\prime}_{\delta}  \nonumber\\
&&     -\sigma_{\alpha\delta} \gamma_{\beta}v^{\prime}_{\gamma}
     +\sigma_{\beta\delta} \gamma_{\alpha}v^{\prime}_{\gamma}) ,\nonumber   \\
 \eta_{\alpha\beta\gamma\delta}(v,&&v^{\prime}) = \eta_{4}(\omega)(g_{\alpha\gamma}
     g_{\beta\delta}-g_{\alpha\delta}g_{\beta\gamma})+\eta_{5}(\omega)
     \sigma_{\gamma\delta}\sigma_{\alpha\beta}  
          +i\eta_{6}(\omega)(g_{\alpha\gamma}
     \sigma_{\beta\delta}-g_{\beta\gamma}\sigma_{\alpha\delta}  \nonumber\\
&&      -g_{\alpha\delta} \sigma_{\beta\gamma} +g_{\beta\delta}\sigma_{\alpha\gamma})  
     +\eta_{7}(\omega)(v^{\prime}_{\gamma} \gamma_{\delta}
     -v^{\prime}_{\delta} \gamma_{\gamma})(v_{\alpha}\gamma_{\beta} 
     -v_{\beta}\gamma_{\alpha}) 
     +\eta_{8}(\omega)(g_{\alpha\gamma} v_{\beta}v^{\prime}_{\delta} \nonumber\\
&&  -g_{\beta\gamma} v_{\alpha} v^{\prime}_{\delta}  
     -g_{\alpha\delta} v_{\beta}v^{\prime}_{\gamma}
     -g_{\beta\delta} v_{\alpha}v^{\prime}_{\gamma}) 
      +\eta_{9}(\omega)(g_{\alpha\gamma} v_{\beta}\gamma_{\delta} 
     -g_{\beta\gamma} v_{\alpha}\gamma_{\delta}
     -g_{\alpha\delta} v_{\beta}\gamma_{\gamma}
\nonumber     \\
&&     +g_{\beta\delta} v_{\alpha}\gamma_{\gamma}           
     +g_{\alpha\gamma} \gamma_{\beta}v^{\prime}_{\delta}
     -g_{\beta\gamma} \gamma_{\alpha}v^{\prime}_{\delta}
     -g_{\alpha\delta} \gamma_{\beta}v^{\prime}_{\gamma}
     +g_{\beta\delta} \gamma_{\alpha}v^{\prime}_{\gamma})    
     +i\eta_{10}(\omega)(v_{\beta} \gamma_{\delta}\sigma_{\alpha\gamma} \nonumber\\
&&     -v_{\alpha} \gamma_{\delta}\sigma_{\beta\gamma}
     -v_{\beta} \gamma_{\gamma}\sigma_{\alpha\delta}     
     +v_{\alpha} \gamma_{\gamma}\sigma_{\beta\delta}
     +\sigma_{\alpha\gamma}\gamma_{\beta}v^{\prime}_{\delta}
     -\sigma_{\beta\gamma}\gamma_{\alpha}v^{\prime}_{\delta} 
     -\sigma_{\alpha\delta}\gamma_{\beta}v^{\prime}_{\gamma}
     +\sigma_{\beta\delta}\gamma_{\alpha}v^{\prime}_{\gamma}).
    \end{eqnarray}

\section{Meson form factors}\label{app:factor}

Since all form factors are functions of $\omega$, we will neglect the 
variable $\omega$ in following formulae for simplicity. 
To second order in the new framework of HQEFT, the meson form factors defined in 
eq.(\ref{eq:delhi}) are given by

\newcommand{\FYE}[2]
{(\frac{1}{4 #1^2}+\frac{1}{4 #2^2})(-\bar{\Lambda} \kappa_{1}
 +\varrho_{1}+\frac{1}{\bar{\Lambda}}\chi_{1}) }

\newcommand{\FYEB}
{(-\bar{\Lambda} \kappa_{1}+\varrho_{1}+\frac{1}{\bar{\Lambda}}\chi_{1}) }

\newcommand{\FEE}[1]
{\frac{1}{4 #1^2}(\bar{\Lambda} \kappa_{2}
  -\varrho_{2}-\frac{1}{\bar{\Lambda}}\chi_{2}) }  

\newcommand{\FEEB}
{(\bar{\Lambda} \kappa_{2}-\varrho_{2}-\frac{1}{\bar{\Lambda}}\chi_{2}) }   

\newcommand{\FSE}[1]{\frac{1}{4 #1^2}(\bar{\Lambda} \kappa_{3}
  -\varrho_{3}-\frac{1}{\bar{\Lambda}}\chi_{3}) }    

\newcommand{\FSEB}{(\bar{\Lambda} \kappa_{3}-\varrho_{3}-\frac{1}{\bar{\Lambda}}\chi_{3}) }     

\newcommand{\MM}[3]{\frac{1}{#3 #1}+\frac{1}{#3 #2} }

\newcommand{\MME}[3]{\frac{1}{#3 #1^2}+\frac{1}{#3 #2^2} }
  
\begin{eqnarray}
h_{+}&=&\xi+(\xi-1)[\frac{1}{2\bar{\Lambda}}(\frac{1}{m_b}+\frac{1}{m_c})
  (\kappa_{1}+3\kappa_{2})+\frac{1}{4\bar{\Lambda}}(\MME{m_b}{m_c}{})
  (c_1-3c_2-\frac{1}{\bar{\Lambda}} c_4)]\nonumber\\
&&  +\xi[-\frac{1}{8\bar{\Lambda}^2}c_6
   +\frac{1}{8\bar{\Lambda}^2}(\frac{3}{m_b^2}+\frac{3}{m_c^2} 
  +\frac{2}{m_b m_c})(\kappa_1+3\kappa_2)^2 ]+\frac{1}{\bar{\Lambda}}(1-\omega)\nonumber\\
&&\times  [(\MM{m_b}{m_c}{})\kappa_3-(\MME{m_b}{m_c}{2}) c_3] 
+\frac{1}{4\bar{\Lambda}^2}(\MME{m_b}{m_c}{}) [(4\chi_7+2\chi_8)
 (1-\omega^2)\nonumber\\
&& -4(\chi_9+\chi_{12}) (1-\omega)]  
  +\frac{1}{4m_b m_c \bar{\Lambda}^2}[\eta_1
  +6\eta_2 -4\eta_3(1-\omega) -\eta_4(1+2\omega)\nonumber\\
&&  -9\eta_5-6\eta_6+(4\eta_7+2\eta_8)(1-\omega^2) -8(\eta_9+2\eta_{10})(1-\omega)]  \nonumber\\
&&   -\frac{1}{4\bar{\Lambda}^2}(\MM{m_b}{m_c}{})^2 (\kappa_1+3\kappa_2)^2 
  +\frac{1}{2\bar{\Lambda}^2}(\MM{m_b}{m_c}{})^2 
\kappa_3(\kappa_1+3\kappa_2)(1-\omega) ,\\
h_{-}&=&0 ,\\
h_{1}&=&\xi+(\xi-1)[\frac{1}{2\bar{\Lambda}}(\MM{m_b}{m_c}{})
  (\kappa_{1}-\kappa_{2})+\frac{1}{4\bar{\Lambda}} (\MME{m_b}{m_c}{})
  (c_1+c_2-\frac{1}{\bar{\Lambda}} c_5)]  \nonumber\\
&&  +\xi [-\frac{1}{8\bar{\Lambda}^2}(\MME{m_b}{m_c}{})c_7
  +\frac{1}{8\bar{\Lambda}^2}(\frac{3}{m_b^2}+\frac{3}{m_c^2}  
  +\frac{2}{m_b m_c})   (\kappa_1-\kappa_2)^2 ]    \nonumber\\
&&  +\frac{1}{4m_b^2 \bar{\Lambda}^2}[2\chi_8(1-\omega^2)-(2\chi_9+2\chi_{10}
  -6\chi_{11}-2\chi_{12})
  (1-\omega)] \nonumber \\
&&  +\frac{1}{4m_c^2 \bar{\Lambda}^2}[2\chi_8(1-\omega^2)-(2\chi_{10}-6\chi_{11}-2\chi_{12})
  (1-\omega)] +\frac{1}{4m_b m_c \bar{\Lambda}^2}[\eta_1\nonumber\\
&& -2\eta_2 -\eta_4(1+2\omega)
  -\eta_5-2\eta_6(1-2\omega)+ 2\eta_8(1-\omega^2)
  -4(\eta_9-\eta_{10})(1-\omega)]  \nonumber\\
&& -\frac{1}{4\bar{\Lambda}^2}(\MM{m_b}{m_c}{})^2 (\kappa_1-\kappa_2)^2, \\
h_{2}&=&0 ,\\
h_{3}&=&\xi+(\xi-1)[\frac{1}{2\bar{\Lambda}}(\MM{m_b}{m_c}{})
  (\kappa_{1}-\kappa_{2})+\frac{1}{4\bar{\Lambda}} (\MME{m_b}{m_c}{})
  (c_1+c_2-\frac{1}{\bar{\Lambda}} c_5) ]\nonumber\\
&&  +\xi[-\frac{1}{8\bar{\Lambda}^2}(\MME{m_b}{m_c}{}) c_7
   +\frac{1}{8\bar{\Lambda}^2}(\frac{3}{m_b^2}+\frac{3}{m_c^2}
   +\frac{2}{m_b m_c})   (\kappa_1-\kappa_2)^2 ]\nonumber\\
&&    -\frac{1}{m_c \bar{\Lambda}}\kappa_3 (1-\omega)
   +\frac{1}{2m^2_c \bar{\Lambda}}(1-\omega) c_3  
     +\frac{1}{4m_b^2 \bar{\Lambda}^2}[2\chi_8(1-\omega^2)\nonumber \\
&& -(2\chi_9+2\chi_{10}-6\chi_{11}-2\chi_{12})
  (1-\omega)] +\frac{1}{4m_c^2 \bar{\Lambda}^2}[(4\chi_7+2\chi_8)(1-\omega^2)\nonumber\\
&&  -(4\chi_{10}+4\chi_{11}+8\chi_{12})
  (1-\omega)] +\frac{1}{4m_b m_c \bar{\Lambda}^2}[\eta_1-2\eta_2+2\eta_3(1-\omega)  \nonumber\\
&& +\eta_4 -\eta_5-2\eta_6-2(\eta_9+\eta_{10})(1-\omega)]
  -\frac{1}{4\bar{\Lambda}^2}(\MM{m_b}{m_c}{})^2 (\kappa_1-\kappa_2)^2   \nonumber\\
&& -\frac{1}{2 m_c \bar{\Lambda}^2}(\MM{m_b}{m_c}{})\kappa_3(\kappa_1-\kappa_2)(1-\omega),\\
h_{4}&=&\xi+(\xi-1)[\frac{1}{2\bar{\Lambda}}(\MM{m_b}{m_c}{})
  (\kappa_{1}-\kappa_{2})+\frac{1}{4\bar{\Lambda}} (\MME{m_b}{m_c}{})
  (c_1+c_2-\frac{1}{\bar{\Lambda}} c_5)]   \nonumber\\
&&
  +\xi [-\frac{1}{8\bar{\Lambda}^2}(\MME{m_b}{m_c}{}) c_7
  +\frac{1}{8\bar{\Lambda}^2}(\frac{3}{m_b^2}+\frac{3}{m_c^2}
  +\frac{2}{m_b m_c})   (\kappa_1-\kappa_2)^2 ]  \nonumber\\
&&
  -\frac{1}{m_b \bar{\Lambda}}\kappa_3 (1-\omega) +\frac{1}{2 m^2_b \bar{\Lambda}}c_3(1-\omega)
   +\frac{1}{4m_b^2 \bar{\Lambda}^2}[(4\chi_7+2\chi_8)(1-\omega^2)  \nonumber\\
&&-(4\chi_{10}+4\chi_{11}+8\chi_{12})
  (1-\omega)]  +\frac{1}{4m_c^2 \bar{\Lambda}^2}[2\chi_8(1-\omega^2)
   -(2\chi_{10}-6\chi_{11}\nonumber\\
&&  -2\chi_{12})(1-\omega)]  
  +\frac{1}{4m_b m_c \bar{\Lambda}^2}[\eta_1-2\eta_2+2\eta_3(1-\omega)
   +\eta_4-\eta_5-2\eta_6\nonumber\\
&&-2(\eta_9+\eta_{10})(1-\omega)] -\frac{1}{4\bar{\Lambda}^2}
    (\MM{m_b}{m_c}{})^2 (\kappa_1-\kappa_2)^2  \nonumber\\
&& -\frac{1}{2 m_b \bar{\Lambda}^2}(\MM{m_b}{m_c}{})\kappa_3(\kappa_1-\kappa_2)(1-\omega),\\
h_{5}&=&\frac{1}{m_c \bar{\Lambda}} \kappa_3 -\frac{1}{2 m^2_c \bar{\Lambda}} c_3
  -\frac{1}{4m_c^2 \bar{\Lambda}^2}[4\chi_7(1+\omega)-2\chi_{10}-10\chi_{11}   -10\chi_{12}]\nonumber \\
&&
 -\frac{1}{4m_b m_c \bar{\Lambda}^2}[2\eta_3-2\eta_4+4\eta_6+4\eta_7+2\eta_8(1-\omega)
  +2\eta_9-6\eta_{10}]  \nonumber\\
&&  +\frac{1}{2m_c \bar{\Lambda}^2}(\MM{m_b}{m_c}{})
  \kappa_3 (\kappa_1-\kappa_2) ,\\
h_{6}&=&\frac{1}{m_b \bar{\Lambda}} \kappa_3 -\frac{1}{2m_b^2\bar{\Lambda}} c_3
  -\frac{1}{4m_b^2 \bar{\Lambda}^2}[4\chi_7(1+\omega)+2\chi_9-2\chi_{10}
  -10\chi_{11}-10\chi_{12}]\nonumber\\
&&   -\frac{1}{4m_b m_c \bar{\Lambda}^2}[2\eta_3-2\eta_4+4\eta_6+4\eta_7
  +2\eta_8(1-\omega)+2\eta_9 -6\eta_{10}]  \nonumber\\
&&  +\frac{1}{2m_b \bar{\Lambda}^2}(\MM{m_b}{m_c}{})\kappa_3 (\kappa_1-\kappa_2),\\
h_{7}&=&\xi+(\xi-1)[\frac{1}{2\bar{\Lambda}}(\MM{m_b}{m_c}{})
  \kappa_{1}+\frac{1}{4\bar{\Lambda}} (\MME{m_b}{m_c}{}) c_1 
  +\frac{1}{4m_c^2\bar{\Lambda}} (c_2-\frac{1}{\bar{\Lambda}} c_5)]\nonumber\\
&&  +\xi [-\frac{1}{2\bar{\Lambda}}(\MM{m_b}{m_c}{}) \kappa_2 
   +\frac{1}{4m_b^2\bar{\Lambda}}(c_2-\frac{1}{\bar{\Lambda}} c_5)
   -\frac{1}{8\bar{\Lambda}^2}(\MME{m_b}{m_c}{}) c_7\nonumber\\
&& +\frac{1}{8\bar{\Lambda}^2}(\kappa_1-\kappa_2)^2(\frac{3}{m_b^2}+\frac{3}{m_c^2}
   +\frac{2}{m_b m_c})]+\frac{1}{2\bar{\Lambda}}
      [\frac{1}{m_c} -\frac{1}{m_b}(1-2\omega)]\kappa_2\nonumber\\
&&   +\frac{1}{4m_b^2\bar{\Lambda}} c_2 (1-2\omega) 
   +\frac{1}{4m_b^2 \bar{\Lambda}^2}[3\chi_4  +\chi_5(5-4\omega)+2\chi_6(1-2\omega)    \nonumber\\
&&   +2(\chi_7+\chi_8)(1-\omega^2)
   -\chi_9(3-\omega-2\omega^2)  -\chi_{10}(1+\omega-2\omega^2) \nonumber\\
&& +\chi_{11}(3-5\omega+2\omega^2)-\chi_{12}(1+5\omega-6\omega^2)] 
  +\frac{1}{4m_c^2 \bar{\Lambda}^2}[2\chi_8(1-\omega^2)\nonumber\\
&&   -(2\chi_9+2\chi_{10}-6\chi_{11}
   -2\chi_{12})(1-\omega)] +\frac{1}{4m_b m_c \bar{\Lambda}^2}[\eta_1-2\omega\eta_2+\omega\eta_4  \nonumber\\
&&   +(\eta_5+2\eta_6)(1-2\omega)-\eta_8(1-\omega^2)
    +(3\eta_9-\eta_{10})(1-\omega)]-\frac{1}{4\bar{\Lambda}^2}(\MM{m_b}{m_c}{})^2   \nonumber\\
&& \times \kappa_1(\kappa_1-\kappa_2) 
  +\frac{1}{4\bar{\Lambda}^2}(\MM{m_b}{m_c}{})
  [\frac{1}{m_c}-\frac{1}{m_b}(1-2\omega)]\kappa_2(\kappa_1-\kappa_2) ,\\
h_{8}&=&
   -\frac{1}{m_b\bar{\Lambda}}(1+\omega)(\kappa_2-\kappa_3)+\frac{1}{2m_b^2\bar{\Lambda}}
   (1+\omega) (c_2-c_3)  +\frac{1}{4m_b^2\bar{\Lambda}^2}[4(\chi_5+\chi_6)(1+\omega)\nonumber\\  
&& +2\chi_7(1+2\omega+\omega^2)
   -\chi_9(1+3\omega+2\omega^2) +\chi_{10}(1 -\omega-2\omega^2)  
   -\chi_{11}(3+5\omega+2\omega^2)\nonumber\\
&&  -3\chi_{12}(1+3\omega+2\omega^2)]+\frac{1}{4m_b m_c \bar{\Lambda}^2}
[(2\eta_2-2\eta_3 +\eta_4+2\eta_5-\eta_9 +3\eta_{10})(1+\omega)\nonumber\\
&&  +\eta_8(1+2\omega+\omega^2)]  
  -\frac{1}{2m_b\bar{\Lambda}^2}(\MM{m_b}{m_c}{})
  (1+\omega)(\kappa_1-\kappa_2)(\kappa_2-\kappa_3),\\
h_{9}&=& \frac{2}{m_b\bar{\Lambda}}\kappa_2-\frac{1}{m_b^2\bar{\Lambda}} c_2
   -\frac{1}{\bar{\Lambda}}(\MM{m_b}{m_c}{})\kappa_3  
   +\frac{1}{2\bar{\Lambda}}(\MME{m_b}{m_c}{}) c_3
   -\frac{1}{4m_b^2\bar{\Lambda}^2}[8\chi_5+8\chi_6\nonumber\\
&& +4\chi_7(1+\omega)-2\chi_9(1+2\omega) 
   +2\chi_{10}(1-2\omega)
   -\chi_{11}(6+4\omega)-6\chi_{12}(1+2\omega)]  \nonumber\\
&&  -\frac{1}{4m_c^2\bar{\Lambda}^2}[-4\chi_7(1+\omega)-2\chi_9
   +2\chi_{10}+10\chi_{11}+10\chi_{12}] 
   -\frac{1}{m_b m_c \bar{\Lambda}^2}[\eta_2-\eta_3\nonumber\\
&&  +\eta_5+\eta_6+\eta_9+\eta_{10}]+\frac{1}{m_b\bar{\Lambda}^2}
    (\MM{m_b}{m_c}{})\kappa_2(\kappa_1-\kappa_2)  
    -\frac{1}{2\bar{\Lambda}^2}(\MM{m_b}{m_c}{})^2\nonumber\\
&&\times \kappa_3(\kappa_1-\kappa_2)  ,\\
h_{10}&=&
   \frac{1}{4m_b^2\bar{\Lambda}^2}[4\chi_7(1+\omega)+2\chi_9-2\chi_{10}-10\chi_{11}
   -10\chi_{12}]  
  +\frac{1}{4m_b m_c \bar{\Lambda}^2}[2\eta_4-4\eta_6\nonumber\\
&&  +2\eta_8(1+\omega)-6\eta_9+2\eta_{10}] , \\
h_{11}&=&
  \frac{1}{m_b\bar{\Lambda}}(-2\kappa_2+\kappa_3)+\frac{1}{m_b^2\bar{\Lambda}} 
  (c_2-\frac{1}{2}c_3) 
   +\frac{1}{4m_b^2\bar{\Lambda}^2}[8\chi_5+8\chi_6+4\chi_7(1+\omega)\nonumber\\
&&   -2\chi_9(1+2\omega)+2\chi_{10}(1-2\omega) -\chi_{11}(6+4\omega)
   -6\chi_{12}(1+2\omega)]\nonumber\\
&&  +\frac{1}{4m_b m_c \bar{\Lambda}^2}[4\eta_2-2\eta_3+2\eta_4+4\eta_5
  +2\eta_8(1+\omega) -2\eta_9+6\eta_{10}] \nonumber\\
&&  +\frac{1}{2m_b\bar{\Lambda}^2}(\MM{m_b}{m_c}{})(-2\kappa_2+\kappa_3)(\kappa_1-\kappa_2) ,\\
h_{12}&=&
  \frac{1}{m_b\bar{\Lambda}}\kappa_3-\frac{1}{2m_b^2\bar{\Lambda}} c_3
   -\frac{1}{2m_b m_c \bar{\Lambda}^2}\eta_3  
   +\frac{1}{2m_b\bar{\Lambda}^2}(\MM{m_b}{m_c}{})\kappa_3(\kappa_1-\kappa_2),\\
h_{V}&=&-\xi-(\xi-1)[\frac{1}{2m_b \bar{\Lambda}}(\kappa_{1}+3\kappa_{2})
  +\frac{1}{2m_c \bar{\Lambda}} (\kappa_{1}-\kappa_{2}) +\frac{1}{4\bar{\Lambda}}  
  (\MME{m_b}{m_c}{})c_1\nonumber\\
&&   -\frac{1}{4\bar{\Lambda}}(\frac{3}{m_b^2}-\frac{1}{m_c^2}) c_2
   -\frac{1}{4m_b^2 \bar{\Lambda}^2} c_4 
  -\frac{1}{4m_c^2 \bar{\Lambda}^2} c_5 ] 
  +\xi [\frac{1}{8m_b^2 \bar{\Lambda}^2} c_6 +\frac{1}{8m_c^2 \bar{\Lambda}^2} c_7 \nonumber\\
&&  -\frac{1}{8\bar{\Lambda}^2}(\frac{3}{m_b^2}(\kappa_1+3\kappa_2)^2 
  +\frac{3}{m_c^2}(\kappa_1-\kappa_2)^2
  +\frac{2}{m_b m_c}(\kappa_1+3\kappa_2)(\kappa_1-\kappa_2))] \nonumber\\
&& +\frac{1}{2m_b^2\bar{\Lambda}} c_3(1-\omega) 
  -\frac{1}{m_b \bar{\Lambda}}\kappa_3 (1-\omega)
  -\frac{1}{4m_b^2 \bar{\Lambda}^2}[(4\chi_7+2\chi_8)(1-\omega^2)\nonumber\\
&& -4(\chi_{9}+\chi_{12}) (1-\omega)]  
   -\frac{1}{4m_c^2 \bar{\Lambda}^2}[2\chi_8(1-\omega^2)-(2\chi_9+2\chi_{10}-6\chi_{11} \nonumber\\
&&  -2\chi_{12})(1-\omega)] -\frac{1}{4m_b m_c \bar{\Lambda}^2}[\eta_1+2\eta_2-2\eta_3(1-\omega) 
   +\eta_4+3\eta_5 +2\eta_6\nonumber\\
&& +2(\eta_9+\eta_{10})(1-\omega)] 
  +\frac{1}{4\bar{\Lambda}^2}[\frac{1}{m_b}(\kappa_1+3\kappa_2)+\frac{1}{m_c} 
  (\kappa_1-\kappa_2)]^2 \nonumber\\
&&-\frac{1}{2m_b \bar{\Lambda}^2}\kappa_3(1-\omega)[\frac{1}{m_b}(\kappa_1+3\kappa_2)
+\frac{1}{m_c}(\kappa_1-\kappa_2)] ,\\
h_{A_{1}}&=&\xi+(\xi-1)[\frac{1}{2m_b \bar{\Lambda}}(\kappa_{1}+3\kappa_{2})
  +\frac{1}{2m_c \bar{\Lambda}} (\kappa_{1}-\kappa_{2}) +\frac{1}{4\bar{\Lambda}}
  (\MME{m_b}{m_c}{}) c_1\nonumber\\
&& -\frac{1}{4\bar{\Lambda}}(\frac{3}{m_b^2}-\frac{1}{m_c^2}) c_2 
  -\frac{1}{4m_b^2 \bar{\Lambda}^2} c_4
 -\frac{1}{4m_c^2 \bar{\Lambda}^2} c_5 ]  
  +\xi [-\frac{1}{8m_b^2 \bar{\Lambda}^2} c_6-\frac{1}{8m_c^2 \bar{\Lambda}^2} c_7\nonumber\\
&&   +\frac{1}{8\bar{\Lambda}^2}(\frac{3}{m_b^2}(\kappa_1+3\kappa_2)^2 
 +\frac{3}{m_c^2}(\kappa_1-\kappa_2)^2 
  +\frac{2}{m_b m_c}(\kappa_1+3\kappa_2)(\kappa_1-\kappa_2))]  \nonumber\\
&&  +\frac{1}{m_b \bar{\Lambda}}\kappa_3 (1-\omega)
  -\frac{1}{2m_b^2\bar{\Lambda}} c_3 (1-\omega)
  +\frac{1}{4m_b^2 \bar{\Lambda}^2}[(4\chi_7+2\chi_8)(1-\omega^2) \nonumber\\
&& -4(\chi_{9}+\chi_{12})(1-\omega)]
 +\frac{1}{4m_c^2 \bar{\Lambda}^2}[2\chi_8(1-\omega^2)
 -(2\chi_9+2\chi_{10}-6\chi_{11} \nonumber\\
&& -2\chi_{12})(1-\omega)]   
  +\frac{1}{4m_b m_c \bar{\Lambda}^2}[\eta_1
  +2\eta_2-2\eta_3(1-\omega)
  +\eta_4+3\eta_5+2\eta_6\nonumber\\
&&  +2(\eta_9+\eta_{10})(1-\omega)]  
  -\frac{1}{4\bar{\Lambda}^2}[\frac{1}{m_b}(\kappa_1+3\kappa_2)
  +\frac{1}{m_c}
  (\kappa_1-\kappa_2)]^2\nonumber\\
&&  +\frac{1}{2m_b \bar{\Lambda}^2}\kappa_3(1-\omega) 
 [\frac{1}{m_b}(\kappa_1+3\kappa_2)+\frac{1}{m_c}(\kappa_1-\kappa_2)] ,\\
h_{A_{2}}&=&\frac{1}{m_c \bar{\Lambda}}\kappa_3 -\frac{1}{2m_c^2\bar{\Lambda}} c_3
  -\frac{1}{4m_c^2 \bar{\Lambda}^2}[4\chi_7(1+\omega)+2\chi_9-2\chi_{10} 
-10\chi_{11}-10\chi_{12}]\nonumber\\
&& +\frac{1}{4m_b m_c \bar{\Lambda}^2}[-2\eta_3+2\eta_4
   +(4\eta_7+2\eta_8)(1+\omega)-2\eta_9 -10\eta_{10}]\nonumber\\
&& -\frac{1}{2m_c \bar{\Lambda}^2}\kappa_3[\frac{1}{m_b}(\kappa_1+3\kappa_2)
  +\frac{1}{m_c}  (\kappa_1-\kappa_2)],\\
h_{A_{3}}&=&\xi+(\xi-1)[\frac{1}{2m_b \bar{\Lambda}}(\kappa_{1}+3\kappa_{2})
  +\frac{1}{2m_c \bar{\Lambda}} (\kappa_{1}-\kappa_{2}) +\frac{1}{4\bar{\Lambda}}
  (\MME{m_b}{m_c}{})c_1 \nonumber\\
&&  -\frac{1}{4\bar{\Lambda}}(\frac{3}{m_b^2}-\frac{1}{m_c^2}) c_2 
  -\frac{1}{4m_b^2 \bar{\Lambda}^2} c_4  
  -\frac{1}{4m_c^2 \bar{\Lambda}^2} c_5 ]  
  +\xi [-\frac{1}{8m_b^2 \bar{\Lambda}^2} c_6-\frac{1}{8m_c^2 \bar{\Lambda}^2} c_7  \nonumber\\
&&   +\frac{1}{8\bar{\Lambda}^2}(\frac{3}{m_b^2}(\kappa_1+3\kappa_2)^2   
  +\frac{3}{m_c^2}(\kappa_1-\kappa_2)^2
   +\frac{2}{m_b m_c}(\kappa_1+3\kappa_2)(\kappa_1-\kappa_2))]\nonumber\\
&& +\frac{1}{m_b \bar{\Lambda}}\kappa_3 (1-\omega)-\frac{1}{2\bar{\Lambda}}
  [\frac{1}{m_b^2}(1-\omega)-\frac{1}{m_c^2}] c_3
  -\frac{1}{m_c \bar{\Lambda}}\kappa_3+\frac{1}{4m_b^2 \bar{\Lambda}^2}[(4\chi_7+2\chi_8)\nonumber\\
&& \times (1-\omega^2)-4(\chi_{9}+\chi_{12})(1-\omega)]
  +\frac{1}{4m_c^2 \bar{\Lambda}^2}[4\chi_7(1+\omega)+2\chi_8(1-\omega^2)\nonumber\\
&& +2\chi_9\omega -2\chi_{10}(2-\omega)-2\chi_{11}(2+3\omega)-2\chi_{12}(4+\omega)]  
  +\frac{1}{4m_b m_c \bar{\Lambda}^2}[\eta_1  \nonumber\\
&& +2\eta_2+2\eta_3\omega -\eta_4+3\eta_5
  +2\eta_6-(4\eta_7+2\eta_8)(1+\omega)+2\eta_9(2-\omega)\nonumber\\
&& +2\eta_{10}(6-\omega)]
    -\frac{1}{4\bar{\Lambda}^2}[\frac{1}{m_b}(\kappa_1+3\kappa_2)+\frac{1}{m_c}
  (\kappa_1-\kappa_2)]^2 \nonumber  \\
&&   +\frac{1}{2\bar{\Lambda}^2}[\frac{1}{m_b}(1-\omega)-\frac{1}{m_c}]\kappa_{3}  
 [\frac{1}{m_b}(\kappa_1+3\kappa_{2})+\frac{1}{m_c}(\kappa_1-\kappa_{2})],
\end{eqnarray}
where
\begin{eqnarray}
  c_1&=&-\bar{\Lambda} \kappa_{1}+\varrho_{1}+\frac{1}{\bar{\Lambda}}\chi_{1}, \\
  c_2&=&\bar{\Lambda} \kappa_{2}-\varrho_{2}-\frac{1}{\bar{\Lambda}}\chi_{2} ,\\
  c_3&=&\bar{\Lambda} \kappa_{3}-\varrho_{3}-\frac{1}{\bar{\Lambda}}\chi_{3} ,\\
  c_4&=&3\chi_4+9\chi_5+6\chi_6 ,\\
  c_5&=&3\chi_4+\chi_5-2\chi_6 , \\
  c_6&=&\eta_1+6\eta_2-3\eta_4-9\eta_5-6\eta_6 ,\\
  c_7&=&\eta_1-2\eta_2-3\eta_4-\eta_5+2\eta_6 .
\end{eqnarray}



\newcommand{\PICL}[2]
{
\begin{picture}(170,170)(0,0)
\put(2,2){
\epsfxsize=6.5cm
\epsfysize=6.5cm
\epsffile{#1} }
\put(95,12){\makebox(0,0){#2}}
\end{picture}
}

\newcommand{\PICR}[2]
{
\begin{picture}(0,0)(0,0)
\put(240,19){
\epsfxsize=6.5cm
\epsfysize=6.5cm
\epsffile{#1} }
\put(330,27){\makebox(0,0){#2}}

\end{picture}
}

\newpage
\large{\centerline{FIGURES}}
\small
\mbox{}
{\vspace{1.4cm}}

\PICL{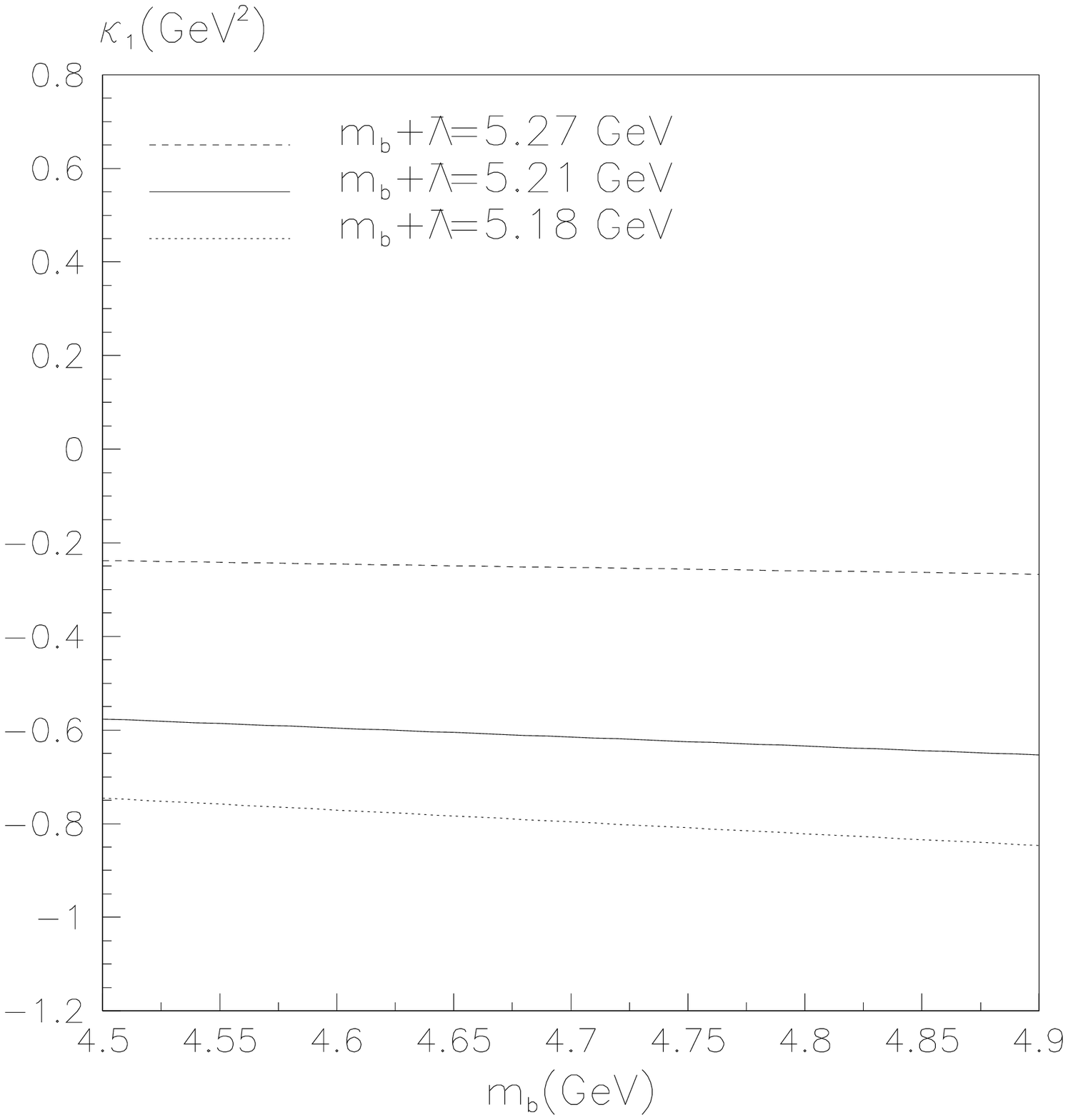}{(a)}

\PICR{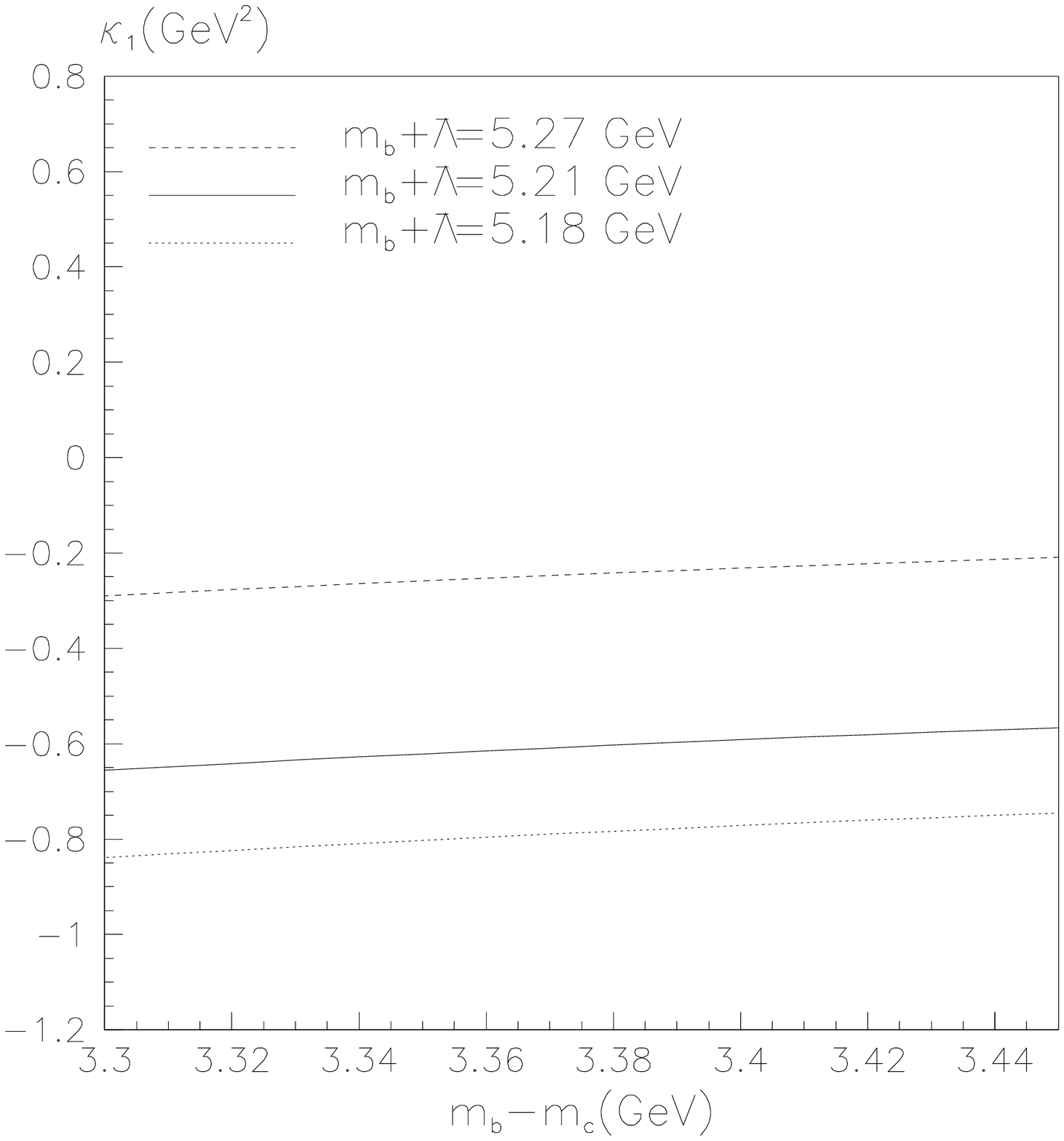}{(b)}

\PICL{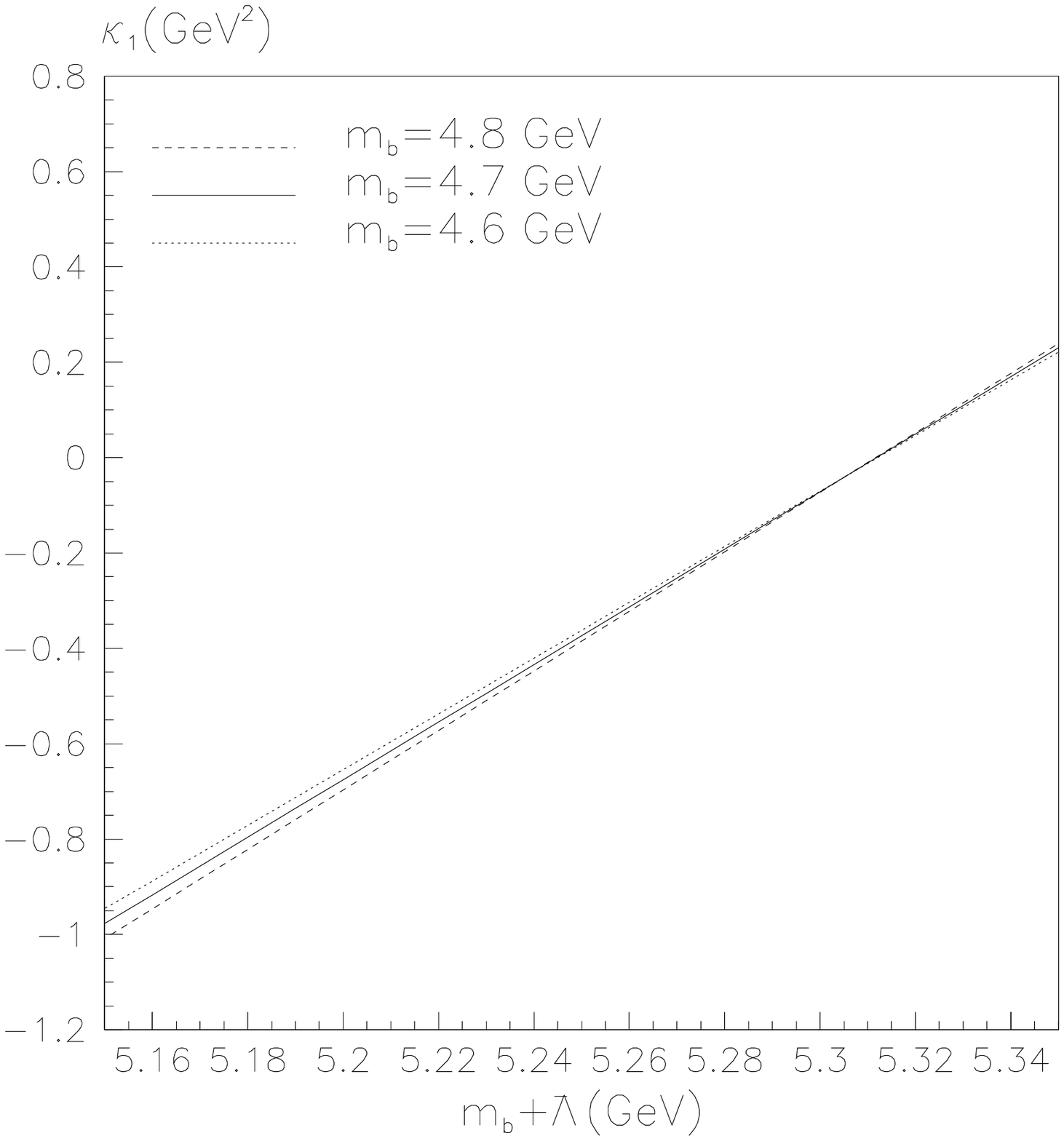}{(c)}

\PICR{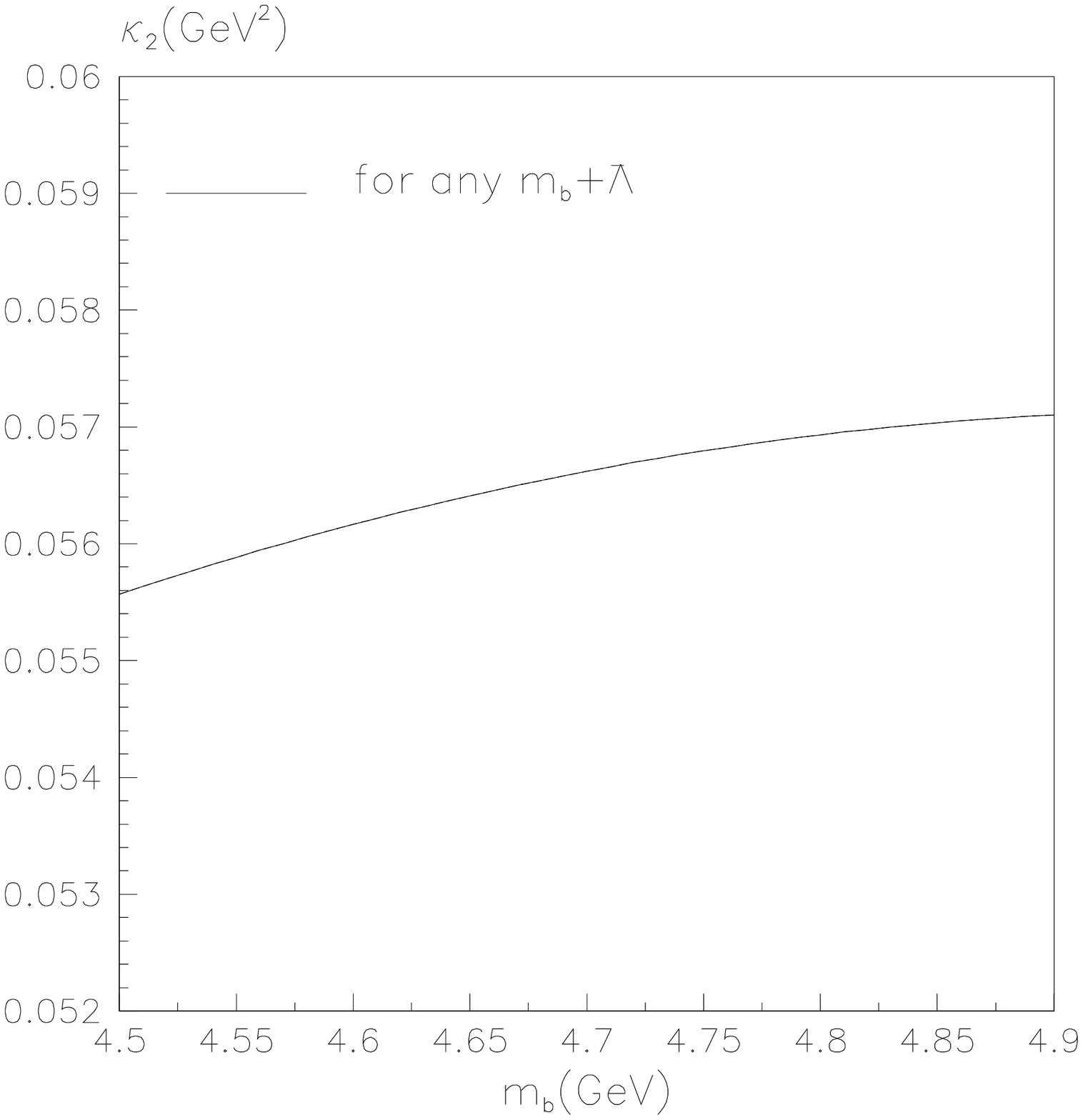}{(d)}

\PICL{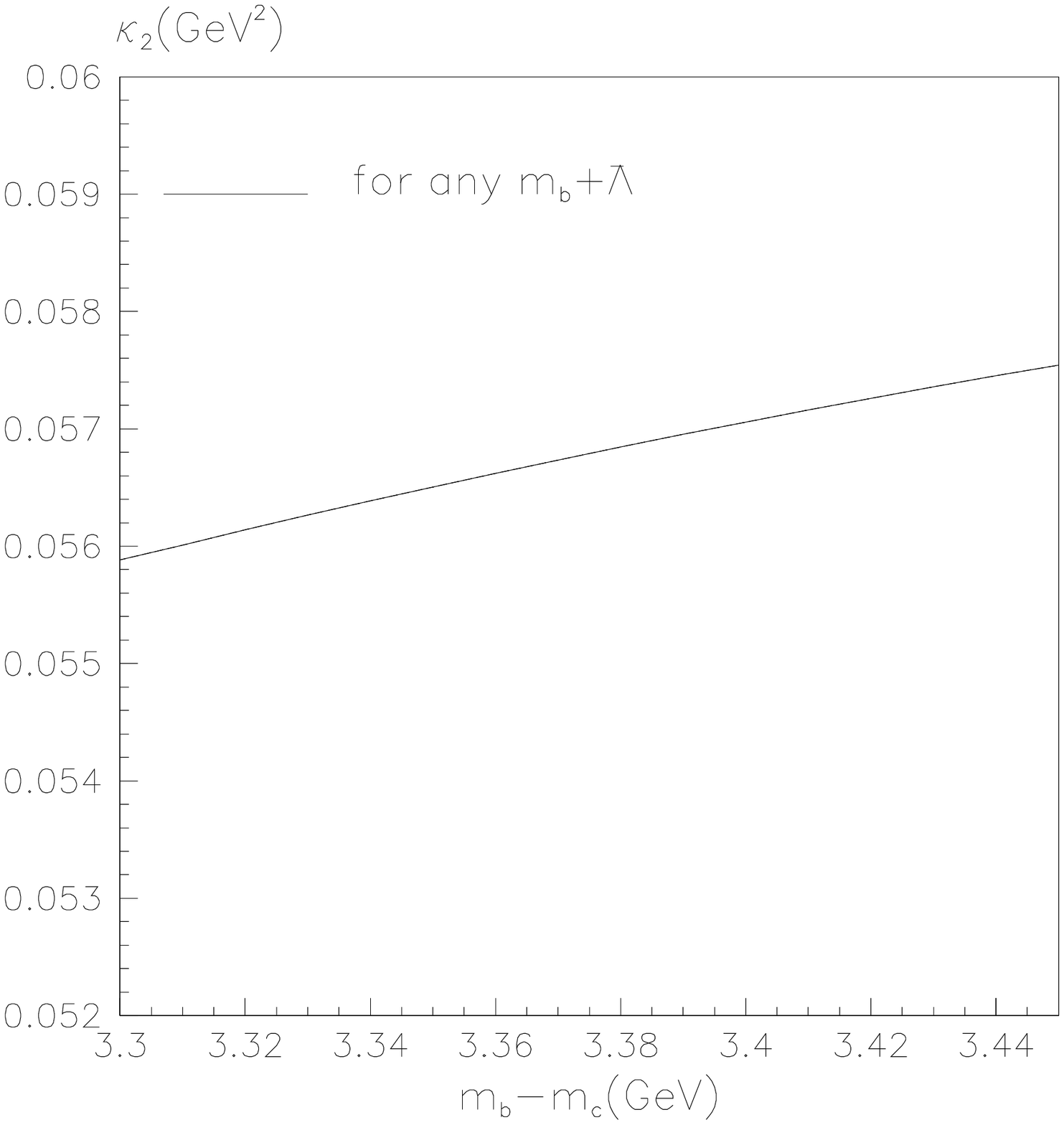}{(e)}

\PICR{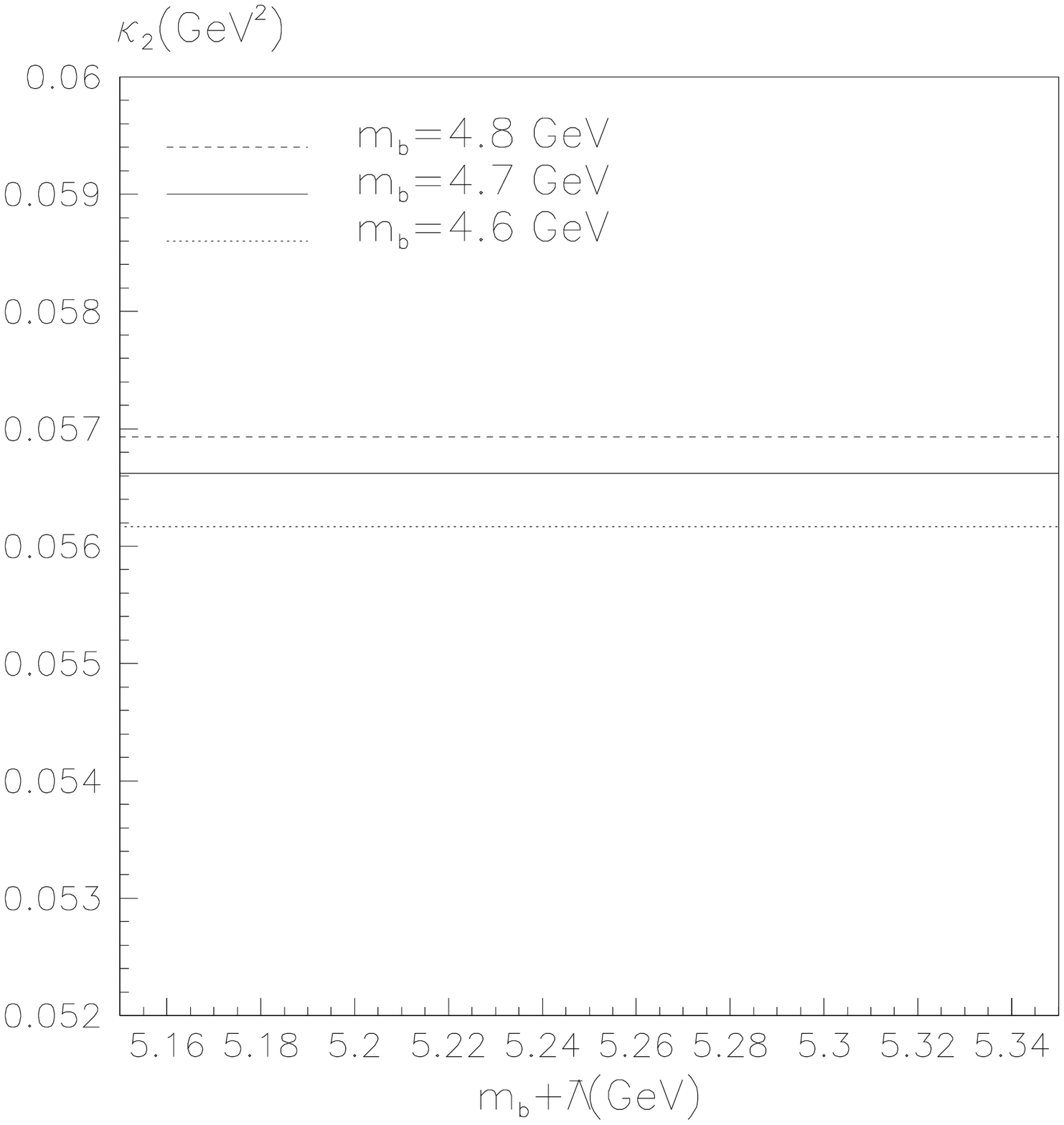}{(f)}

\newpage
\mbox{}
{\vspace{1.7cm}}

\PICL{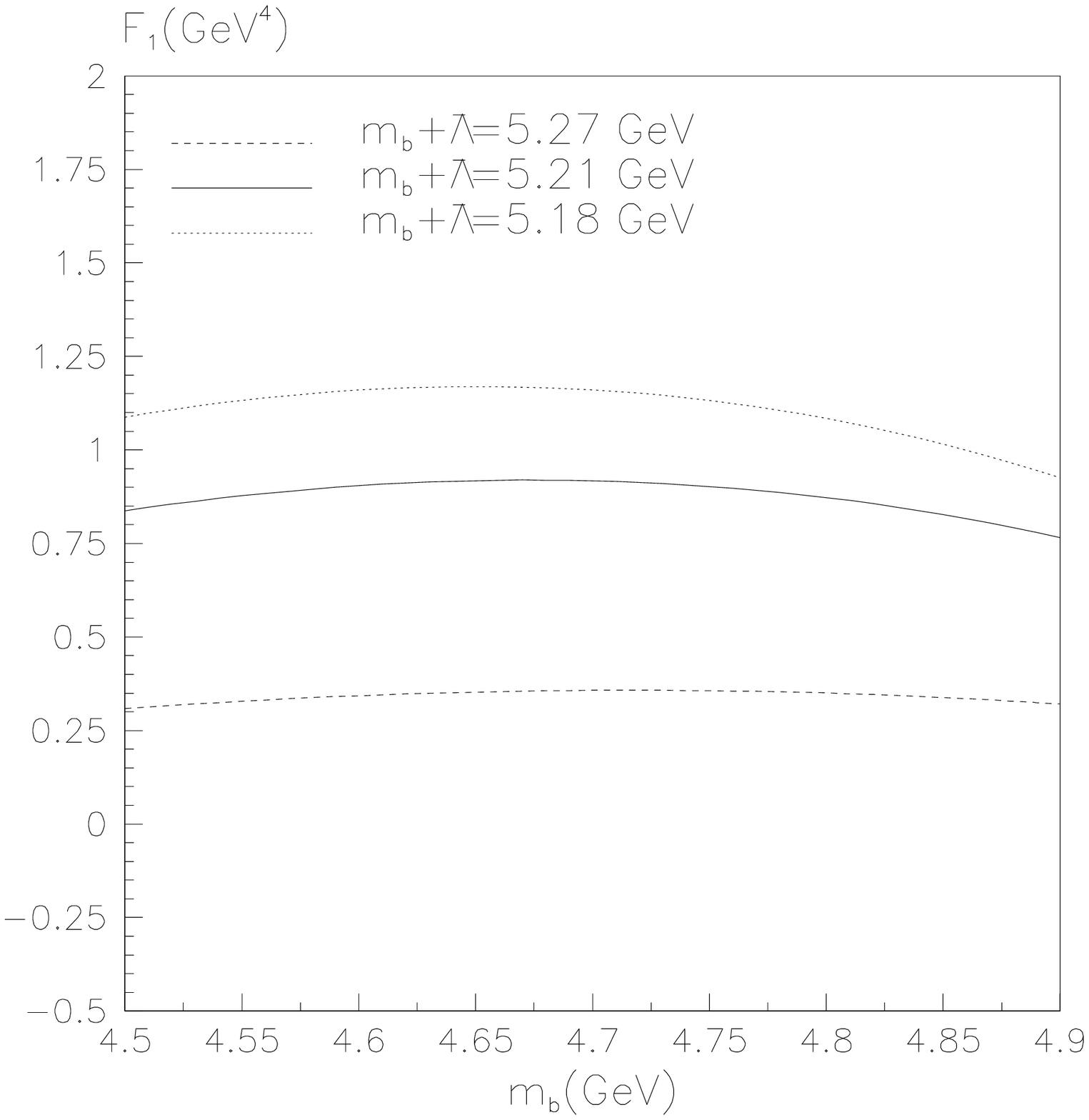}{(g)}

\PICR{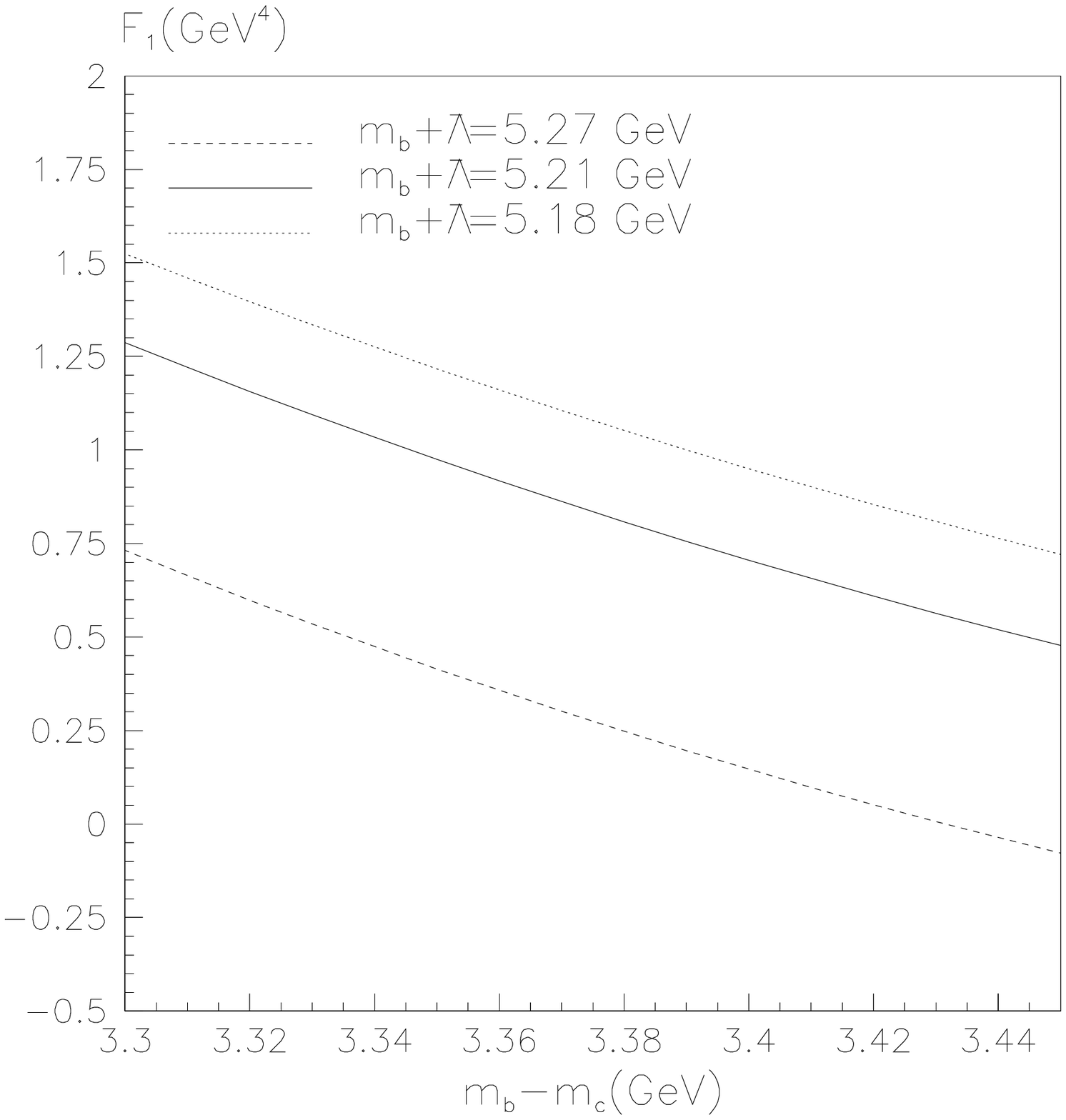}{(h)}

\PICL{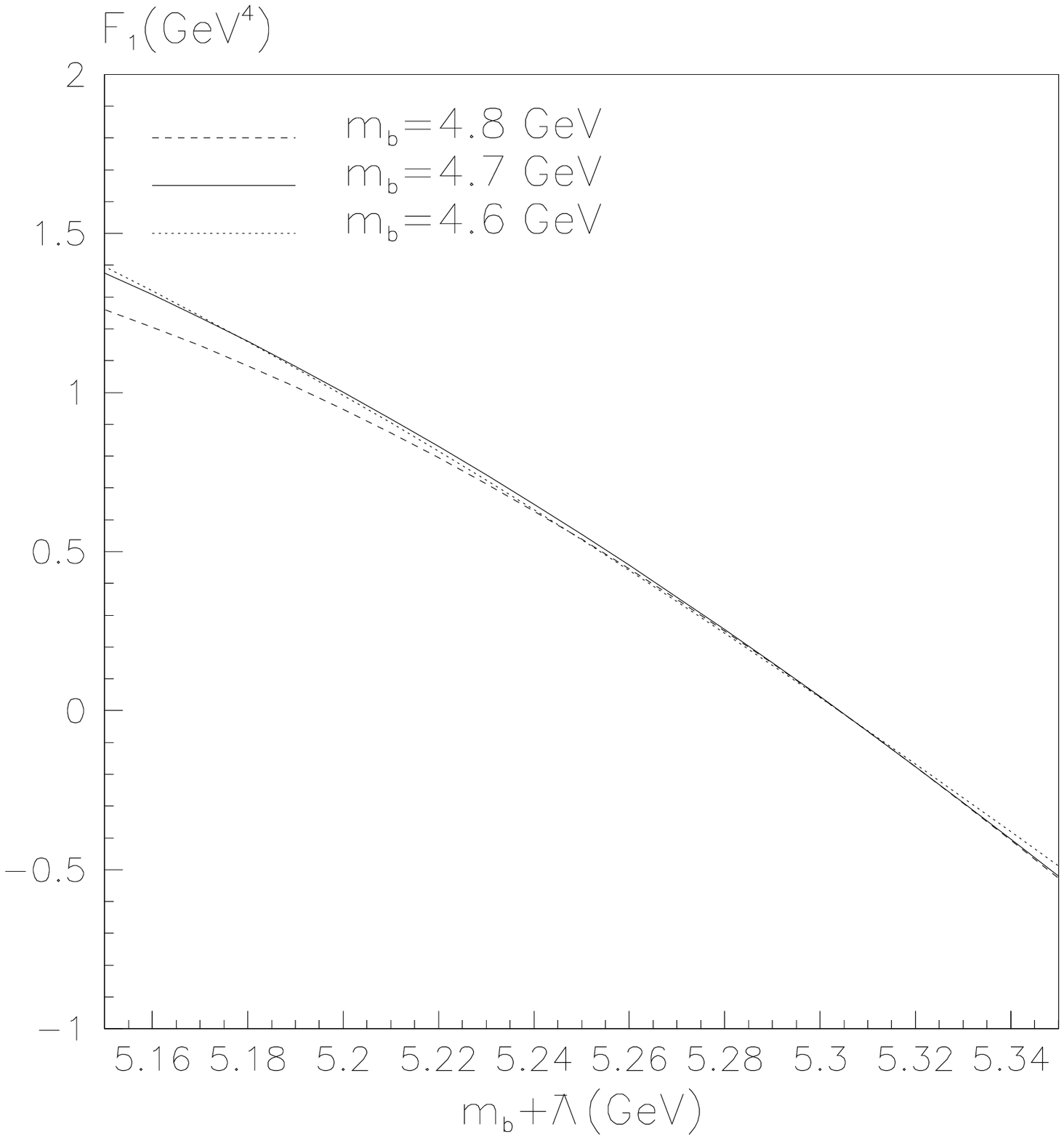}{(i)}

\PICR{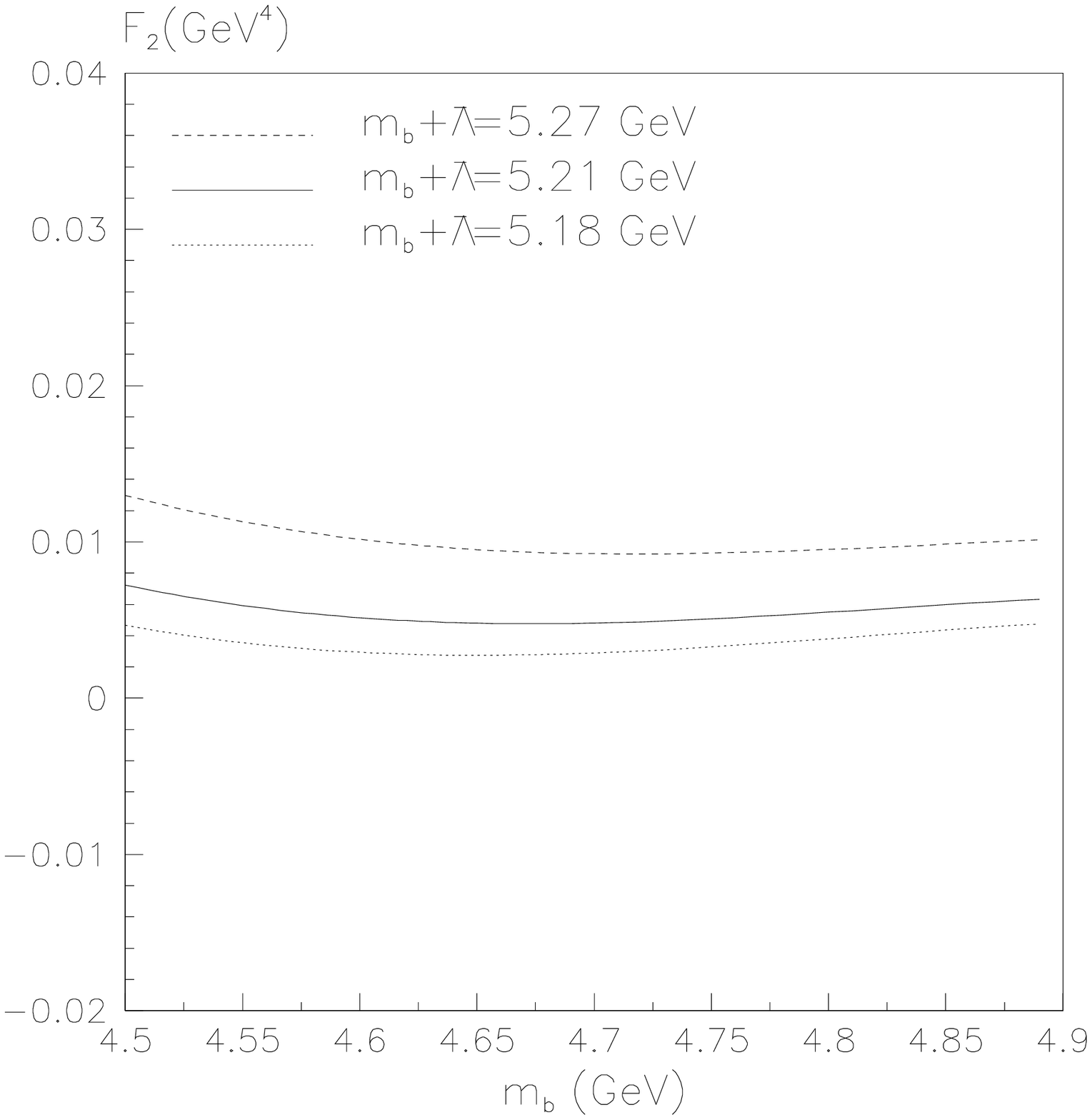}{(j)}

\PICL{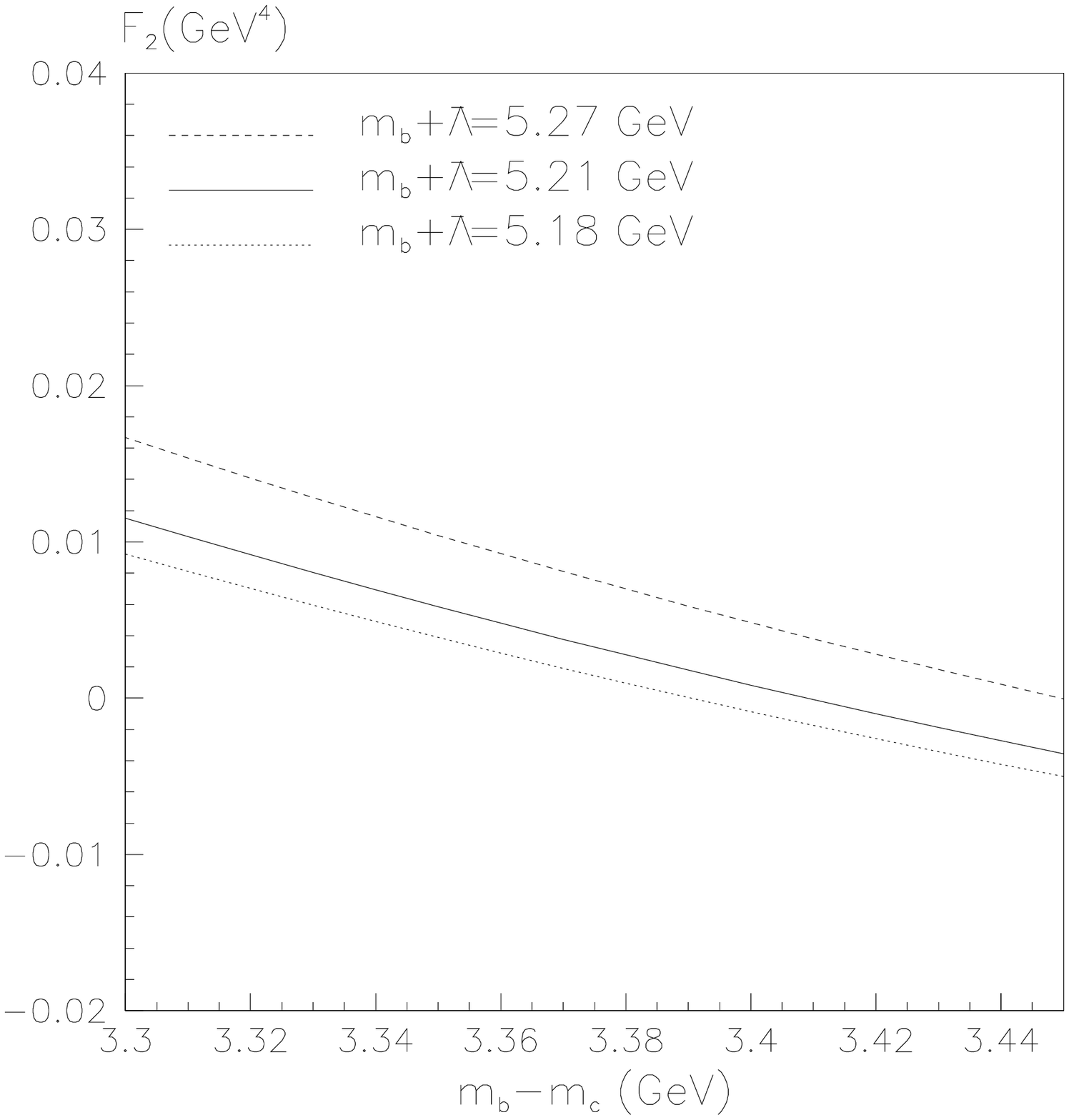}{(k)}

\PICR{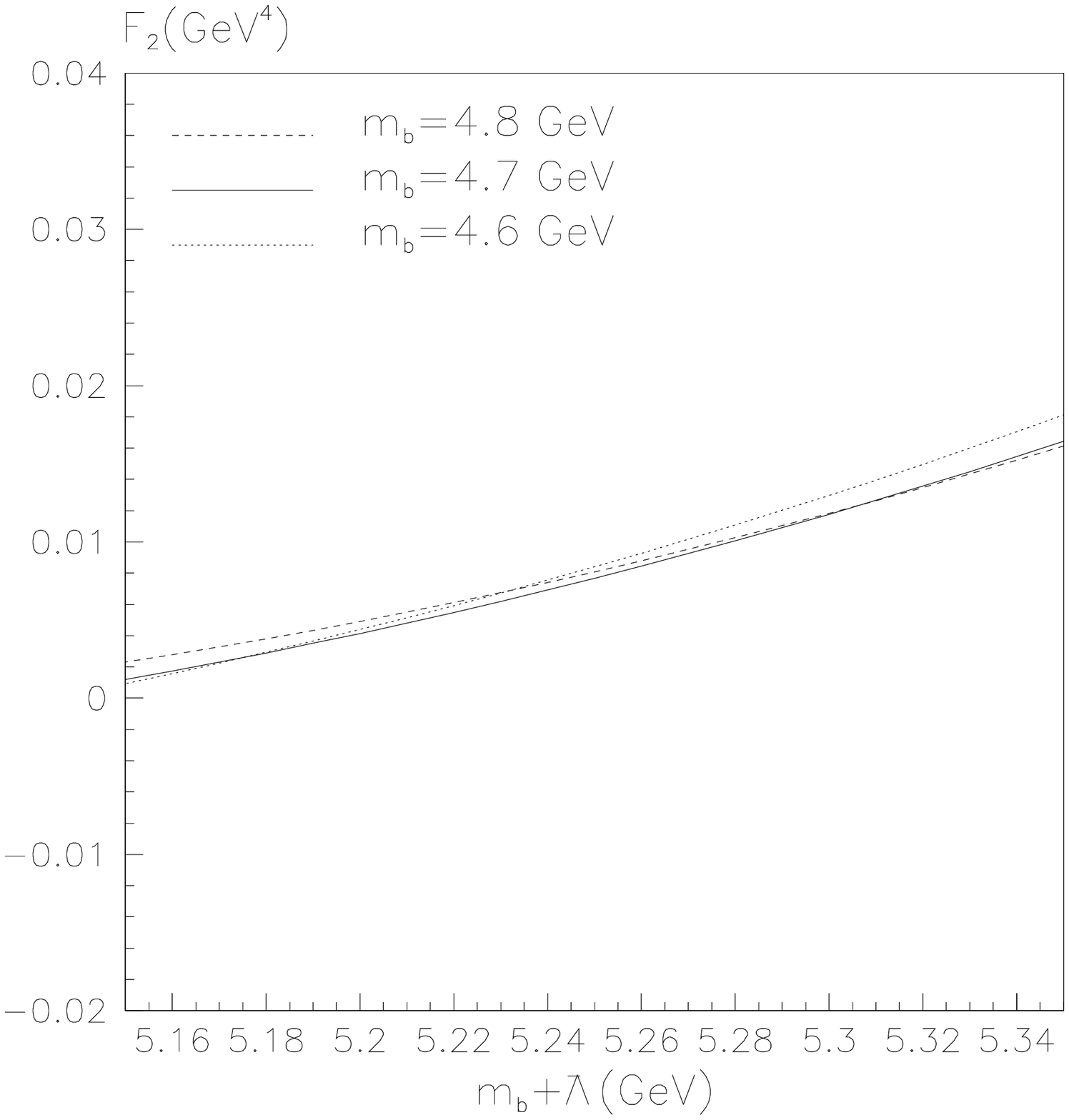}{(l)}

Fig.1 $\kappa_{1}$, $\kappa_{2}$, $F_1$ and $F_2$ as functions of $m_b$, 
  $m_b-m_c$ and $m_b+\bar{\Lambda}$ for 
  (a), (d), (g), (j): $m_b-m_c=3.36$GeV; (b), (e), (h), (k): $m_b=4.7$GeV; 
  and (c), (f), (i), (l): $m_b-m_c=3.36$GeV.

\newpage 
\mbox{}
{\vspace{1.7cm}}

\PICL{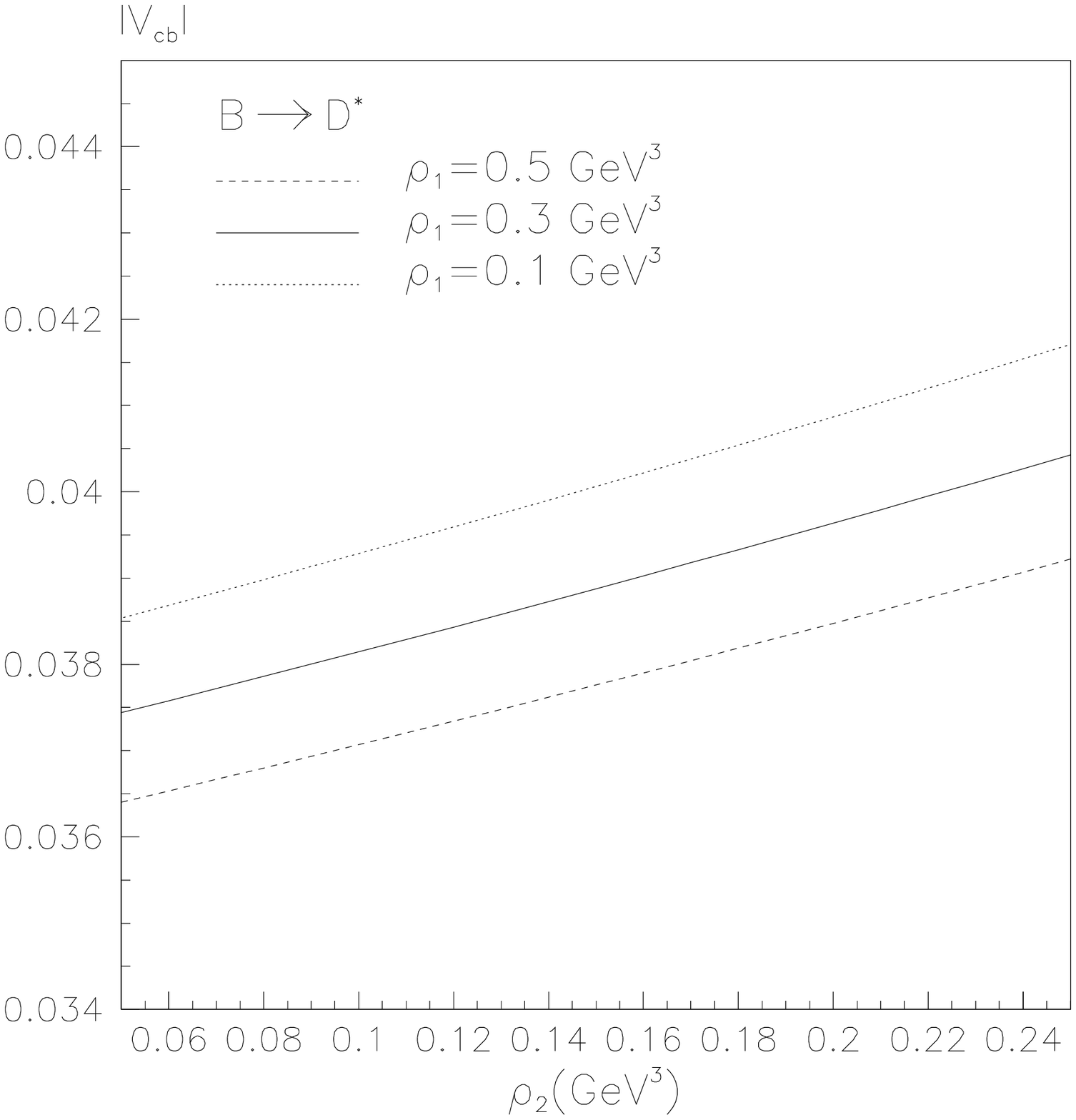}{(a)}

\PICR{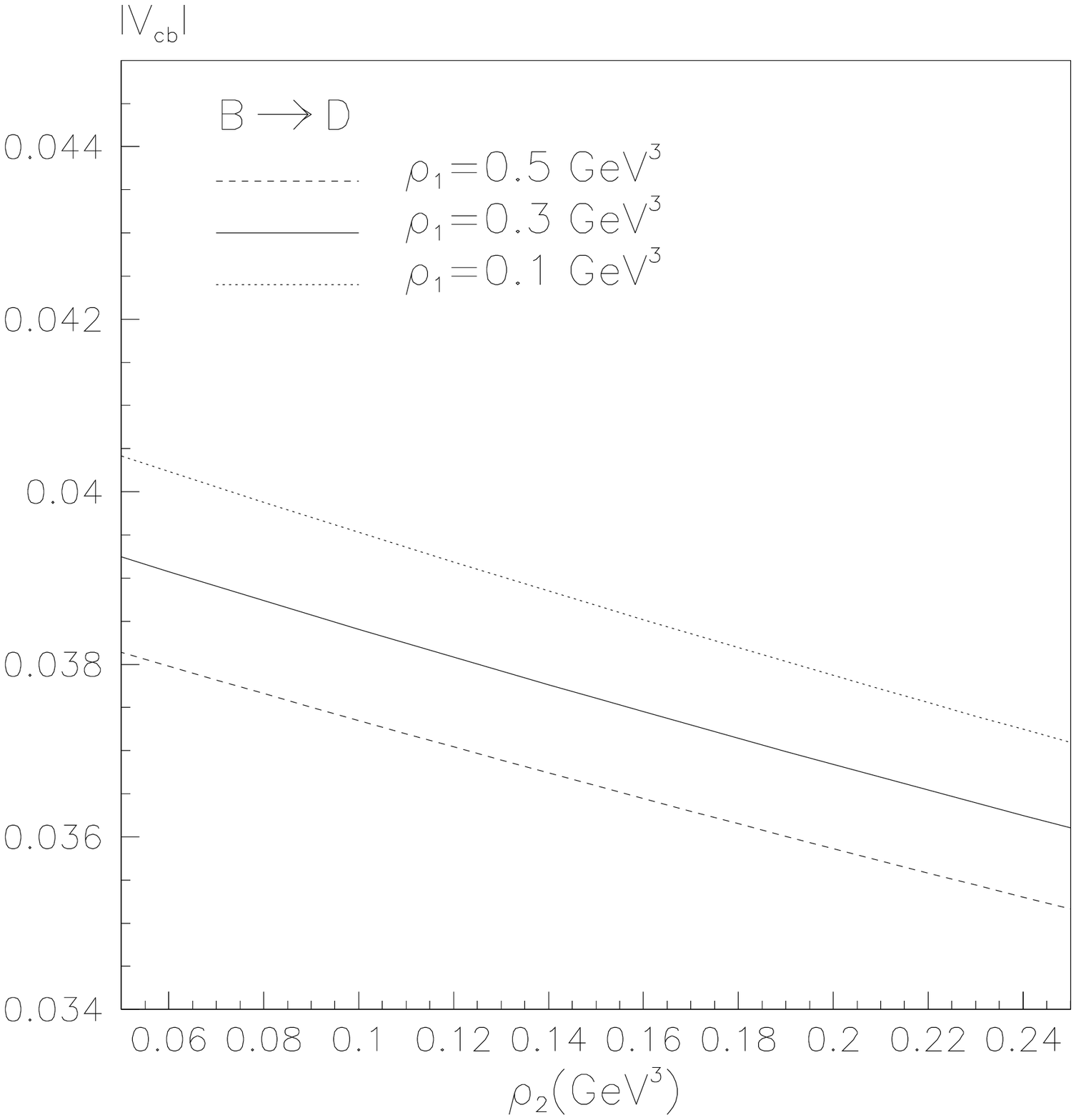}{(b)}

\PICL{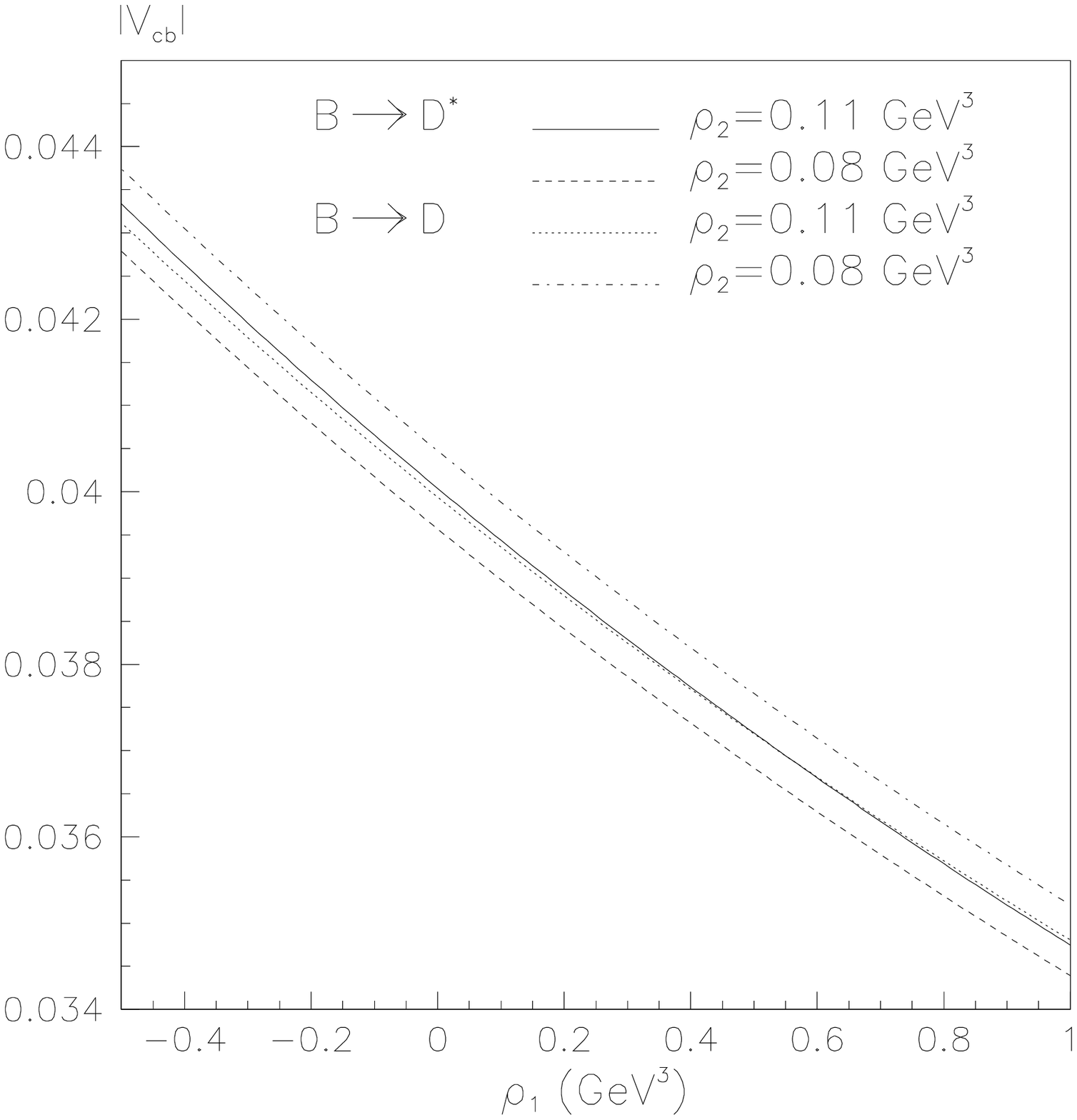}{(c)}

\PICR{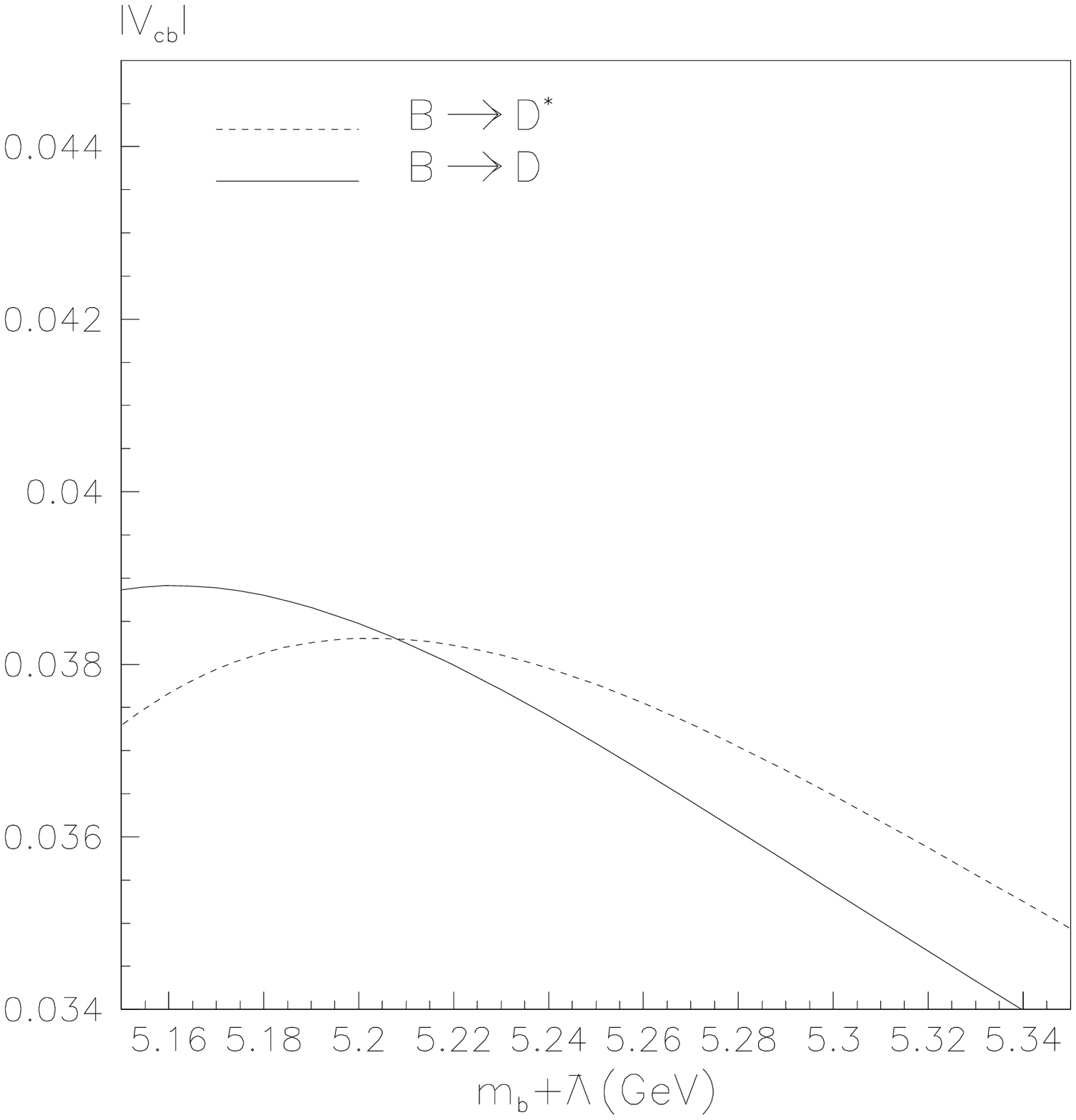}{(d)}

Fig.2 $\vert V_{cb}\vert$ extracted from the two channels $B\rightarrow D^{\ast}l\nu$ 
and $B\rightarrow Dl\nu$ as function of $\varrho_1$, $\varrho_2$ and $m_b+\bar{\Lambda}$. 
(a)$-$(c): $m_b=4.7\mbox{GeV}$, $m_b-m_c=3.36\mbox{GeV}$ and 
$m_b+\bar{\Lambda}=5.21\mbox{GeV}$; (d): $m_b=4.7\mbox{GeV}$, 
$m_b-m_c=3.36\mbox{GeV}$, $\varrho_1=0.3\mbox{GeV}^3$ and $\varrho_2=0.11
\mbox{GeV}^3$.

\newpage
\mbox{}
{\vspace{1.7cm}}

\PICL{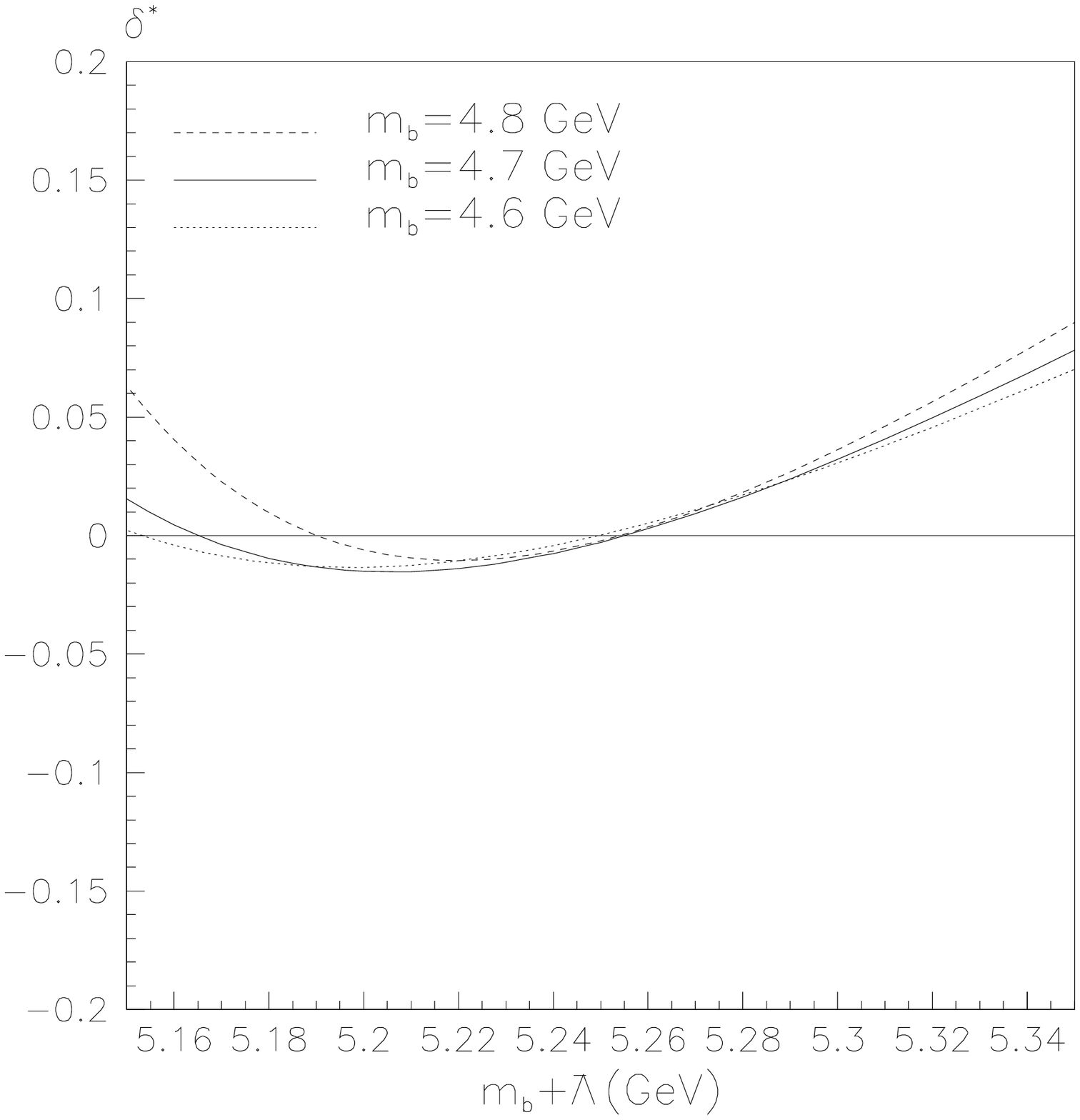}{(a)}

\PICR{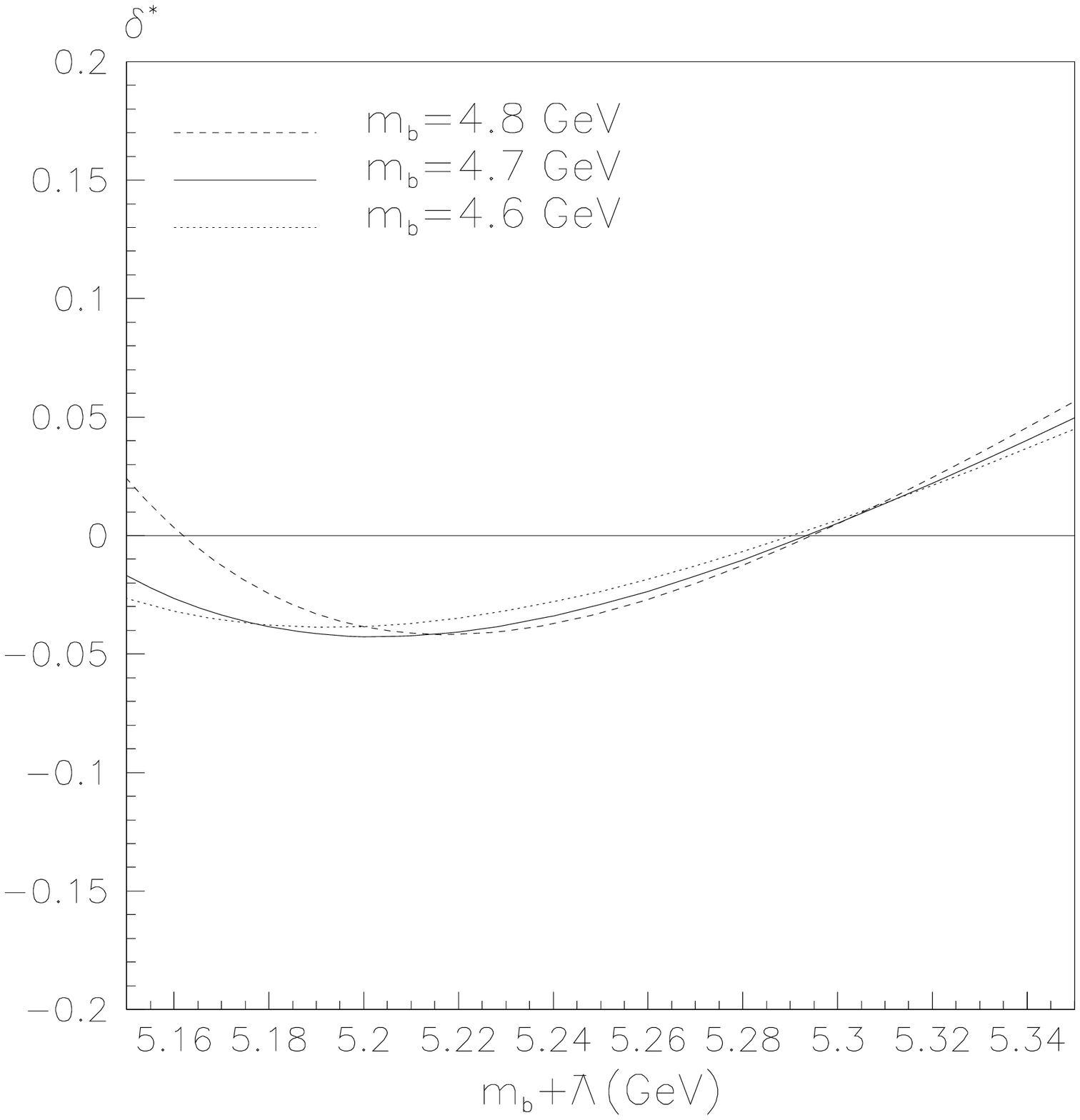}{(b)}

\PICL{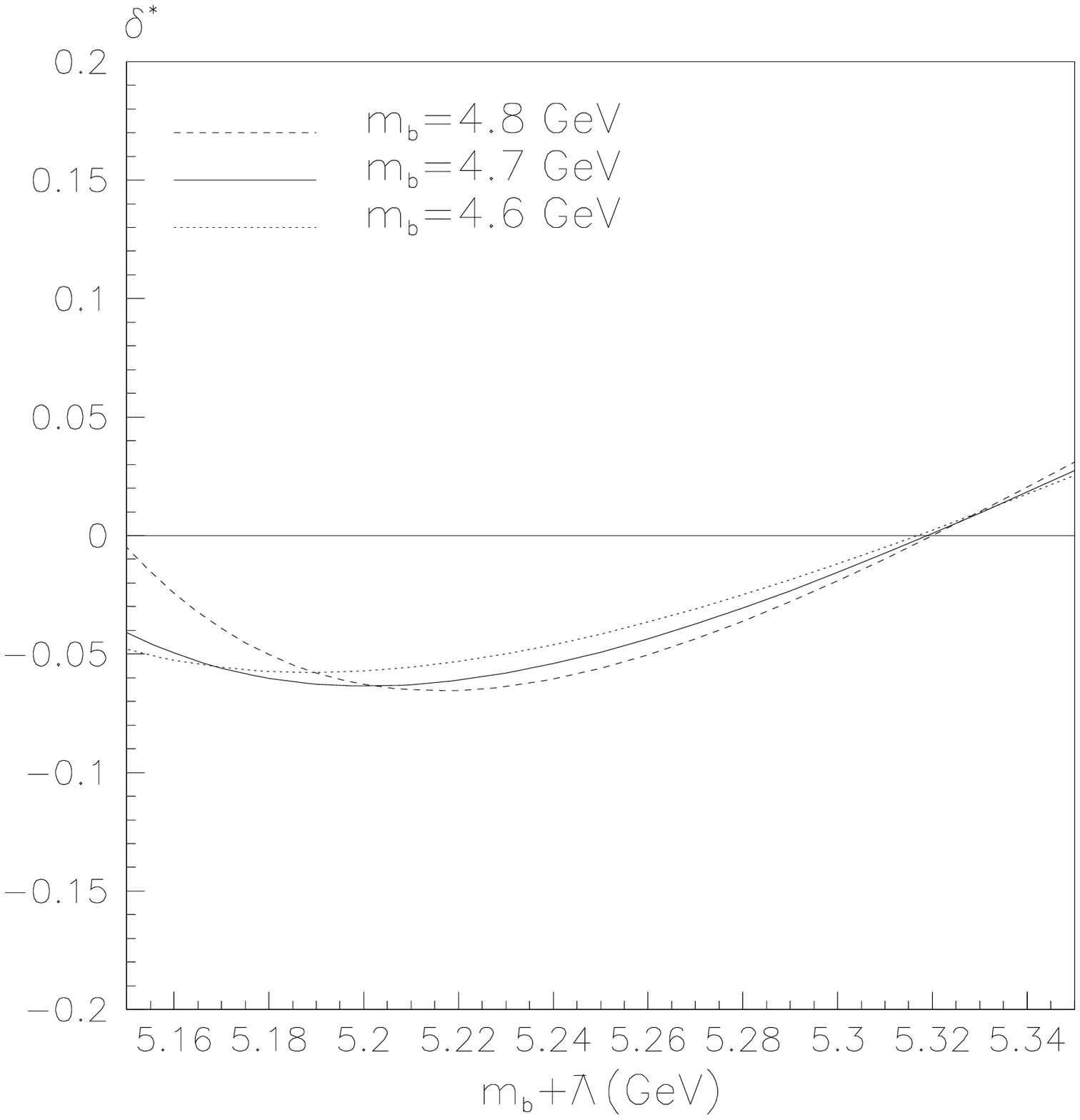}{(c)}

Fig.3 $\delta^{\ast}$ as function of $m_b+\bar{\Lambda}$ for 
(a): $m_b-m_c=3.41\mbox{GeV}$; (b): $m_b-m_c=3.36\mbox{GeV}$; 
(c): $m_b-m_c=3.32\mbox{GeV}$.

\newpage
\mbox{}
{\vspace{1.7cm}}

\PICL{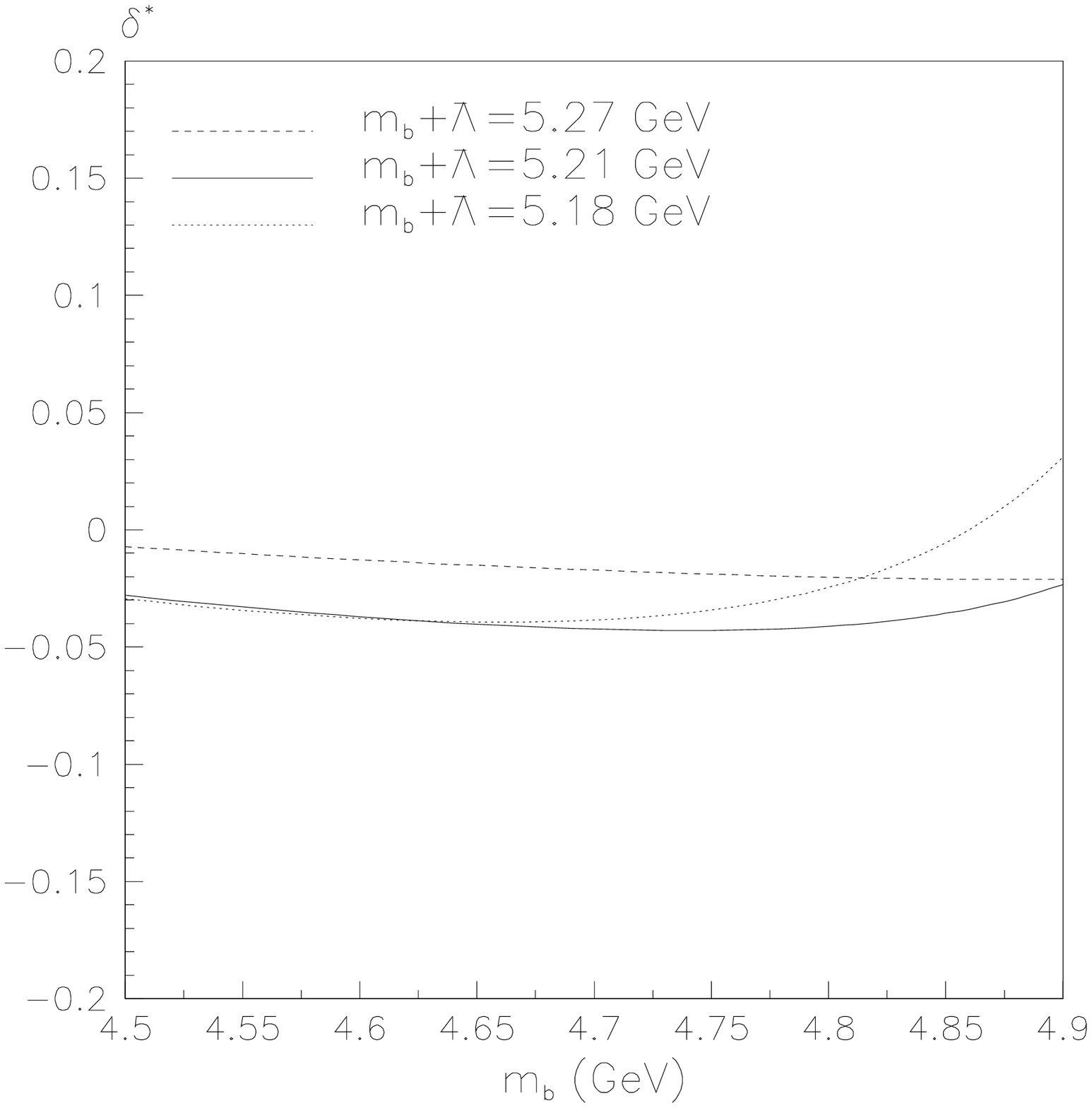}{(a)}

\PICR{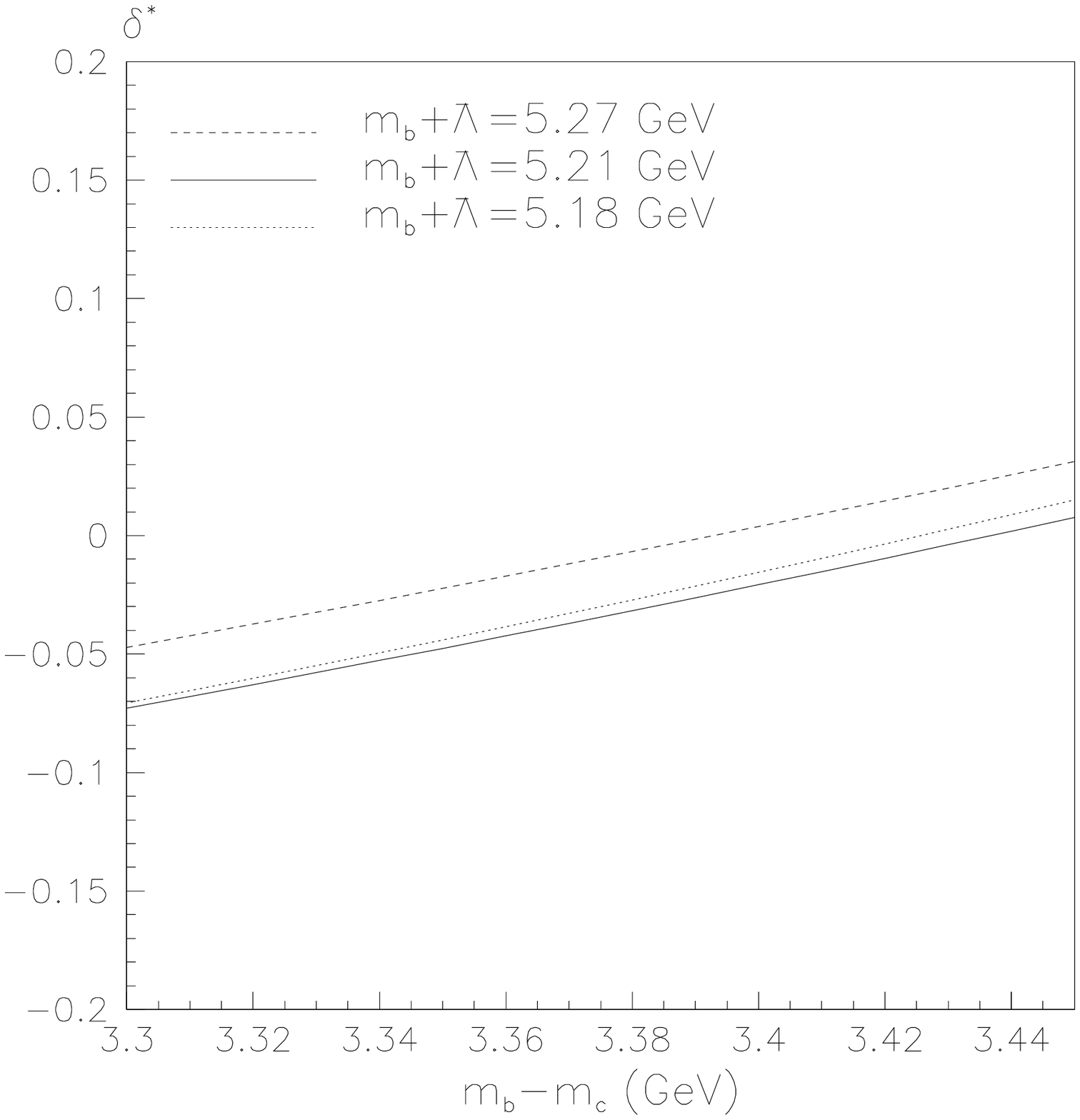}{(b)}

Fig.4 $\delta^{\ast}$ as function of $m_b$ and $m_b-m_c$ for 
(a): $m_b-m_c=3.36\mbox{GeV}$; (b): $m_b=4.7\mbox{GeV}$.

\newpage
\mbox{}
{\vspace{1.7cm}}

\PICL{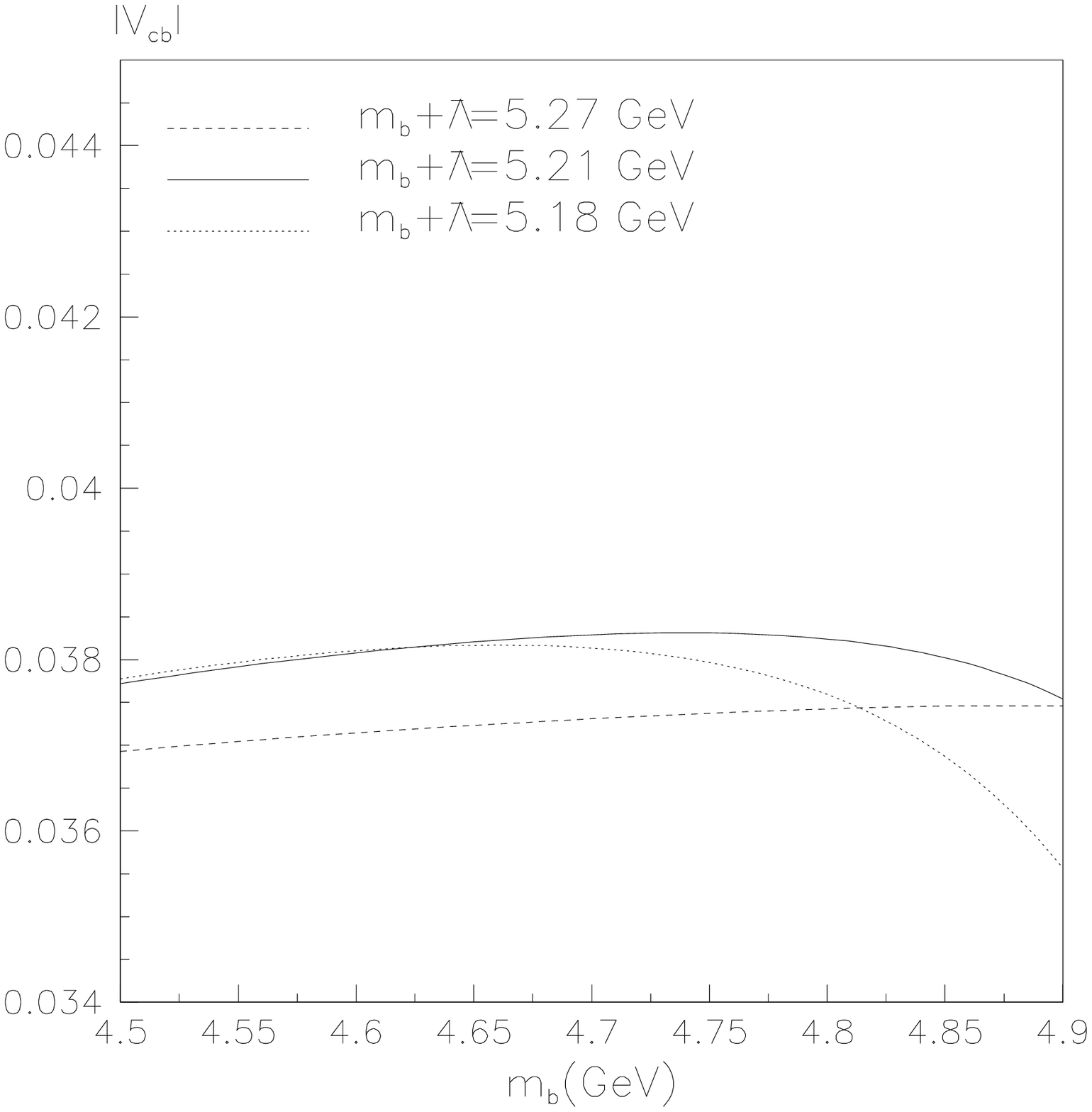}{(a)}

\PICR{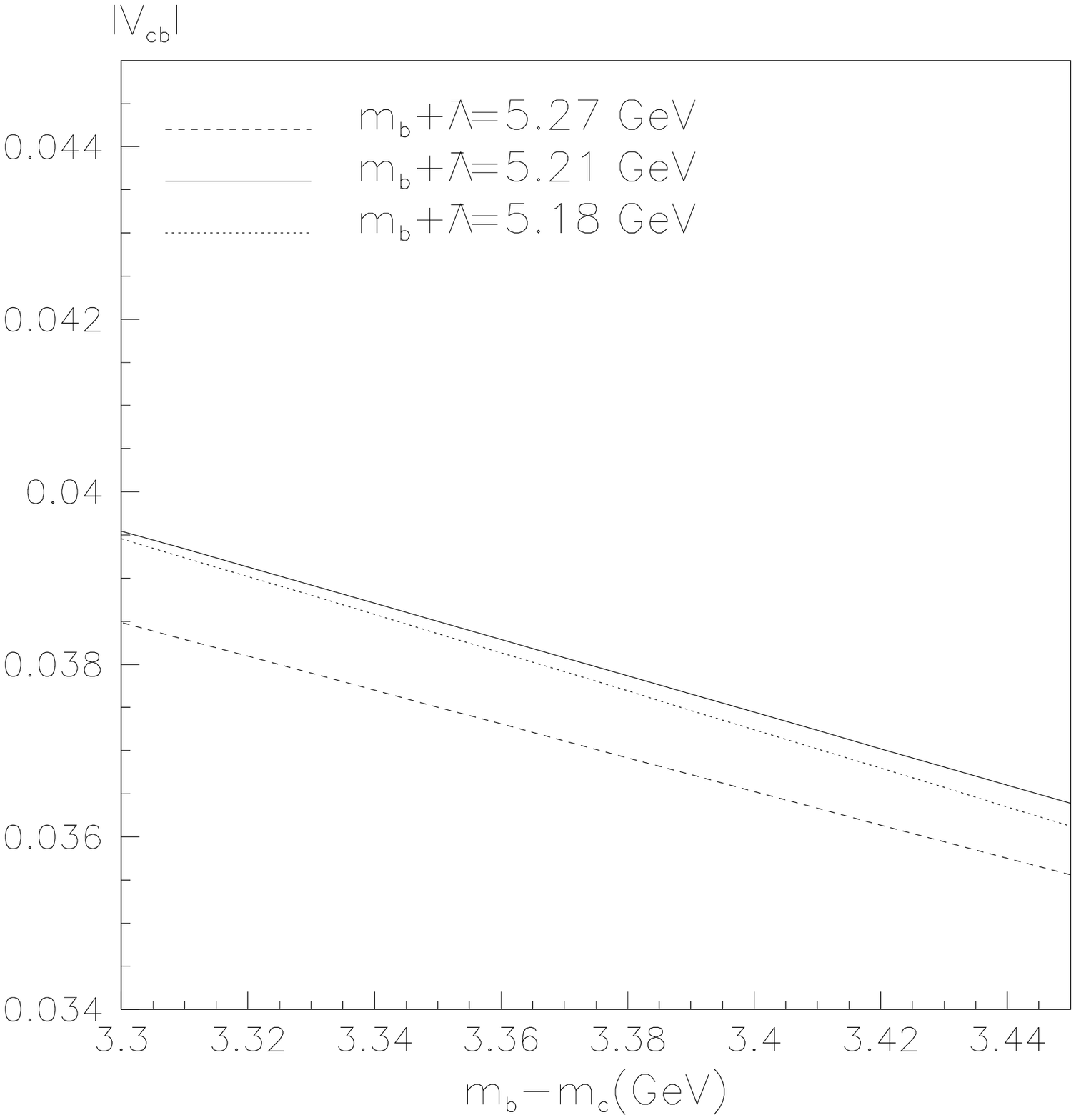}{(b)}

\PICL{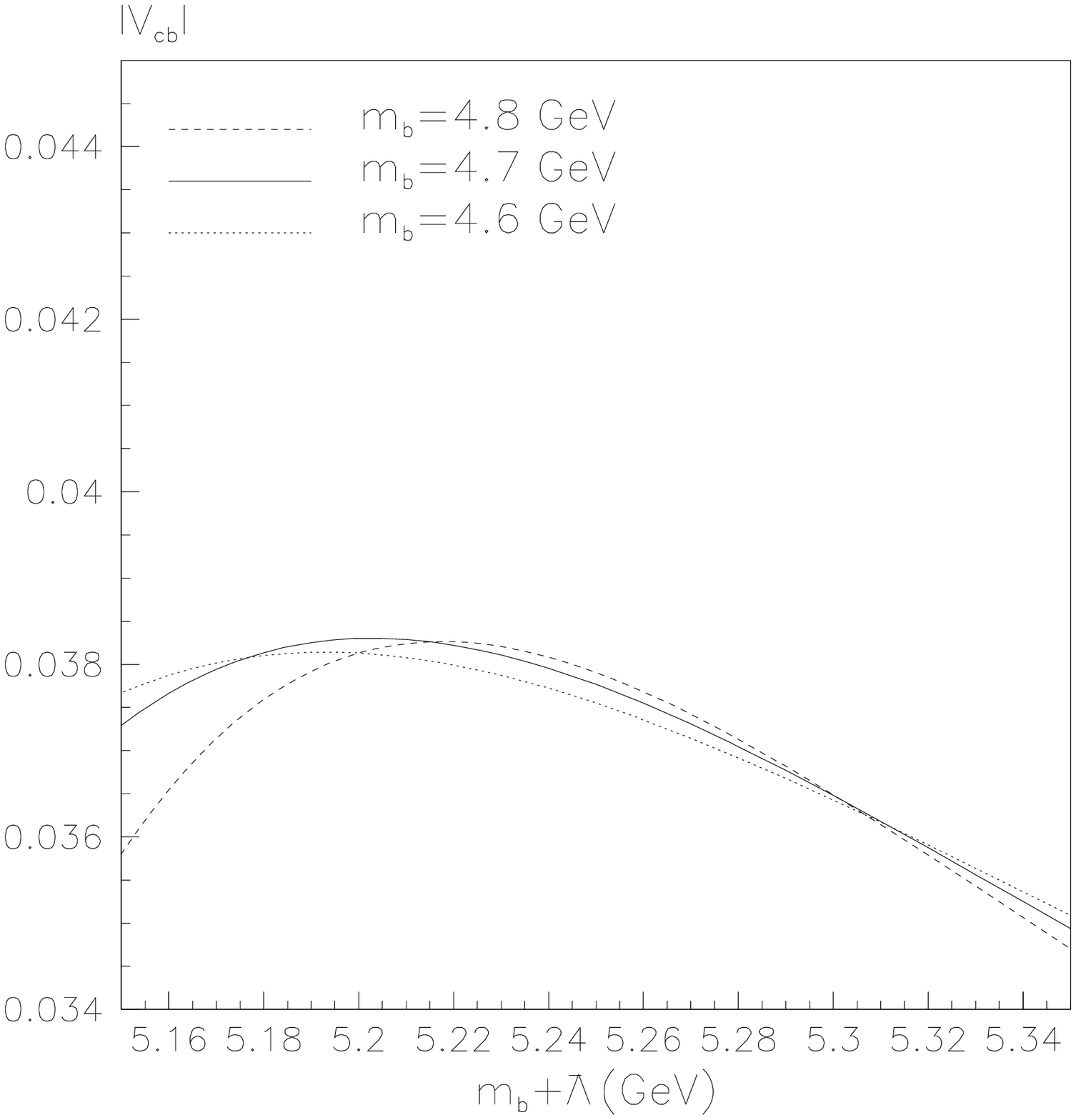}{(c)}

Fig.5 $\vert V_{cb} \vert$ extracted from $B\rightarrow 
D^{\ast}l\nu$ decay as function of $m_b$, $m_b-m_c$ and 
$m_b+\bar{\Lambda}$ for (a), (c): $m_b-m_c=3.36\mbox{GeV}$; 
(b): $m_b=4.7\mbox{GeV}$.

\newpage
\mbox{}
\vspace{2cm}

\centerline{\vbox{Table.1 Some results extracted when $m_b=4.6$GeV, $\varrho_1=0.3\mbox{GeV}^3$ 
and $\varrho_2=0.11\mbox{GeV}^3$. Here $\vert V_{cb}\vert_{BD^{\ast}}$ ($\vert V_{cb}\vert_{BD}$) 
refers to the value of $\vert V_{cb}\vert$ extracted from $B\rightarrow D^{\ast}l\nu$ 
($B\rightarrow Dl\nu$) decay. 
  } }
\vspace{0.2cm}
\centerline{
\begin{tabular}{|c|c|c|c|c|c|c|c|c|c|}
  \hline 
 $m_b-m_c$  &  $m_b+\bar{\Lambda}$ & $\kappa_1$& $\kappa_2$ & 
 $F_1$ & $F_2 $ & $\delta^{\ast}$ & $ \delta $ & $\vert V_{cb}\vert_{BD^{\ast}} $ 
 & $\vert V_{cb}\vert_{BD}$ \\
  \hline 
 3.32 &  5.18 &  -0.7954 &  0.0558 &  1.4042 &  
      0.0070 &  -0.0573 &  -0.0354 & 0.0389  & 0.0392  \\ 
  \cline{2-10} 
  &  5.20 &  -0.6778 &  0.0558 &  1.2356 &  
      0.0085 &  -0.0572 &  -0.0270 & 0.0389  & 0.0388  \\ 
  \cline{2-10} 
  &  5.21 &  -0.6190 &  0.0558 &  1.1480 &  
      0.0094 &  -0.0556 &  -0.0217 & 0.0388  & 0.0386  \\ 
  \cline{2-10} 
  &  5.24 &  -0.4426 &  0.0558 &  0.8730 &  
      0.0120 &  -0.0461 &  -0.0026 & 0.0384  & 0.0379  \\ 
  \cline{2-10} 
  &  5.27 &  -0.2662 &  0.0558 &  0.5811 &  
      0.0148 &  -0.0310 &  0.0202 & 0.0378  & 0.0370  \\ 
  \hline
 3.36 &  5.18 &  -0.7709 &  0.0562 &  1.1598 &  
      0.0029 &  -0.0377 &  -0.0097 & 0.0381  & 0.0381  \\ 
  \cline{2-10} 
  &  5.20 &  -0.6541 &  0.0562 &  0.9916 &  
      0.0044 &  -0.0384 &  -0.0016 & 0.0381  & 0.0378  \\ 
  \cline{2-10} 
  &  5.21 &  -0.5957 &  0.0562 &  0.9045 &  
      0.0051 &  -0.0371 &  0.0036 & 0.0381  & 0.0376  \\ 
  \cline{2-10} 
  &  5.24 &  -0.4205 &  0.0562 &  0.6315 &  
      0.0076 &  -0.0280 &  0.0228 & 0.0377  & 0.0369  \\ 
  \cline{2-10} 
  &  5.27 &  -0.2453 &  0.0562 &  0.3429 &  
      0.0102 &  -0.0128 &  0.0461 & 0.0371  & 0.0361  \\ 
 \hline
 3.41 &  5.18 &  -0.7442 &  0.0566 &  0.8931 &  
      -0.0015 &  -0.0116 &  0.0248 & 0.0371  & 0.0369  \\ 
  \cline{2-10} 
  &  5.20 &  -0.6284 &  0.0566 &  0.7266 &  
      -0.0002 &  -0.0134 &  0.0324 & 0.0372  & 0.0366  \\ 
  \cline{2-10} 
  &  5.21 &  -0.5705 &  0.0566 &  0.6406 &  
      0.0005 &  -0.0125 &  0.0374 & 0.0371  & 0.0364  \\ 
  \cline{2-10} 
  &  5.24 &  -0.3968 &  0.0566 &  0.3721 &  
      0.0027 &  -0.0042 &  0.0566 & 0.0368  & 0.0357  \\ 
  \cline{2-10} 
  &  5.27 &  -0.2231 &  0.0566 &  0.0894 &  
      0.0051 &  0.0109 &  0.0802 & 0.0363  & 0.0350  \\ 
\hline
\end{tabular}  }

\vspace{2cm}
\centerline{Table.2 Some results extracted when $m_b=4.7$GeV, $\varrho_1=0.3\mbox{GeV}^3$ and 
  $\varrho_2=0.11\mbox{GeV}^3$.}
\vspace{0.2cm}
\centerline{
\begin{tabular}{|c|c|c|c|c|c|c|c|c|c|}
  \hline 
 $m_b-m_c$  &  $m_b+\bar{\Lambda}$ & $\kappa_1$& $\kappa_2$ & 
 $F_1$ & $F_2 $ & $\delta^{\ast}$ & $ \delta $ & $\vert V_{cb}\vert_{BD^{\ast}} $ 
 & $\vert V_{cb}\vert_{BD}$ \\
 \hline 
 3.32 &  5.18 &  -0.8238 &  0.0561 &  1.3967 &  
      0.0070 &  -0.0603 &  -0.0545 & 0.0390  & 0.0399  \\ 
  \cline{2-10} 
  &  5.20 &  -0.7022 &  0.0561 &  1.2398 &  
      0.0084 &  -0.0635 &  -0.0459 & 0.0392  & 0.0396  \\ 
  \cline{2-10} 
  &  5.21 &  -0.6414 &  0.0561 &  1.1567 &  
      0.0092 &  -0.0629 &  -0.0400 & 0.0391  & 0.0393  \\ 
  \cline{2-10} 
  &  5.24 &  -0.4590 &  0.0561 &  0.8901 &  
      0.0115 &  -0.0541 &  -0.0179 & 0.0388  & 0.0385  \\ 
  \cline{2-10} 
  &  5.27 &  -0.2766 &  0.0561 &  0.5987 &  
      0.0141 &  -0.0374 &  0.0090 & 0.0381  & 0.0374  \\ 
 \hline
 3.36 &  5.18 &  -0.7962 &  0.0566 &  1.1605 &  
      0.0029 &  -0.0385 &  -0.0266 & 0.0381  & 0.0388  \\ 
  \cline{2-10} 
  &  5.20 &  -0.6754 &  0.0566 &  1.0014 &  
      0.0041 &  -0.0427 &  -0.0182 & 0.0383  & 0.0385  \\ 
  \cline{2-10} 
  &  5.21 &  -0.6150 &  0.0566 &  0.9175 &  
      0.0048 &  -0.0423 &  -0.0125 & 0.0383  & 0.0382  \\ 
  \cline{2-10} 
  &  5.24 &  -0.4338 &  0.0566 &  0.6491 &  
      0.0069 &  -0.0339 &  0.0098 & 0.0380  & 0.0374  \\ 
  \cline{2-10} 
  &  5.27 &  -0.2526 &  0.0566 &  0.3572 &  
      0.0092 &  -0.0172 &  0.0372 & 0.0373  & 0.0364  \\ 
 \hline
 3.41 &  5.18 &  -0.7660 &  0.0572 &  0.9015 &  
      -0.0017 &  -0.0096 &  0.0106 & 0.0370  & 0.0374  \\ 
  \cline{2-10} 
  &  5.20 &  -0.6462 &  0.0572 &  0.7411 &  
      -0.0007 &  -0.0151 &  0.0184 & 0.0372  & 0.0371  \\ 
  \cline{2-10} 
  &  5.21 &  -0.5863 &  0.0572 &  0.6568 &  
      -0.0001 &  -0.0152 &  0.0240 & 0.0372  & 0.0369  \\ 
  \cline{2-10} 
  &  5.24 &  -0.4066 &  0.0572 &  0.3884 &  
      0.0018 &  -0.0076 &  0.0464 & 0.0369  & 0.0361  \\ 
  \cline{2-10} 
  &  5.27 &  -0.2269 &  0.0572 &  0.0981 &  
      0.0038 &  0.0092 &  0.0745 & 0.0363  & 0.0352  \\ 
\hline
\end{tabular}
}

\newpage
\mbox{}
\vspace{2cm}

\centerline{Table.3 Some results extracted when $m_b=4.8$GeV, $\varrho_1=0.3\mbox{GeV}^3$ and 
  $\varrho_2=0.11\mbox{GeV}^3$.}
\vspace{0.2cm}
\centerline{
\begin{tabular}{|c|c|c|c|c|c|c|c|c|c|}
  \hline 
 $m_b-m_c$  &  $m_b+\bar{\Lambda}$ & $\kappa_1$& $\kappa_2$ & 
 $F_1$ & $F_2 $ & $\delta^{\ast}$ & $ \delta $ & $\vert V_{cb}\vert_{BD^{\ast}} $ 
 & $\vert V_{cb}\vert_{BD}$ \\
 \hline 
 3.32 &  5.18 &  -0.8524 &  0.0564 &  1.3013 &  
      0.0077 &  -0.0502 &  -0.0716 & 0.0386  & 0.0407  \\ 
  \cline{2-10} 
  &  5.20 &  -0.7268 &  0.0564 &  1.1690 &  
      0.0091 &  -0.0628 &  -0.0650 & 0.0391  & 0.0404  \\ 
  \cline{2-10} 
  &  5.21 &  -0.6640 &  0.0564 &  1.0969 &  
      0.0097 &  -0.0651 &  -0.0592 & 0.0392  & 0.0401  \\ 
  \cline{2-10} 
  &  5.24 &  -0.4756 &  0.0564 &  0.8574 &  
      0.0119 &  -0.0606 &  -0.0345 & 0.0390  & 0.0391  \\ 
  \cline{2-10} 
  &  5.27 &  -0.2872 &  0.0564 &  0.5851 &  
      0.0144 &  -0.0437 &  -0.0027 & 0.0383  & 0.0379  \\ 
  \hline
 3.36 &  5.18 &  -0.8214 &  0.0569 &  1.0843 &  
      0.0038 &  -0.0246 &  -0.0396 & 0.0376  & 0.0393  \\ 
  \cline{2-10} 
  &  5.20 &  -0.6966 &  0.0569 &  0.9469 &  
      0.0049 &  -0.0384 &  -0.0334 & 0.0381  & 0.0391  \\ 
  \cline{2-10} 
  &  5.21 &  -0.6342 &  0.0569 &  0.8725 &  
      0.0055 &  -0.0412 &  -0.0277 & 0.0382  & 0.0388  \\ 
  \cline{2-10} 
  &  5.24 &  -0.4470 &  0.0569 &  0.6271 &  
      0.0074 &  -0.0372 &  -0.0028 & 0.0381  & 0.0379  \\ 
  \cline{2-10} 
  &  5.27 &  -0.2598 &  0.0569 &  0.3502 &  
      0.0095 &  -0.0202 &  0.0296 & 0.0374  & 0.0367  \\ 
 \hline
 3.41 &  5.18 &  -0.7873 &  0.0576 &  0.8453 &  
      -0.0007 &  0.0096 &  0.0030 & 0.0363  & 0.0377  \\ 
  \cline{2-10} 
  &  5.20 &  -0.6635 &  0.0576 &  0.7033 &  
      0.0002 &  -0.0061 &  0.0084 & 0.0369  & 0.0375  \\ 
  \cline{2-10} 
  &  5.21 &  -0.6016 &  0.0576 &  0.6268 &  
      0.0007 &  -0.0096 &  0.0139 & 0.0370  & 0.0373  \\ 
  \cline{2-10} 
  &  5.24 &  -0.4159 &  0.0576 &  0.3764 &  
      0.0022 &  -0.0066 &  0.0389 & 0.0369  & 0.0364  \\ 
  \cline{2-10} 
  &  5.27 &  -0.2302 &  0.0576 &  0.0961 &  
      0.0040 &  0.0104 &  0.0720 & 0.0363  & 0.0352  \\ 
\hline
\end{tabular}
}

\end{document}